\documentclass[twoside]{article}
\usepackage{epsfig}
\begin{document}
\setlength{\textheight}{8.0truein}    

\centerline{\bf ON THE TRANSPORT OF ATOMIC IONS}
\vspace*{0.035truein}
\centerline{\bf  IN LINEAR AND MULTIDIMENSIONAL ION TRAP ARRAYS 
} \vspace*{0.37truein}  \centerline{\footnotesize
D. HUCUL\footnote{Electronic mail: dhucul@mit.edu}\, , M. YEO, S.
OLMSCHENK, C. MONROE} \vspace*{0.015truein}
\centerline{\footnotesize\it FOCUS Center and Department of Physics,
University of Michigan, Ann Arbor, Michigan, 48109}
\baselineskip=10pt \vspace*{10pt} \centerline{\footnotesize W.K.
HENSINGER} \vspace*{0.015truein} \centerline{\footnotesize\it
Department of Physics and Astronomy, University of Sussex}
\baselineskip=10pt \centerline{\footnotesize\it Falmer, Brighton,
East Sussex, BN1 9QH, UK} \baselineskip=10pt \vspace*{10pt}
\centerline{\footnotesize J. RABCHUK} \vspace*{0.015truein}
\centerline{\footnotesize\it Department of Physics, Western Illinois
University} \baselineskip=10pt \centerline{\footnotesize\it Macomb,
Illinois 61455} \vspace*{0.225truein} \centerline{\today}

\vspace*{0.21truein}

\abstract{Trapped atomic ions have become one of the most promising
architectures for a quantum computer, and current effort is now
devoted to the transport of trapped ions through complex segmented
ion trap structures in order to scale up to much larger numbers of
trapped ion qubits.  This paper covers several important issues
relevant to ion transport in any type of complex multidimensional rf
(Paul) ion trap array.  We develop a general theoretical framework
for the application of time-dependent electric fields to shuttle
laser-cooled ions along any desired trajectory, and describe a
method for determining the effect of arbitrary shuttling schedules
on the quantum state of trapped ion motion.  In addition to the
general case of linear shuttling over short distances, we introduce
issues particular to the shuttling through multidimensional
junctions, which are required for the arbitrary control of the
positions of large arrays of trapped ions.  This includes the
transport of ions around a corner, through a cross or T junction,
and the swapping of positions of multiple ions in a laser-cooled
crystal.  Where possible, we make connections to recent experimental
results in a multidimensional T junction trap, where arbitrary
2-dimensional transport was realized.}

\vspace*{10pt}

\vspace*{1pt}

\section{\label{Intro}Introduction}
Trapped ion systems serve as a promising direction toward realizing
an operational quantum computer \cite{cirac:1995}-\cite{T paper}.
Many experiments in ion trap systems have been performed to show
entanglement \cite{turchette:1998}-\cite{steane:2006}, fundamental
logic gates \cite{steane:2006}-\cite{haljan:2005}, and teleportation
\cite{barrett:2004, riebe:2004}. Algorithms have even been performed
on a small number of trapped ions
\cite{gulde:2003}-\cite{reichle:2006}. One of the remaining challenges toward
realizing a useful quantum information processor is that of scaling
up these proof-of-principle experiments.

One proposal for scaling up a trapped ion quantum computer is to
create an integrated array of linear rf Paul ion traps, divided into
regions for storage and entanglement. Such a device would carry out
logical operations by generating two-particle entanglement between
any pair of ions by shuttling the ions out from storage into the
entanglement zones, and bringing them back into storage as required
for the completion of the algorithm \cite{NIST bible, monroe
nature}. This quantum computing architecture requires arbitrary
two-dimensional control of trapped ions that may consist of four key
protocols: linear shuttling, corner shuttling, separation and
recombination. These key protocols may be combined to produce other
necessary operations such as a swapping protocol to switch the
positions of two trapped ions \cite{T paper}.

The process of shuttling ions from a storage region to an
entanglement region and back requires sophisticated, accurate and
detailed knowledge of the time-dependent electric fields in order to
control the ions' dynamics in the trap arrays. For trap arrays
containing many ions, the cost of calculating the necessary electric
fields for each intermediate set of voltages during a shuttling
operation is prohibitive. An alternative approach is to develop a
set of numerically-obtained ``basis functions," that represent the
contribution to the electric potential seen by the ion due to a unit
voltage applied to each of the dc electrodes in the trap array, the
others being held at zero voltage. The electric potential produced
by an arbitrary set of voltages on the electrodes is calculated by
multiplying the basis function for each electrode with the actual
applied voltage, and then adding up the corresponding potentials at
all points in space.

In order to shuttle ions in an array of linear ion traps, the
control voltages are varied in time and the basis function technique
is used to calculate the potential as a function of time. To choose
the appropriate methods to simulate the ions' motion in the trap, it
is important to determine the purpose of the simulation to be
carried out. Typically we will be interested in moving ions between
points inside the array successfully while minimizing the kinetic
energy that the ion acquires during the shuttling process. This can
be simulated by solving the classical equations of motion using the
calculated potential. The question arises whether there are
important corrections if one considers the full quantum evolution of
the system. Berman and Zaslavsky \cite{berman:1978} showed that the
breakdown of quantum-classical correspondence occurs on a time scale
at which the quantum wave function spreads sufficiently over a
macroscopic part of phase space to feel anharmonicities in the
potential. This is because the quantum evolution of the Wigner
function may be expressed as the sum of the Poisson bracket
(describing classical evolution) and quantum correction terms that
contain higher order spatial potential derivatives
\cite{karkuszewski:2002}. These quantum corrections will be
negligible if the ion is shuttled adiabatically (or such that it
remains in the Lamb Dicke regime) as the ion remains close to the
bottom of the well and the potential may be approximated well as a
harmonic potential. Quantum corrections may become important if the
ion samples anharmonic parts of the potential. In that case we
expect quantum corrections to be important if the shuttling process
occurs on timescales that are of order
$t_\hbar=\frac{1}{\lambda}\ln\left(\frac{A}{\hbar}\right)$
 where $\lambda$ is the Lyapunov exponent
for the dynamic evolution of the system and $A$ is the action of
motion \cite{karkuszewski:2002}. Nevertheless, it may be that
corrections in the calculated electric potential due to the finite
accuracy of the numerical solver will weigh stronger than the
appearance of quantum-classical divergence. We also point out that
the quantum bit of a single ion is always encoded in the internal
state of the ion, and we may only require the ion to remain inside
the Lamb-Dicke regime after the shuttling process in order to allow
the execution of further quantum gate operations.  Preserving the
actual motional quantum state of the ion during the shuttling
process is therefore not likely to be a criterion for the
development of shuttling protocols that move ions between
interaction and entanglement zones. Finally the ion may also be
cooled via sympathetic cooling
\cite{larson:1986}-\cite{demarco:2003} after the shuttling
operation. Indeed such cooling may also accommodate shuttling
operations that fail to confine ions within the Lamb-Dicke regime.
Therefore the primary function of the simulation is to provide a
highly reliable transport protocol of the ion through the
complicated potential inside the array.

This paper is organized in the following way.  In the next section,
we first discuss the derivation of the electric field inside an ion
trap.  We then consider the numerical calculation of the resultant
classical motion of an ion in this field.  In section 3, by
determining the quantum mechanical state of the ion after shuttling,
we derive constraints and figures of merit that may be used to
design and characterize shuttling protocols.  In section 4, we
compare and contrast salient features of various two-dimensional ion
trap architectures, paying particular attention to the junction
regions. In section 5, using the T-junction ion trap array as a case
study, we consider the practical design and implementation of key
ion shuttling protocols.  This culminates in the swapping of two
ions in a linear chain.  In section 6, we briefly consider ion
transport and storage in a 3 dimensional array and present
conclusions in section 7.


\section{\label{proof}Simulation of Trapped Ion Dynamics Via Basis Functions}
\subsection{Justification of the Basis Function Technique}
It is possible to simulate the potential in any complex, multi-zone
ion trap by developing electric potential basis functions for a given trap geometry. The
electric potential for any arbitrary voltage configuration of the
trap electrodes can then be built up as a linear combination of the
basis functions. The electric potential of any arbitrary charge
configuration with Dirichlet boundary conditions can be written as
\cite{Jackson}:

\begin{equation}\label{eq:potential}
\Phi ( {\bf x}) = \frac{1}{4\pi \epsilon _{0}}\int_{V}
    \rho ({\bf x}\,') G({\bf x},{\bf x},') d\,^{3}{\bf x}\,' -
\frac{1}{4\pi}\oint_{S}
    \Phi ({\bf x}\,')\frac{\partial G({\bf x},{\bf x}\,')}{\partial n'}da'
\end{equation}

In Eq. \ref{eq:potential}, the first integral is an integral over
the volume interior to the boundary with the appropriate symmetric
Green function \begin{math}G({\bf x},{\bf x}\,')\end{math}. Inside
of an empty ion trap, there is no free charge so
\begin{math}\rho ({\bf x}\,')=0\end{math} making the first term of Eq.
\ref{eq:potential} zero. The second integral is an integral over the
surface of each electrode \begin{math} \Phi ({\bf x}')
\end{math} multiplied by the outward normal derivative of the Green function
with respect to the surface \begin{math} n'\end{math}. It is
possible to write the potential that is specified on every trap
electrode as a sum of potentials on each individual electrode with all other electrodes held at ground.

\begin{equation}\label{eq:surfaces}
\Phi ({\bf x}') = \sum_i \Phi_i ( {\bf x}\,')
\end{equation}
This changes Eq. \ref{eq:potential} to

\begin{eqnarray}
\label{eq:sumint} \Phi( {\bf x}) =  -\frac{1}{4\pi}\sum_i
\oint_{S_{i}} \Phi_i ({\bf x},') \frac{\partial G_i({\bf x},{\bf
x}\,')}{\partial n_{i}'}da'
\end{eqnarray}

As can be seen in Eq. \ref{eq:sumint}, the total electric potential
\begin{math}\Phi (\mathbf x) \end{math} is a sum of the
potentials produced by each electrode surface individually when all
other electrodes and boundaries are held at zero potential, as is expected
from the linear nature of Laplace's equation. Since
the voltage is constant over each electrode surface, we can rewrite
Eq. \ref{eq:sumint} as a sum of the constant voltage $V_i$ times the
surface integral only for electrode  {\em i} in the trap.

\begin{equation}
\label{eq:basiseqn} \Phi( \mathbf x) = \sum_i
\frac{-V_i}{4\pi}\Bigg{(} \oint_{S'} \frac{\partial G(\mathbf
x,\mathbf x\,')}{\partial n'}da' \Bigg{)}_i = \sum_i V_i \Theta_i,
\end{equation} where
\begin{equation}\label{eqn:basisfn}
\Theta_i = -\frac{1}{4\pi}\oint_{S'} \frac{\partial G(\mathbf x,
\mathbf x\,')}{\partial n'}da'
\end{equation} is the basis function for the electric potential
produced by the $i$-th electrode held at 1 volt, all others held at
ground. The basis functions, as solutions of Laplace's equation, are
strictly valid only for static voltage configurations. However, they
are perfectly satisfactory for describing the rf potential and
switching potentials used in rf Paul traps, because the shortest
wavelengths ($\approx 10^1$ m) associated with the time-dependent
fields at these frequencies ($\approx 10^6$ Hz) will be much greater
than the corresponding trap dimensions ($> 10^{-3}$ m), allowing us
to calculate the fields and potentials in the problem
quasi-statically. Effectively, we are considering any changes of the
potential in the trap region to be uniform throughout, and
essentially simultaneous with the change in the voltage on the
electrodes. Therefore, we can introduce time dependence in the
switching potentials simply by treating the voltages on the
electrodes, $V_i$, as functions of time.

The basis function $\Theta_{rf}$ obtained in this manner for the rf
electrodes can be used to obtain the potential energy resulting from
the rf electrodes in the pseudopotential approximation
\cite{Pseudopotential}. The formula for the rf pseudopotential is
given by
\begin{equation}\label{eqn:pseudopot}
\Psi_{rf}(\mathbf{x},t)= \frac{e^2 V_{rf}^2}{4 m \Omega_{T}^{2}}
|\mathbf{\nabla}\Theta_{rf}(\mathbf{x})|^{2},
\end{equation} where $V_{rf}$ is the amplitude of the rf voltage applied to the
electrodes, $m$ is the mass of the trapped ion, $e$ is the charge on
the ion, and $\Omega_T$ is the rf angular frequency. Therefore, the
rf pseudopotential is found by calculating the square of the
electric field amplitude corresponding to the electric potential,
$\Theta_{rf}.$ Alternatively, if information about the micromotion
of the trapped ion is sought, the time-dependent coefficient of
$\Theta_{rf}$ would then become
\begin{equation}\label{eqn:rfmicropot}
V_{rfmicro} = V_{rf} \cos(\Omega_T t).
\end{equation}


\subsection{\label{basisdevelopment}Numerical Techniques for Developing Basis Functions}
The use of basis functions in the calculation of time-dependent
potentials in complex ion trap arrays requires an accurate
calculation of each basis function. This basis function is given by
the potential produced by each electrode when it is held at $1.0$ V
while all other surfaces are held at $0.0$ V. Typically, these
functions must be obtained using numerical methods. A
well-established and accurate method of obtaining electrostatic
potentials produced by a realistic arrangement of electrodes is the
finite element method (FEM), which is used in many
commercially-available software packages for electromagnetic field
simulations. This method requires that the entire bounded problem
domain be discretized into a mesh, consisting of nodes and elements.
The nodes are related to one another by simple (linear or quadratic)
functions, and the solver uses an iterative approach such as energy
minimization to obtain the potential at each node so that the
boundary conditions are still satisfied. The interpolating functions
for each element relating nodal solutions are then used to find the
solution throughout the entire solution domain.

The Boundary Element Method (BEM) is an alternative numerical
analysis method to the FEM. The BEM starts from the integral
equation formulation of the relevant differential equation
(Laplace's equation, in this case). Since there are no charges
present in the empty ion trap, only the surface integrals are
non-zero. This results in a problem formulation, much like that
given in equations 1 through 4, for which the potential within the
problem domain is defined by the surface values of the potential and
the appropriate Green's function. If the problem domain is
unbounded, then the free space Green's function for Laplace's
equation can be used. For ion traps, the potential on the surface is
prescribed by the applied voltage. The fields at the surface are
then found by discretizing the surface with nodes and elements and
solving the resulting set of linear equations. This is equivalent to
finding the charge density over each element on the surface. The
solution at an arbitrary point, {\em P}, within the problem domain
is found by evaluating the integrals describing the contribution to
the potential at {\em P} from each charge element on the surface.

A major advantage of the BEM in obtaining basis functions for ion
trap arrays is the fact that the discretization of the problem is
confined to the boundary surfaces, so that the potential and
electric field within the problem domain will be continuous
functions. A second advantage is the reduction in dimensionality of
the problem (i.e., from a volume to a surface) in the BEM. As larger
and larger trap arrays are considered, the bounding box volume for a
finite element model will grow more rapidly than the corresponding
trap surface area. In these cases, the BEM can prove much more
efficient in calculating the basis functions for ion trap arrays.
Because the BEM is restricted to linear problems for which an
analytic form of the free space Green's function exists, it is not
as commonly used in commercially available software. Several
non-commercial (including CMISS) and commercial (SIS's
CPO\footnote{Charged Particle Optics , by Electronoptics:
http://www.electronoptics.com/ }\, and IES's Coulomb
3D\footnote{Coulomb, by Integrated Engineering Software:
http://www.integratedsoft.com/}\,\,) codes use the BEM exclusively
or in conjunction with the FEM.

Most commercially available software for calculating electrostatic
potentials and fields, such as Tosca from Vector
Fields\footnote{http://www.vectorfields.com/}\,\, or Maxwell 3D from
Ansoft\footnote{Maxwell 3D, by Ansoft: http://www.ansoft.com/}\, ,
uses the FEM because of its nearly universal applicability for
solving differential equations in physics. In the particular case of
ion traps, the FEM provides several advantages, including the
ability to account for non-linear material properties of the trap
electrodes, its ability to determine mechanical and thermal effects
on the trap electrodes during trap operation, and having a simple
means for estimating errors in the simulation. Nevertheless, care
must be taken when using it for analyzing ion traps. In particular,
{\em hexahedral elements} should be used with {\em quadratic
interpolating functions}. While triangular and tetrahedral elements
are preferable for ease of meshing the problem volume, they require
a far greater number of nodes to achieve the same accuracy as can be
obtained with hexahedral elements, or bricks.  This is so because
hexahedral elements are more easily lined up along the equipotential
lines in the relevant problem domain. In addition, the regular
spacing of hexahedral elements helps avoid serious discretization
errors when calculating the potential in regions where competing
fields largely cancel. When calculating the rf pseudopotential, the
field amplitude is important. Linear interpolating functions will
give a constant value of the field throughout the element, a value
most accurate at the element's centroid. Quadratic elements give a
more accurate picture of the field throughout each element, although
they are costly in terms of computational effort.

In general, a finer mesh and quadratic elements help avoid
discretization errors, while larger problem domains are needed to
avoid undue influence from the bounding box. These competing needs
result in a rapidly growing cost in memory requirements and
computational time as the trap arrays increase in complexity.
Computational costs can be reduced through the use of symmetry and
strategic meshing.

A symmetric linear Paul trap array will typically have a plane of
symmetry in the plane of the rf electrode layer, and another plane
perpendicular to the first along the linear trap axis. The $z-$axis
is taken to be directed out of the plane of the two-dimensional trap
array, and the trap axis is taken to lie along the $y$-axis. Since
in the calculation of $\Theta_{rf}$ all electrodes except the
rf-layer are set to ground, the boundary conditions on the
electrodes preserve the symmetry of the trap, and it becomes
possible to reduce the computational domain volume for the rf fields
by using the $yx$ and $xz$ symmetry planes as external boundaries of
the problem. If the boundary conditions along these planes are set
so that the resulting electric field is tangent to these planes,
then the calculated potential in the reduced volume corresponds to
the potential resulting from a symmetric arrangement of electrodes
and voltages across the symmetry planes.

\begin{figure}[tb!]
\centerline{\epsfig{file=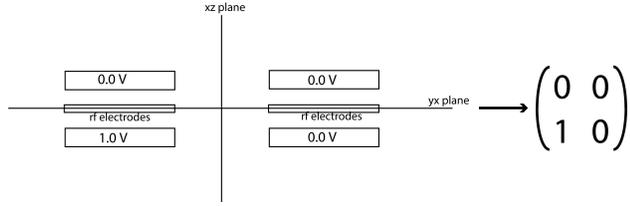, width=8.2cm}} 
\vspace*{13pt} \caption[]{\label{symmetricfig}A cross-section view of
a 3-layer linear rf ion trap, with four identical dc electrodes
placed symmetrically about the two symmetry planes, {\em yx} and
{\em yz}.}
\end{figure}

The calculations of $\Theta_i$ for the control electrodes are not so
easily reduced, since the requirement that only the single control
electrode be set to 1 volt with all other electrodes held at 0 volts
breaks the symmetry of the trap. However, it is possible to use
solutions for the potential which do preserve the symmetry of the
trap to obtain the desired non-symmetric potential by using linear
superposition. Consider a three-layer trap with four control
electrodes arranged symmetrically about the trap center as
illustrated for a linear trap in Fig. \ref{symmetricfig}, where the
basis function $\Theta_i$ is sought for the lower left electrode.

We can again reduce the computational volume of the problem by
imposing boundaries along the $yx$ and $yz$ symmetry planes. If
tangential boundary conditions are applied along both planes, the
resulting solution for the potential in the reduced volume
corresponds to the case when all four control electrodes in
Fig.~\ref{symmetricfig} are held at 1 volt for the full domain. The
solution for the full volume is therefore obtained by adding the
solution of the reduced volume and appropriate reflections of this
reduced volume solution. We identify this solution as
\begin{equation}\label{eqn:tt}
tt = \left( \begin{array}{cc}
      +1 & +1 \\
      +1 & +1
    \end{array} \right).
\end{equation} The array identifies the effective voltages on each of the
control electrodes when both the symmetry planes have tangential
boundary conditions applied. In contrast to tangential boundary
conditions, normal boundary conditions on the symmetry planes
require that the resulting electric field be normal to the boundary,
giving rise to an antisymmetric arrangement of electrodes and an
antisymmetric potential. For example, if both symmetry planes had
normal boundary conditions, the solution in the reduced problem
domain would correspond to the case for the full domain when each
neighboring control electrode is of opposite sign, so that
\begin{equation}\label{eqn:nn}
nn = \left( \begin{array}{cc}
      -1 & +1 \\
      +1 & -1
    \end{array} \right).
\end{equation}
The other two cases involving mixed boundary conditions on the
symmetry planes are identified as
\begin{equation}
    nt = \left( \begin{array}{cc}
      -1 & -1 \\
      +1 & +1
    \end{array} \right), \hspace{.1in} \mbox{and } tn = \left( \begin{array}{cc}
      +1 & -1 \\
      +1 & -1
    \end{array}\right).
\end{equation}

As we have shown in Sec. \ref{proof}, each of these potentials can
be decomposed into sums of four potentials corresponding to the
contribution from each electrode separately. Each solution, $tt, nn,
tn,$ and $nt$, contains a mixture of those contributions.  By
combining the four solutions in the appropriate manner and dividing
by 4, it should therefore be possible to extract the contribution
from any one of the single control electrodes. This can be shown
symbolically by adding the four solutions, as shown below. The use
of the arrays to symbolize the solutions for each symmetry case
makes it clear that this process corresponds to adding the boundary
conditions on the four electrodes together. The result is a solution
for the whole space potential that is produced solely by a unit
voltage on the lower left electrode, all other electrodes being held
at ground.
\begin{equation}\label{1}
    \frac{tt+nn+tn+nt}{4} = \left( \begin{array}{cc}
      0 & 0 \\
      1 & 0
    \end{array}\right) = \Theta_{\mbox{lower left}}
\end{equation}
The basis functions for the other three electrodes are easily
obtained by the appropriate coordinate reflections of the first
solution. This approach, although more time-consuming, is necessary
when modeling and meshing the entire problem domain becomes
prohibitive due to memory restrictions. It enables the experimenter
to mesh the model at a higher density for improved accuracy.

The use of hexahedral elements for meshing an ion trap model places
a much greater constraint on node spacing than would be the case if
tetrahedral elements are used. In the case of Vector Fields' Opera
suite, this means that node placement must be done manually, and
then checked for suitability for hexahedral meshing when placement
is complete. In particular, the number of nodes on opposing faces of
the model must match, so that the elements are able to completely
fill the space in the problem domain. Nevertheless, it is possible
to concentrate node placement along the channels through which ions
will be expected to be shuttled, and along the electrode surfaces
near which the potential is expected to exhibit the greatest
variation. There will generally be some wasted node density in
regions above and below the trap and along the channels beyond the
end electrodes, due to the restrictions on the consistency of the
hexahedral elements.

Ion traps are generally constructed from good conductors and
dielectrics, which exhibit linear behavior under the voltages
typically applied in these traps. In such cases, the accuracy of the
electrostatic potentials and fields obtained using the FEM (assuming
the model is a correct representation of the physical problem) is
primarily a function of the local mesh spacing, and only weakly a
function of the overall mesh density in the problem definition. In
particular, for a local mesh size {\em h} in one dimension
(corresponding to the mesh point spacing) and using quadratic
elements, the error in the calculated potential scales as ${\cal
O}(h^3),$ while the error in the fields will scale as ${\cal
O}(h^2)$ \cite{Gilbert:1973}.

Therefore, a reasonable estimate of the error in the FEM solution
can be made by halving the mesh point spacing throughout the model,
if memory permits, or otherwise, halving the mesh point spacing in
the region requiring greatest accuracy, and running the model again.
Percentage changes in the calculated potential and field will then
give an estimate of the error in the calculation. Thus, if the field
calculation at mesh spacing $h$ gives a result $E$ with unknown
error $\Delta E$, and a calculation at mesh spacing $h/2$ gives a
different result $E'$ with unknown error $\Delta E'$ then, we can
compare the two unknown errors, since error scales with the square
of the mesh spacing, that is,
\begin{equation} \Delta E' = \Delta E /4.\end{equation} Roughly speaking, we can
identify the difference in the two solutions at each point as some
function of the uncertainties in each solution. The most
conservative assumption would be that the two solutions erred in the
same sense from the true value, so that their difference is equal to
the difference of the two uncertainties, that is
\begin{equation} (E-E') \geq
(\Delta E - \frac{\Delta E}{4}) = \frac{3}{4} \Delta E.
\end{equation} Thus, we have a loose upper bound on the error in the
original solution,
\begin{equation} |\Delta E| \leq \frac{4}{3}|E-E'|.\end{equation}

Once the models have been meshed and analyzed, it is still necessary
to evaluate the potential and/or field at each point of interest in
the problem domain. In the interest of carrying out simulations of
ion trajectories it is desirable, therefore, to obtain beforehand a
grid of potential or field amplitude values covering the problem
domain volume corresponding to locations where ion trapping and
shuttling will take place. The grid spacing used for the array
should be at least as small as the nodal spacing used in the
numerical simulation. There will be diminishing returns for using
even denser arrays of points, since the potentials between the nodes
of the finite element mesh are already calculated using quadratic
interpolating functions. Since the potentials are solutions of
Laplace's equation and thus smoothly varying functions of position,
it is possible to generate splined, interpolating functions from
these data grids at the accuracy of the finite element solution to
serve as the basis functions $\Theta_i$ for subsequent calculations
of the ion dynamics.


\subsection{\label{CaluculatingIonDynamics}Trapped Ion Dynamics}
We now consider the desired potential by suitably superposing the
basis functions multiplied by the time varying potential
\begin{equation}\label{Exact Potential}
U(\mathbf{x},t)= eV_{rf}\cos(\Omega_{T}t)\Theta_{rf}(\mathbf{x})
+e\sum_{i}V_{i}(t)\Theta_{i}(\mathbf{x})
\end{equation}
where $e$ is the charge of the ion, $\mathbf{x}$ is the position
vector, $\Omega_{T}/2\pi$ and $V_{rf}$ are the applied rf frequency
and amplitude, $V_{i}(t)$ is the time varying potential applied on
the $i$th control electrode and $\Theta_{i}(x)$ is the basis
function of the $i$th electrode.  Notice here that the coefficient
for all the basis functions have explicit time dependence.

The ion's motion due to the electric potential $\Phi$ will consist
of the low amplitude micro-motion with frequency to the order of
$\Omega_{T}$ and the slower but larger amplitude secular motion.
Very often, we only need to calculate the secular motion of the ion
and ignore the micro-motion. Therefore we may approximate Eq.
\ref{Exact Potential} with a ponderomotive pseudopotential given by
\cite{Pseudopotential}:
\begin{equation}\label{Pseudo-approx}
\Psi(\mathbf{x},t)= \frac{e^{2}V_{rf}^{2}}{4 m \Omega_{T}^{2}}\vert
\mathbf{\nabla}\Theta_{rf}(\mathbf{x})\vert ^{2}
+e\sum_{i}V_{i}(t)\Theta_{i}(\mathbf{x})
\end{equation}
Finally, if there are k ions in the trap, the resultant force on
each ion $\mathbf{F_{j}}$ is given by
\begin{equation}\label{defF}
\mathbf{F_{j}}(\mathbf{x_{1}}, ...,\mathbf{x_{k}},t) = \left\{
\begin{array}{ll}
-\mathbf{\nabla}U(\mathbf{x_{j}},t)+\frac{1}{4 \pi
\epsilon_0}\sum_{i\not=
j}\frac{e^{2}}{|\mathbf{x_{j}}-\mathbf{x_{i}}|^{3}}(\mathbf{x_{j}}-\mathbf{x_{i}}) & \textrm{complete ion motion}\\
-\mathbf{\nabla}\Psi(\mathbf{x_{j}},t)+\frac{1}{4 \pi
\epsilon_0}\sum_{i\not=
j}\frac{e^{2}}{|\mathbf{x_{j}}-\mathbf{x_{i}}|^{3}}(\mathbf{x_{j}}-\mathbf{x_{i}})&
\textrm{ion secular motion only}
\end{array} \right.
\end{equation}
Therefore, to calculate the dynamics of $k$ ions in a trap we need
to solve the set of $k$ coupled second order ordinary differential
equations(ODEs):
\begin{equation}\label{Equation to be solved}
\mathbf{\ddot{x}_{j}}=\frac{\mathbf{F_{j}}}{m}(\mathbf{x_{1}},...,\mathbf{x_{n}},t)
\equiv\mathbf{a_{j}}(\mathbf{x_{1}},...,\mathbf{x_{n}},t)
\end{equation}
where $j$ is an integer from 1 to $k$. Determining the classical motion of
trapped ions plays an important role in calculating the energy gained
during shuttling as will be seen in section 3.


\subsection{\label{sec:overall comparison}Numerical Methods for Obtaining Trapped Ion Dynamics}

In general, there is no analytic solution for the electric field in
an ion trap, so Eq. \ref{Equation to be solved} must be solved
numerically. As will be seen in Section \ref{sec:theory}, the classical motion of trapped ions during shuttling protocols will play an important role in calculating the amount of heating the ions undergo from an arbitrary initial quantum state. The design of shuttling protocols requires high
accuracy solutions of Eq. \ref{Equation to be solved} and as such
the numerical evaluation of Eq. \ref{Equation to be solved} can be
slow. High accuracy solutions are needed to optimize shuttling
protocols by minimizing the acquired kinetic energy from shuttling.
Using an AMD dual core 1.8GHz processor with 2 GB of memory to
calculate the trajectory of the ion with a shuttling sequence that
shuttles an ion around a corner of a T-junction ion trap array, the
computer time taken to obtain the ion trajectory depends on the ODE
solver method ranges from 5 hours to a full week. In complex
shuttling operations where hundreds of ions may be shuttled
throughout an ion trap array, one must make a judicious choice of
ODE solver in order to reach the required accuracy in a feasible
amount of time.

Explicit extrapolation class methods are good for efficiently
(minimal computing time) solving ODE's to high accuracy
\cite{Bulirsch-Stoer}. However, a caveat when using this class of
methods is that the calculated electric field has to be smooth. If
the electric field is rough, Explicit Runge-Kutta (ERK) methods
may be a better choice \cite{numRecERK}. In addition, if a low
accuracy solution is sufficient, single step methods tend to be more
efficient than the extrapolation class methods \cite{numRecERK}.
This section outlines the reasons why the Bulirsch-Stoer method
effectively simulates ion motion in ion trap arrays while Appendix A
discusses how the Bulirsch-Stoer (B.-S.) method works.

The ODE system of Eq. \ref{Equation to be solved} can be stiff if
the requirement of the stability of the solution is more stringent
than the accuracy of the ODE solver \cite{Leader}. One way for a
system to be stiff is if the solution has some components that are
rapidly varying and some other components that are varying much more
slowly (see Appendix A). The reason for the computational
inefficiency is that in order for the solver to be stable, the time
steps that the ODE solver uses must be much shorter than the time
scale of the fastest changing component of the solution. Stiffness
may be a significant problem in ion trap simulations as there are
several time scales involved in the ion's motion.  The dynamical
evolution in ion trap systems has several important time scales; for
example, the rf micromotion has frequency of order 10-100 MHz (0.01
- 0.1 $\mu$s), secular motion of order 100-1000 kHz (1-10 $\mu$s)
while shuttling times may be of order 10-1000 $\mu$s. When
simulating the motion of an ion during complex shuttling operations,
computational resources may be eaten up while the numerical solver
calculates miromotion and secular motion. Stiffness may also appear
as a result of the numerical simulation of the electric potential.
Roughness in the electric potential may result in artificially large
forces on the ions that slows down the simulation. Though explicit
ODE solvers such as extrapolation class and ERK methods are usually
inefficient at numerically evaluating such systems, there are ODE
solver methods known as ``stiff solvers" that are well suited to
handle these systems \cite{Shampine Gear SIAM review}.

We consider ODEs of the form:
\begin{equation}\label{ODEform}
\frac{dx}{dt}=f(t,x)
\end{equation}
The output of any numerical ODE solver is a series of discrete
points called nodes.  A node is of the form $(t_{i},\ x_{i})$ where
$x_{i}$ is an approximation of the exact solution $x(t=t_{i})$.  The
first node is given by the initial conditions. Subsequently every
step that the ODE solver takes calculates one more node.  The size
of every step that the ODE solver takes, i.e. $(t_{i}-t_{i-1})$ is
known as the step-size.  The step size need not be uniform and will
change adaptively in order to maximize efficiency (i.e. minimize
computing time without an undue sacrifice in accuracy).

We define the local error to be the error introduced due to one step
of the ODE solver (for example see equation
\ref{Taylorx}-\ref{error2}). Note that since in general, we do not
know the exact solution \emph{a priori}, the numerical ODE solver
will always generate an estimate for the local error for every step.
Finally, if we require the local error to be arbitrarily small, the
ODE solver step-sizes would then be also arbitrarily small and thus
the computation time would be extremely long. Therefore, we need to
set a practical limit for the local error of every step. This limit
is known as the local error-goal and is specified by the quantities
$a$: the accuracy goal, and $p$: the precision goal. The local error
goal $\epsilon$ is then \cite{mathbook}
\begin{equation}\label{error goal}
\epsilon=10^{-a}+|x|*10^{-p}
\end{equation}
A numerical ODE solver will adaptively change the step-size such
that each step has a local error estimate that is smaller than the
user defined local error goal. Adaptive step size algorithms are
further discussed in Appendix A.

Table \ref{Steptab} shows the computing time, number of steps taken
and average step size between nodes while simulating shuttling an
ion around the corner of a T-junction ion trap without using the pseudo-potential approximation as reported by
Hensinger et al. \cite{T paper} for a fixed local error tolerance
using three different types of ODE solvers. The three ODE solver
methods are the Bulirsch-Stoer method with adaptive step size, the
Explicit Runge-Kutta (ERK) Method with adaptive step size and
adaptive order, and the Backward Difference Formulae (BDF) methods
with adaptive step size and adaptive order. More details about each
method are given in Appendix A.

\begin{table}[tb!]\label{Steptab}
\centering
\begin{tabular}{|c|c|c|c|}
\hline ODE solv.&Computing&Number&Ave step
\\method&Time&of steps&size[s]
\\\hline
Bulirsch-&$5h 54m$&$35392$&$8.8*10^{-10}$
\\Stoer&&&
 \\\hline
$ERK$&$37h 9m$ &$1546660$&$2.0*10^{-11}$
\\\hline
 BDF&$5h 44m$&$408403$&$7.6*10^{-11}$
 \\\hline
\end{tabular}
\caption[]{\label{Sim FOMs}We tabulate quantities that indicate the
performance of our three numerical ODE solvers.  The number of steps
indicate the accuracy of the solution whilst the computing time
indicates the efficiency of the numerical method. The local error
tolerance for all three simulations had an accuracy goal of a = 8
and precision of goal of p = 8 in Eq. \ref{error goal}. The simulations include micromotion.}
\end{table}

For fixed local error goals at each step, the error on average
increases with the number of steps taken. We therefore conjecture
that given two numerical ODE solvers, the ODE solver that takes less
steps will usually be more accurate than the ODE solver that takes
more steps. From this consideration we see that the Bulirsch-Stoer
method is the best as the Bulirsch-Stoer method requires an order of
magnitude fewer steps than the BDF method and two orders of
magnitude fewer steps than the ERK method. In addition, the
Bulirsch-Stoer method takes only about 3\% more computing time than
the BDF method to reach a solution (see Table \ref{Sim FOMs}). The
ERK method takes far too much computing time and this shows that it
is probably impractical for large-scale simulations of ion dynamics
in an ion trap array.

\begin{figure}[tb!]
\centerline{\epsfig{file=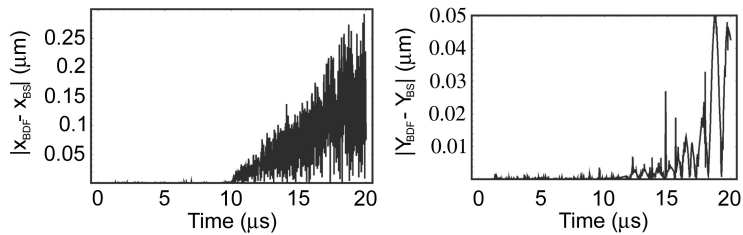, width=10cm}} 
\vspace*{5pt} \caption[]{\label{xyresid}This figure shows the
absolute value of the difference between the Bulirsch-Stoer and BDF
numerical estimates of the x- and y-components of the position of an
ion shuttled along a linear path.  As we can see, the disagreement
in the numerical estimates increase with time.}
\end{figure}

There is a significant difference between the calculated ion motion
using the Bulirsch-Stoer and BDF methods when linearly shuttling an
ion, as can be seen in Fig. \ref{xyresid}. To figure out the
absolute accuracy of each method, it is necessary to compare the
calculated numerical method with a benchmark solution- an extremely
high accuracy solution. However, our modest computing resources do
not permit us to find a reasonable benchmark solution as the
computing time required was several weeks. Because the potentials in
ion trap systems can be approximated by a harmonic oscillator
potential, we compared the absolute accuracy of the Bulirsch-Stoer
and BDF methods to the known solution of a harmonic oscillator.

\begin{figure}[tb!]
\centerline{\epsfig{file=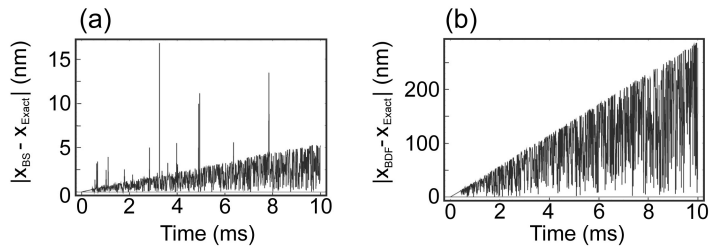, width=10cm}} 
\vspace*{5pt} \caption[]{\label{twoResidualPlot} (a) depicts the
absolute value of the deviation of the numerical estimate derived
from the Bulirsch-Stoer method and the exact solution of the
equations of motion of a simple harmonic oscillator. There are
several spurious peaks in the graph and these are due to
inaccuracies in the interpolation process to fit the generated nodes
and are not errors from the numerical solution. If we ignore these
spurious peaks, we note that the average error increases linearly
with time and is due to the quadratic potential. (b) depicts the absolute value of the deviation of the
numerical estimate derived from the BDF method and the exact
solution. Unlike (a), there are no spurious peaks because the BDF
method generates more nodes which implies that the nodes are closer
together and thus the interpolation process is more accurate. We
note that in both figures the average error increases monotonically
with time and the error of the solution derived from the
Bulirsch-Stoer method is much smaller than the error derived from
the BDF method.}
\end{figure}

We use the Bulirsch-Stoer Method and the Backward Difference
Formulae to numerically evaluate the solution to a simple harmonic
oscillator differential equation for the time interval $(t=0$ s,
$t=0.01$ s$)$ with $\omega/2\pi = 1$ MHz. We first observe that the
BDF method takes more steps than the Bulirsch-Stoer Method; 958331
steps as compared to 207422. The second observation is that the
average error increases monotonically with the number of steps taken
with fixed error goals. This result is shown in Fig.
\ref{twoResidualPlot} as plots of the absolute difference between
the ODE solver method and the exact solution as a function of time.
From Fig. \ref{twoResidualPlot}, if we ignore the spurious
errors\footnote{As the BS method produces less nodes than the BD
method, the polynomial interpolation of the nodes derived from the
BS method is less reliable.  However, the error in the polynomial
interpolation has no impact on the behavior of the nodes and
therefore the overall behavior of the numerical solution.}\,\,\, due
to the interpolation process, the error of the Bulirsch-Stoer method
is much smaller than that of the BDF method and supports our
conjecture that an ODE solver that can cross the interval in less
steps will be more accurate than an ODE solver that crosses the
interval in more steps.

We used the Bulirsch-Stoer method to simulate ion motion during
shuttling because of the superior accuracy of this method for
obtaining a numerical solution for an ion's trajectory and the
superior computational efficiency of this method. Note here that our
observations pertain specifically to our particular ion trap
geometry (see section \ref{sec:T example}) and our specific local
error tolerances. It is possible that some other ODE solver may be
more effective depending on the ion trap geometry as well as the
computational resources available. Although the above analysis
implies that an ODE solver with fewer steps has superior accuracy,
the intermediate motion of the ion between nodes is not accessible.


\section{\label{sec:theory}Theoretical Description of Shuttling Atomic Ions}

So far, we have described the means by which it is possible to
calculate the effective electric potential at the position of the
ions in an ion trap array, and also the classical trajectories that
those ions will take when the voltages on the control electrodes are
changed with time. The goal is to develop a system that allows ions
to be moved to arbitrary locations within the trap array in a
perfectly reliable manner. In addition, the ions should carry and
store quantum information both before and after each shuttling
operation. This indicates the need to identify those shuttling
operations which keep the ions trapped and cold enough to perform
quantum gate operations, all the while providing maximum speed of
operation. In this section,
we develop a general theoretical model of the shuttling process. Our model
focuses on the case in which the motion of the ion in the trap along the
pathway of the ion is affected. We have worked out the model in a rather
complete fashion as a reference for future work and have applied it to
several possible shuttling time profiles. Rather than just treating a simple
model considering only the first vibrational state, we calculate the general
case that may be applied to a much broader context.  The model is then used
to identify those constraints that ensure reliable transport of
ions. By identifying such parameters the reader can construct effective
shuttling protocols for a variety of situations. For those who wish to skip
the details of the theoretical analysis, the key results are presented in
section 3.2.5, just prior to the section detailing how these results can be
used to evaluate various shuttling procedures. A similar theoretical analysis of
shuttling has recently been given by Ref. \cite{reichle2006}.
Furthermore, Ref. \cite{schulz:2006} discussed the application of
control theory to single ion transport. The analysis given here
emphasizes the importance of the inertial forcing of shuttled ions
at the beginning and end of the protocol, as well as the possibility
of significant parametric heating of the ion even for slow shuttling
speeds.

\subsection{\label{sec:shuttlingprotocols}The Shuttling Process}

The rf Paul linear ion trap works by creating an effective potential
near the center of the trap that is quadratic in all three
coordinate directions. The transverse trap is produced by the rf
ponderomotive potential and is symmetric and perfectly harmonic near
the trap minimum, while the trap along the shuttling pathway is
created by applying voltages to segmented control electrodes. This
potential is also harmonic to a very good approximation. A shuttling
operation involves changing the voltages on the control electrodes
in time, so that the potential minimum along the ion pathway is
translated from the initial ion position to the desired final
position.

It is helpful to begin by considering the electric field along the
ion pathway in the vicinity of the ion. The one-dimensional harmonic
potential along the trap axis corresponds to a linearly-varying
field,
\begin{equation}
E(x) = -{\cal E} x,
\end{equation} with its stable equilibrium point at $x = 0.$
This field is the result of the potential difference between the
nearest control electrodes that are held at or below ground, and the
neighboring control electrodes. The shuttling operation described
above corresponds to introducing a potential difference between the
control electrodes. This voltage difference results in a nearly
spatially-uniform, time-dependent electric field superimposed on the
original trapping field and pointing in the direction of shuttling.
The resulting electric field,
\begin{equation} \label{eqn:forcedoscillatorfield}
E(x,t) = - {\cal E} x + E(t) = - {\cal E} (x - x_0(t)),
\end{equation}
now has a stable equilibrium point $x_0(t) = \frac{E(t)}{{\cal E}},$
that varies with time. The resulting electric potential is given by
\begin{equation} \label{eqn:forcedoscillatorpotential}
V(x,t) = {\cal E} \left( \frac{1}{2}x^2 - x_0(t)x\right) + V_0(0,t),
\end{equation}
where $V_0$ represents the (time-varying) potential at $x = 0.$ We
choose the zero of the electric potential to be located at $x_0(t)$,
\mbox{i.e.} $V(x = x_0(t),t) = 0,$ and therefore,
\begin{equation}
V_0(0,t) = \frac{1}{2} {\cal E} x_{0}^{2}(t) = \frac{E^2(t)}{2 {\cal
E}}.
\end{equation}
In practice, this time-dependent potential can be introduced during
the shuttling process by continually raising (lowering) the voltages
on the electrodes behind (ahead of) the moving ion (see
Sec.~\ref{Linear Shuttling}).

\begin{figure}[tb!]
\centerline{\epsfig{file=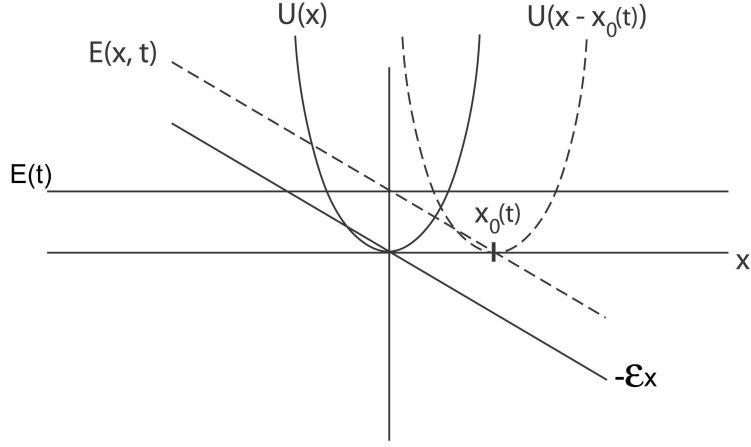, width=10cm}} 
\vspace*{13pt} \caption[]{\label{fig:efield} Schematic drawing
showing the relationship between the initial electric field $-{\cal
E} x$ creating the axial trap, the time-dependent forcing field,
$E(t)$, and the resulting field $E(x,t),$ along with the potentials
at times 0 and $t$ and the location of the minimum, $x_0(t)$.}
\end{figure}

Finally, we obtain the expression for the potential energy in the
trap frame as a function of $x$ and $t$,
\begin{eqnarray} \label{eqn:forcedoscillatorPE}
U(x,t) &=& \frac{1}{2} e {\cal E} \left( x^2 - 2x_0(t)x +
x_{0}^{2}(t) \right)
\\ & = & \frac{1}{2} m \omega^2 (x - x_0(t))^2,
\end{eqnarray}
where we have identified ${\cal E} = \frac{m \omega^2}{e}$ (See
Fig.~\ref{fig:efield}).
This translating potential can be thought of as a moving bowl for
the purpose of carrying a marble from place to place. Quantum
mechanically, the last term in Eq.~\ref{eqn:forcedoscillatorPE} does
not induce transitions between states and merely produces an overall
phase factor in the quantum state because it is independent of the
position operator $x$ (see Eq. \ref{eqn:phiofsandt}). Therefore, the
problem of shuttling atomic ions and determining the effect of
shuttling on their motional states is equivalent to the problem of
solving for the transitions induced in a harmonic oscillator being
forced by a uniform field, $e E(t) = m\omega^2 x_0(t).$ The forcing
field determines the location of the instantaneous potential minimum
of the moving ion trap.

\begin{figure}[tb!]
\centerline{\epsfig{file=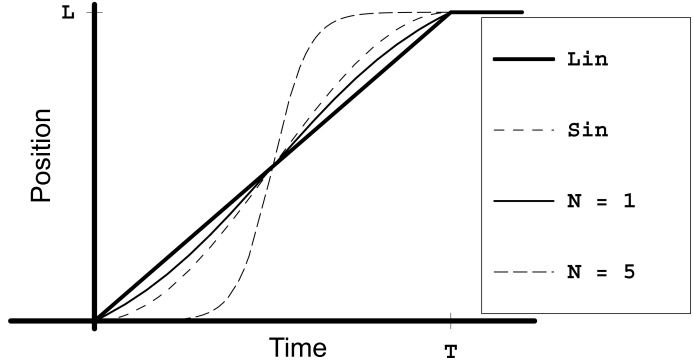, width=10cm}} 
\vspace*{13pt} \caption[]{\label{fig:protocols} Plot of the position
of the potential minimum of the trap versus time, when the potential
minimum time profile is linear, sinusoidal and  a hyperbolic tangent
(N =1 and N = 5). The variables $L$ and $T$ represent the total
shuttling distance and time, respectively.}
\end{figure}

We now examine the case when a cooled ion is shuttled a distance $L$
over a time $T$, so that $x_0(t) = 0$ for $t<0$ and $x_0(t) = L$ for
$t>T$. The trajectory of the potential minimum $x_0(t)$ is directly
related to the time-dependent voltage difference $\Delta V_{cap}(t)$
applied to the relevant control electrodes. That is, we expect that
\begin{equation}\label{eqn:movingmin}
x_0(t) = E(t) \frac{e }{m \omega^2} = \eta(x_0(t)) \frac{\Delta
V_{cap}(t)}{d} \frac{q} { m \omega^2},
\end{equation}
where $\eta(x)$ is a unitless geometrical function relating the
control electrode voltage difference to the electric field at
position $x$, and $d$ is the characteristic center-to-center
distance between neighboring electrodes. Therefore,
Eq.~\ref{eqn:movingmin} tells us that from a knowledge of the
desired functional form for the trajectory $x_0(t)$ and the position
dependent geometrical function $\eta(x)$ the required voltage
differences $\Delta V_{cap}(t)$ across the control electrodes can be
determined. Functional forms for the trajectories of the potential
minimum include piecewise linear functions, sinusoids, and other
transcendental functions such as the hyperbolic tangent function
\cite{T paper}. We will therefore consider the following three
potential minimum time profiles for translating the harmonic
potential: linear ($x_{0l}(t)$), sinusoidal ($x_{0s}(t)$), and
hyperbolic tangent ($x_{0t}(t)$), defined as
\begin{eqnarray}\label{eqn:timeprofiles}
x_{0l}(t) &=& L\frac{t}{T}(H(t) - H(t-T)) + L H(t-T),\\
x_{0s}(t)&=& \frac{L}{2} \left(1-\cos\left(\frac{\pi
t}{T}\right)\right)
(H(t) - H(t-T)) + L H(t-T),\\
x_{0t}(t)&=& \frac{L}{2} \frac{\left(\tanh \left(N
\frac{2t-T}{T}\right) +\tanh (N)\right)}{\tanh(N)}(H(t) - H(t-T)) +
L H(t-T). \end{eqnarray} In these expressions, $H(t)$ is the
Heaviside step function, and the parameter $N$ in the hyperbolic
tangent potential minimum time profile characterizes the translation
rate at the midpoint of the motion and also determines the magnitude
of the discontinuity in the velocity of the potential at the
beginning and end of the protocol (Fig.~\ref{fig:protocols}). For
$N>1$, the time between $10\%$ and $90\%$ of the transition is $\sim
T/N$ and the velocity discontinuity is $\sim (4L/T)Ne^{-2N}$

Any time dependence of $\omega$ will also enter into the functional
form of $x_0(t),$ as can be seen from Eq.~\ref{eqn:movingmin}. We
can better separate the influence of fluctuations in the frequency
from that of the time-dependent electric field, $E(t),$  by
transforming to the rest frame of the moving potential. The position
coordinate becomes $s= x - x_0(t)$ and a pseudo-forcing term, $m
\ddot{x}_0 s,$ is simultaneously introduced into the potential
energy because the reference frame of the moving potential minimum
will not be inertial.  The potential energy from
Eq.~\ref{eqn:forcedoscillatorPE} then becomes
\begin{equation} \label{eqn:forcedoscillatorPEins}
U(s,t) =  \frac{1}{2} m \omega^2(t) s^2 + m \ddot{x}_0(t) s.
\end{equation}
The potential $U(s,t)$ still describes a forced, parametric harmonic
oscillator, but the frequency variation of the potential and the
forcing term due to the translation of the potential are now
separate. What is more, the forcing term no longer includes the net
displacement of the oscillator $x_0(t)$. Instead, it is solely a
result of the inertial force on the ion due to an acceleration in
the transport of the potential. If the potential were simply
accelerating at a constant rate the minimum could be redefined, as
was done for the potential in the lab frame
(Eq.~\ref{eqn:forcedoscillatorPE}). However, a shuttling process
necessarily involves both a start from rest and a bringing to rest
of the harmonic potential. Therefore, the ion will at the least
receive two kicks or pushes away from the instantaneous potential
minimum. This can be seen clearly by examining the second derivative
of the representative time profiles for the potential minimum in the
lab frame, given in Eq.~\ref{eqn:timeprofiles}. After invoking the
properties of the derivative of a delta function, we get:
\begin{eqnarray}\label{eqn:timeprofilesforxddot}
\ddot{x}_{0l}(t) &=& \frac{L}{T}(\delta(t) - \delta(t-T)) ,\\
\ddot{x}_{0s}(t)&=& \frac{L\pi^2}{2T^2} \cos\left(\frac{\pi
t}{T}\right)
(H(t) - H(t-T)), \\
\ddot{x}_{0t}(t)&=& -L\frac{4N^2 }{T^2} \coth (N) \frac{\tanh
\left(N \frac{2t-T} {T}\right) }{\cosh^{2} \left( N
\frac{2t-T}{T}\right)
} (H(t) - H(t-T)) +\\
& &+ L\frac{N}{T} \frac{ \coth (N) }{ \cosh^{2}\left( N
\frac{2t-T}{T}\right)}(\delta(t) - \delta(t-T)).\nonumber
\end{eqnarray}
Here we see that the inertial forcing induced during a typical
shuttling protocol has the general form
\begin{equation}\label{eqn:generalddotx0}
\ddot{x}_0(t) = A(t,T)\frac{L}{T}\left[\delta(t) -
\delta(t-T)\right] + B(t,T) \frac{L}{T^2}\left[H(t)-H(t-T) \right],
\end{equation}
where $A(t,T)$ and $B(t,T)$ are defined by the particular shuttling
protocol. The delta function term, proportional to $L/T$, is
associated with inertial kicks received by the shuttled ion due to
the sudden start-up and completion of the shuttling protocol. The
step function term, proportional to $L/T^2$, is associated with the
inertial forcing due to the acceleration and deceleration of the
shuttling potential during the shuttling protocol. The linear
potential minimum time profile $(B(t,T)=0)$ is seen to provide two
large `kicks' of magnitude $L/T$ but in opposite directions. On the
other hand, it produces no push on the ion except at the start and
finish of the protocol. The sinusoidal potential minimum time
profile $(A(t,T)=0)$ has zero velocity at the start and end of the
shuttling, but it does provide a steady push proportional to $L/T^2$
over the duration of the shuttling, first back and then forward. The
hyperbolic tangent potential minimum time profile has both features
of the other profiles, to a degree controlled by the parameter $N$.
A large value of $N$ results in a smooth beginning and ending to the
process, but a large backwards and then forward pushing in the
middle. A small $N$ produces the opposite result.

\begin{figure}[tb!]
\centerline{\epsfig{file=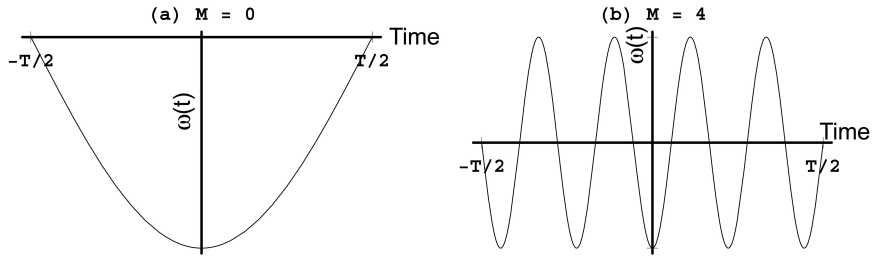, width=12cm}} 
\vspace*{13pt} \caption[]{\label{fig:freqvar} Plots of the
`single-step' (M = 0 in Eq. \ref{eqn:modelfreqvar}) and `long
distance' (M = 4) models for the frequency variation in the
shuttling potential. The vertical axis ranges from $\omega_0^2(1-g)$
to $\omega_0^2(1+g)$.}
\end{figure}

We also need to introduce an appropriate model for frequency
variations of the potential as the ion is carried along during the
shuttling procedure. In general, we want to consider frequency
variations of the type,
\begin{equation}\label{eqn:freqvarformmodel}
\omega(t)^2 = \omega_0^2 (1 - f(t)).
\end{equation}
We will assume in our analysis that the function $f(t)$ is zero at
the beginning and ending of the shuttling process. For convenience,
we will consider perturbations extending from $t=-T/2$ to $t = T/2,$
and then adjust the time scale so that the shuttling and frequency
variation models match. Two types of perturbation will be
considered. First, a `short-step' model is considered where the
trapping potential is weakened by decreasing the voltage on the
electrode in front of the ion and then strengthened by increasing
the voltage on the electrode behind the ion. This will result in a
potential for which the trap frequency will gradually decrease and
then increase. The second type of perturbation to be considered is
that of a fluctuating trap frequency. In this case, the ion can be
thought of as being forced in one direction by a continuously
increasing electric field. As a result, the frequency experienced by
the ion can be modulated due to the fluctuating strength of the
``static'' trapping fields as the ion passes gaps or other changes
in the electrode structure. This fluctuation could affect the trapping potential in either the transverse or longitudinal directions. Another source of frequency variation in
the potential of this type might be low frequency noise from the
control electrodes used to trap and shuttle the ions. Both types of
perturbations can be modeled by the same function,
\begin{equation}\label{eqn:modelfreqvar}
f(t) = g \cos \left\{(M+1/2) \frac{2\pi t}{T}\right\}.
\end{equation}
When the parameter $M $ is set to zero this forcing produces a
decrease and then increase in the frequency over the duration of the
shuttling protocol, as required for the `short-step' model. For $M$
an integer, the sinusoidal variation of the potential has $M+1/2$
``cycles'' throughout the shuttling, which may correspond to the
number of periodic structures in the trap electrode array. These two
models for the frequency variation of the shuttling potential are
illustrated in Fig.~\ref{fig:freqvar}.
The parameter $g$, known as the frequency modulation depth,
characterizes the fractional variation in the square of the
frequency of the potential. In order to optimize the shuttling
process, we will first examine the effect of arbitrary frequency
fluctuations and inertial forcing on the final motional state of
shuttled ions, and then apply the results to the models outlined
above.

\subsection{\label{sec:forcedparam}The Forced Parametric Oscillator}

The problem of the forced harmonic oscillator has been solved
quantum mechanically by Husimi~\cite{husimi} and
Kerner~\cite{Kerner}, independently. Husimi's solution includes the
effects of both inertial forcing and frequency variation on the
oscillator. We seek expressions for the average final motional
state, $\langle n \rangle$ and the variance in the distribution
about the mean, $\langle \Delta n^2 \rangle,$ following Husimi's
solution. In particular, we will first examine the solution to the
time-dependent Schr\"{o}dinger's equation and show that it can be
separated into a solution for the unforced parametric oscillator and
a solution for the forced parametric oscillator. Then we will seek
solutions for those two cases using the method of generating
functions. This approach starts from the basic observation that the
Hermite polynomials from which the eigenfunctions of the harmonic
oscillator are constructed can be used to obtain a power series
expansion of a generating function. A propagator is used to describe
the time evolution of the oscillator system. The generating
functions of the individual wavefunctions are used in conjunction
with this approach to obtain generating functions for the transition
amplitudes and transition probabilities relating the initial and
final states of the system. The method of generating functions is a
powerful method for our purposes, since the desired quantities are
not the individual matrix elements describing the likelihood of
ending up in a particular state, but the average value of $n(T)$,
which is given by the sum over all possible final states at time
$T$. This sum can be obtained by manipulating the generating
function directly.

\subsubsection{\label{sec:solution} Solving Schr\"{o}dinger's equation}

Starting from the one-dimensional Schr\"{o}dinger's equation for
$\psi(s,t)$ in the frame of the potential minimum,
\begin{equation}\label{eqn:Schrodinger}
 \imath \hbar \frac{\partial \psi(s,t) }{\partial t} = -\frac{\hbar^2}{2m}
\frac{\partial^2 \psi(s,t)}{\partial s^2} + \frac{1}{2} m
\omega^2(t) s^2 \psi(s,t) + m\ddot{x}_0(t)s\psi(s,t),
\end{equation}
a second coordinate transformation is introduced, so that
\begin{eqnarray} \label{eqn:coordtrans}
s' &=& s - \xi(t)\nonumber\\
\frac{\partial \psi (s,t)}{\partial t} &=& \frac{\partial \psi
(s',t)} {\partial t} - \dot{\xi}(t)\frac{\partial \psi (s',t)}
{\partial s'} \nonumber\\
\frac{\partial^2 \psi (s,t)}{\partial s^2} &=& \frac{\partial^2 \psi
(s',t)}{\partial s'^2}.
\end{eqnarray} The transformation, $\psi(s',t) = \phi(s',t)
 e^{ims'\dot{\xi}/\hbar}$ is then introduced to eliminate the first order
spatial derivative arising in the second line of
Eq.~\ref{eqn:coordtrans}, and upon substitution into
Eq.~\ref{eqn:Schrodinger} results in the following equation for
$\phi(s',t):$
\begin{eqnarray}\label{eqn:phisprimeandt}
i \hbar \frac{\partial \phi(s',t) }{\partial t} &=&
-\frac{\hbar^2}{2m} \frac{\partial^2 \phi(s',t)}{\partial s'^2} +
\frac{1}{2} m \omega^2(t) s'^2
\phi(s',t) \nonumber\\
& & + m\left((\ddot{\xi} + \ddot{x}_0) + \omega^2(t) \xi \right) s'
\phi(s',t) - \nonumber\\
& & - m/2\left(\dot{\xi}^2 + \dot{x}_0^2 - \omega^2(t)\xi^2 -
2\ddot{x}_0 \xi \right)\phi(s',t).
\end{eqnarray} The first line of Eq.~\ref{eqn:phisprimeandt}
is the wave equation for the unforced parametric harmonic potential,
which has solutions given by $\chi(s',t).$ The coefficient of
$\phi(s',t)$ in the third line on the RHS is independent of
coordinate $s'$, and gives rise to a simple time-dependent phase
factor. Finally, the second line on the RHS can be eliminated by
choosing the transformation coordinate, $\xi$, to be the solution of
the equation,
\begin{equation}\label{eqn:classfho}
\ddot{\xi} + \omega^2(t)\xi  +\ddot{x}_0 = 0.
\end{equation}
This is the classical equation of a forced, parametric harmonic
oscillator, where $\xi$ is identified as the classical position of
an ion relative to the moving potential minimum.

Combining the observations made above, we see that the wave equation
for $\phi(s',t)$ is in fact separable, and its solution can be
written down in terms of the solutions $\chi (s',t)$ of the unforced
parametric oscillator equation, and the phase factor from the
remaining time-dependent terms in Eq.~\ref{eqn:phisprimeandt}:
\begin{equation}\label{eqn:phiofsandt}
\phi(s',t)  = \chi(s',t) \exp\left\{\frac{i}{\hbar} \int_{t_0}^t
m/2\left(\dot{\xi}^2 + \dot{x}_0^2 - \omega^2(t)\xi^2 - 2\ddot{x}_0
\xi \right) dt \right\}.
\end{equation}
Recalling the canonical transformation introduced above, the full
solution to the time-dependent wave function, $\psi(s',t)$, is then
found to be
\begin{equation}\label{eqn:psiofsandt}
\psi(s',t) = \chi(s',t) \exp\left\{\frac{i}{\hbar}ms'\dot{\xi} +
\frac{i}{\hbar}\int_{t_0}^t m/2\left(\dot{\xi}^2 + \dot{x}_0^2 -
\omega^2(t)\xi^2 - 2\ddot{x}_0 \xi \right) dt \right\}.
\end{equation}
Therefore, the problem of finding the wavefunction of the forced,
parametric oscillator as a function of time has been reduced to one
of finding the quantum mechanical solution, $\chi(s',t)$, for the
unforced parametric oscillator and the classical solution, $\xi(t)$,
of the forced, parametric oscillator. As described in the
introduction to this section, we wish to obtain the generating
functions for the matrix elements describing the transition from the
initial to the final state of the ion. This is facilitated by a
propagator approach to describe the time-evolution of the quantum
mechanical state of the ion.

\subsubsection{\label{sec:generatingfunction}The method of generating functions}

We begin our derivation of the generating functions for the
transition probabilities in unforced and forced parametric
oscillators by recalling the propagator for the simple harmonic
oscillator \cite{sakurai}:
\begin{equation}\label{eqn:shogreen}
\psi(x,t) =  \int K_{sho}(x,t|x',t') \psi(x',t') dx'.
\end{equation}
The propagator $K_{sho}(x,t|x',t')$ satisfies the time-dependent
Schr\"{o}dinger's equation for the harmonic oscillator, and is given
by \cite{sakurai}
\begin{eqnarray}\label{eqn:transforsho}
K_{sho}(x,t|x',t') &=& \sqrt{\frac{m\omega_0}{2\pi i\hbar\
\sin (\omega_0 \Delta t)}} \nonumber \\
&\times& \exp \left\{ \frac{i m \omega_0}{2\hbar \sin (\omega_0
\Delta t)} \left(x^2\cos (\omega_0\Delta t) - 2xx' +{x'}^2 \cos
(\omega_0 \Delta t) \right)\right\},
\end{eqnarray}
where $\Delta t = t - t'.$ The probability amplitude for the simple
harmonic oscillator, initially in a pure state $\psi(x',t')$, to be
in the $n$-th eigenstate at time $t$  is then given by a double
integral over $x$ and $x'$
\begin{equation}\label{eqn:matrixelementssho}
b_{n}(t,t') = \int\int \psi_n^*(x)K_{sho}(x,t|x',t')\psi(x',t')\,dx
\,dx' .
\end{equation}
The probabilities for the simple harmonic oscillator to be in the
$n$-th state are then
\begin{equation}\label{eqn:probsho}
P_{n}(t,t') = |b_{n}(t,t')|^2.
\end{equation}

In the case that the particle is initially in the $k$-th eigenstate
of the harmonic oscillator, the expressions for the probability and
probability amplitude of the system being in the $n$-th state in
Eqs. \ref{eqn:probsho} and \ref{eqn:matrixelementsnk} can be thought
of as transition probabilities for the evolving system. Of course,
for a stationary quadratic potential with a fixed frequency, the
transition probabilities would be
\begin{equation}\label{eqn:transprobsho}
P_{nk}(t,t') = |b_{nk}(t,t')|^2 = \delta_{nk}.
\end{equation}
However, when the ion is shuttled and experiences a nonuniform
acceleration and/or a changing trap frequency, we can expect
transitions from one eigenstate of the harmonic oscillator to
another. The key to determining those transition probabilities is
the propagator $K(x,t|x',t')$ for the shuttling potential, which is
a solution of the Schr\"{o}dinger's equation for the shuttling
potential:
\begin{equation}\label{eqn:SchrodingerforGreen}
\imath \hbar \frac{\partial K(x,t|x',t') }{\partial t} =
-\frac{\hbar^2}{2m} \frac{\partial^2 K(x,t|x',t')}{\partial x^2} +
\left(\frac{1}{2} m \omega^2 x^2 + m\ddot{x}_0(t)x\right)
K(x,t|x',t'),
\end{equation}
and satisfies $K(x,t'|x',t') = \delta(x-x')$. Assuming this function
is known, the transition amplitudes for the ion to begin in the
$k$-th state of the harmonic potential at time $t'$ and then after
being shuttled to end up in the $n$-th state of the harmonic
potential at time $t$ are
\begin{equation}\label{eqn:matrixelementsnk}
b_{nk}(t,t') = \int\int \psi_n^*(x)K(x,t|x',t')\psi_k(x') \, dx \,
dx' .
\end{equation}
This expression for the transition amplitudes can be used to
construct a generating function, $B(u,v),$ for the transition
amplitudes of the shuttled ion. We start by using the known
generating function for the eigenfunctions of a simple harmonic
oscillator with frequency $\omega_0$ (\mbox{e.g.},
Husimi~\cite{husimi}, Eq. 4.6):
\begin{equation}\label{eqn:generatingfunctionsholinear}
\sqrt{\alpha} e^ { -u^2 + 2\alpha ux  - \alpha^2x^2 }
 = \sum_{n}\sqrt{\frac{\sqrt{\pi} 2^n} {n!}} u^n \psi_n ( \alpha x),
\end{equation}
where $\alpha = \sqrt{m\omega_0/\hbar}$ and $|u| \leq 1.$ We
multiply both sides of Eq.~\ref{eqn:matrixelementsnk} by
$\sqrt{\frac{\pi 2^{n+k}} {n!k!}}u^kv^n$ and then by summing both
sides over $k$ and $n$, we have
\begin{eqnarray}\label{eqn:generatingfunctionU}
B(u,v) &=& \sum_{k,n}\sqrt{\frac{\pi 2^{n+k}} {n!k!}} u^k v^n b_{nk}(t,t')\nonumber \\
&=& \alpha \int dx \, dx' K(x,t|x',t') \exp\left\{-u^2 - v^2 +
2\alpha (ux'+ vx)  - \alpha^2({x'}^2+x^2)\right\}.
\end{eqnarray}
Notice that the generating function $B(u,v)$ is a function of the
initial and end times of the shuttling protocol as well. Once the
propagator for the potential is known, any particular transition
amplitude can be obtained from this generating function by expanding
it about the parameters $u$ and $v$ and reading off the transition
amplitude $b_{nk}$ from the coefficient of the $u^kv^n$ term in the
expansion. A similar generating function can be developed for the
transition probabilities by treating the probabilities $P_{nk}$ as
coefficients in a double power series in $u$ and $v$, and then using
the integral obtained for $b_{nk}$ in Eq.~\ref{eqn:matrixelementsnk}
so that
\begin{eqnarray}\label{eqn:generatingfunctionP}
P(u,v) &=& \sum_{k,n} u^k v^n P_{nk}(t,t') = \sum_{k,n} u^k v^n
|b_{nk}(t,t')|^2\nonumber \\
&=& \int\int\int\int dx \, dx' \, dy \, dy' K(x,t|x',t')K^*(y,t|y',t') \nonumber\\
&\times & \left\{\sum_n v^n \psi_n^*(\alpha x) \psi_n(\alpha
y)\right\} \left\{\sum_k u^k \psi_k^*(\alpha x') \psi_k(\alpha
y')\right\}.
\end{eqnarray}
Each term in braces in the bottom line of
Eq.~\ref{eqn:generatingfunctionP} can be replaced for $|u|,|v|\leq
1$ by the bilinear generating function (see Husimi~\cite{husimi},
Eq. 4.4)
\begin{equation}\label{eqn:generatingfunctionshobilinear}
\sqrt{\frac{\alpha^2}{\pi (1-u^2)}} \exp \left\{-\alpha^2
\frac{(1+u^2)(x^2+y^2) - 4uxy }{2(1-u^2)} \right\} = \sum_{n} u^n
\psi_n \left( \alpha y \right) \psi_n^* \left(\alpha x \right).
\end{equation}
Making the substitution, we obtain the generating function for the
transition probabilities,
\begin{eqnarray}\label{eqn:generatingfunctionPfinal}
P(u,v) &=& \frac{\alpha^2}{\pi \sqrt{(1-u^2)(1-v^2)}}
\int\int\int\int
dx \, dx' \, dy \, dy' K(x,t|x',t')K^*(y,t|y',t') \nonumber\\
&\times & \exp \left\{-\alpha^2\left( \frac{(1+u^2)(x^2+y^2) - 4uxy
}{2(1-u^2)} + \frac{(1+v^2) ({x'}^2+{y'}^2) - 4vx'y'
}{2(1-v^2)}\right) \right\}.
\end{eqnarray}
Again, once the propagators for the shuttling potentials are known,
the generating function for the transition probabilities can be
obtained by carrying out the fourfold integral on the RHS of
Eq.~\ref{eqn:generatingfunctionPfinal}.

One can obtain directly the average final state of the shuttled ion
by manipulating this expression as follows:
\begin{equation}\label{eqn:avgfinalstate}
\left.\frac{\partial P(u,v)}{\partial v} \right|_{v=1} = \sum_k
u^k\sum_n n P_{nk}(t,t') = \sum_k u^k \langle n_k \rangle.
\end{equation}
By expanding the term on the LHS in powers of $u$, one can read off
for an ion which started in the $k$-th eigenstate of the trapping
potential its average final state, defined to be $\langle n_k
\rangle.$ If we make the further assumption that the oscillator
started out in the ground state, $k = 0$, we only need the term in
the expansion which has no dependence on $u$. We can find this term
easily by setting $u = 0,$ so that only that part of $P(u,v)$ which
doesn't depend on $u$ survives. Thus, the average final state for an
ion which started in the zeroth eigenstate of the trapping
potential, $\langle n_0 \rangle,$ is found by setting $u = 0$ in the
expression on the LHS of Eq.~\ref{eqn:avgfinalstate},
\begin{equation}\label{eqn:avgfinalstatek0}
\langle n_0 \rangle = \sum_n n P_{n0}(t,t') = \left.\frac{\partial
P(0,v)}{\partial v} \right|_{v=1}.
\end{equation}
We can also find the distribution of the wavefunction about the mean
for the final motional state by manipulating the generating
function, $P(u,v)$. For an arbitrary initial state, $k$, we can find
the average value of $n(n-1) = n^2-n$, defined as $\langle n^2_k
\rangle - \langle n_k \rangle$, by taking the second derivative of
the generating function with respect to $v$, and then evaluating it
for $v=1,$
\begin{equation}\label{eqn:avgnnmin1}
\left.\frac{\partial^2 P(u,v)}{\partial v^2} \right|_{v=1}= \sum_k
u^k \sum_n n(n-1) P_{nk}(t,t') = \sum_k u^k \left(\langle
n_k^2\rangle -\langle n_k\rangle \right) ,
\end{equation}
from which we can easily obtain the distribution of the final ion
state about the mean, $\langle n_k^2 \rangle - \langle
n_k\rangle^2$. In the particular case that $k=0$, we can obtain the
distribution about the mean, $\langle \Delta n_0^2 \rangle,$
directly from derivatives with respect to $v$ of the generating
function $P(0,v)$ as in Eq.~\ref{eqn:avgfinalstatek0} above,
\begin{eqnarray}\label{eqn:avgdistrib}
\langle \Delta n_0^2 \rangle &=& \langle n_0^2\rangle -
\langle n_0 \rangle^2 \nonumber \\
&=& \left.\frac{\partial^2 P(0,v)}{\partial v^2} \right|_{v=1} +
\left.\frac{\partial P(0,v)}{\partial v} \right|_{v=1} - \left(
\left.\frac{\partial P(0,v)}{\partial v} \right|_{v=1}\right)^2.
\end{eqnarray}
Thus, the method of generating functions is a powerful way to obtain
the average final state and the distribution about the mean of the
final state of a forced parametric oscillator. In the particular
case that the ion was initially in the ground state, these values
can be obtained directly from first and second order derivatives of
$P(0,v)$ with respect to the parameter $v$, with $v$ subsequently
set equal to 1. These values can be written down in closed form
expressions if the propagator for the forcing potential is known and
the integrals for the generating functions are solvable.

\subsubsection{\label{sec:classical} Classical solutions for the unforced and forced parametric oscillator}

As we showed in Sec.~\ref{sec:solution}, the quantum mechanical
solutions of the unforced and forced harmonic oscillator problem are
expressed in terms of the classical quantities describing the motion
of these systems. Therefore, we turn our attention to the solution
of the classical forced parametric oscillator equation, given in
Eq.~\ref{eqn:classfho}. We specify $\xi$ as that solution of the
forced parametric oscillator for which the initial conditions
$\xi(t_0,t_0) = \dot{\xi}(t_0,t_0) = 0$ hold in the frame of the
moving potential. This will serve to cause the phase factor in
Eq.~\ref{eqn:psiofsandt} to vanish at time $t_0$. This solution can
be obtained by considering first the homogeneous equation, which is
the unforced parametric oscillator equation,
\begin{equation}\label{eqn:classufho}
\ddot{X} =- \omega(t)^2 X.
\end{equation}
There exist two independent solutions $X_1(t)$ and $X_2(t)$ of
Eq.~\ref{eqn:classufho}, which satisfy the initial conditions
\{$X(t_0)=0, \dot{X}(t_0)=1$\} and \{$X(t_0)=1,\dot{X}(t_0)=0$\},
respectively. These solutions have the property that, for any time
$t>t_0$
\begin{equation}\label{eqn:x1x2properties}
X_1X_2 - X_2X_1 = 0; \hspace{0.1 in}\mbox{and}\hspace{0.1 in}
\dot{X}_1X_2 - X_1\dot{X}_2 = 1,
\end{equation} where the first property is obvious and the second property is derived, using Eq.~\ref{eqn:classufho} and the first property, as follows:
\begin{eqnarray}\label{eqn:x1x2combine}
X_2(t)\ddot{X}_1(t) - X_1(t)\ddot{X}_2(t) &=& -\omega(t)^2\left(X_1(t)X_2(t) - X_2(t)X_1(t)\right) = 0 \nonumber\\
\frac{d}{dt} \left(\dot{X}_1(t)X_2(t) - X_1(t)\dot{X}_2(t) \right) &=& 0 \nonumber\\
\dot{X}_1(t)X_2(t) - X_1(t)\dot{X}_2(t) &=& \dot{X}_1(t_0)X_2(t_0) -
X_1(t_0)\dot{X}_2(t_0) = 1.
\end{eqnarray}
Therefore, these solutions can be used to construct a
one-dimensional Green's function,
\begin{equation}\label{eqn:Greendef}
G(t,t') = X_1(t)X_2(t')-X_1(t')X_2(t),
\end{equation} for $t\geq t' \geq t_0$ which has the properties
\begin{equation}\label{eqn:Greenprop}
G(t',t') = 0; \hspace{0.1 in}\mbox{and}\hspace{0.1 in} \frac{d
G(t',t')} {dt} = 1 .
\end{equation}
This Green's function, which has the dimensions of time, can be
shown to be \cite{boas} the solution of the parametric oscillator
equation with delta function forcing,
\begin{equation}\label{eqn:Greeneqn}
\ddot{G}(t,t') + \omega(t)^2 G(t,t') = \delta(t-t'),
\end{equation} where $G(t,t') = 0,$ for $t<t'$ to satisfy causality.
The solution for $G$ when $t>t'$ then represents the response of the
oscillator to a unit impulse occurring at time $t'$. In the simplest
case for which the frequency of the potential is fixed at
$\omega_0$, we have
\begin{eqnarray}\label{eqn:Simplegreen}
X_1(t,t_0; \omega_0 ) &=& \frac{1}{\omega_0}\sin(\omega_0 (t-t_0)),\nonumber\\
X_2(t,t_0; \omega_0) &=& \cos(\omega_0 (t-t_0)),\nonumber\\
G(t,t') &=& \frac{1}{\omega_0}\sin(\omega_0( t - t')).
\end{eqnarray} Once the appropriate $G(t,t')$ has been obtained,
the solution for $\xi(t,t_0)$ satisfying Eq.~\ref{eqn:classfho} can
now be constructed as follows:
\begin{eqnarray}\label{eqn:xiandxo}
\xi(t,t_0) &=& -\int_{t_0}^t \ddot{x}_0(t')G(t,t') dt' \nonumber \\
\dot{\xi}(t,t_0) &=& -\int_{t_0}^t \ddot{x}_0(t')\frac{\partial
G(t,t')} {\partial t} dt'.
\end{eqnarray} The classical energy gain of the ion due to forcing,
relative to the characteristic energy of the harmonic potential, is
therefore
\begin{equation}\label{eqn:E}
\Upsilon(t,t_0) = \frac{m}{2\hbar \omega(t)}
\left(\omega(t)^2\xi(t,t_0)^2 + \dot{\xi}(t,t_0)^2\right).
\end{equation} In order to isolate the influence of the frequency
variation on the shuttled ion's energy, we switch the role of $t$
and $t_0$ in Eq.~\ref{eqn:xiandxo}. This can be understood as the
motion of the ion when the sequence of frequency variation is
reversed. The need for this arises from the fact that for the forced
motion the energy gain is not only a function of the end time, but
also the initial time. Therefore, we have \cite{husimi},
\begin{eqnarray}\label{eqn:etaandxo}
\xi(t_0,t) &=& -\int_t^{t_0} \ddot{x}_0(t') G(t_0,t') dt' \\
\dot{\xi}(t_0,t) &=& \frac{\partial \xi (t_0,t)}{\partial t_0} =
-\int_t^{t_0} \ddot{x}_0(t')\frac{\partial G(t_0,t')}{\partial t_0}
dt',
\end{eqnarray} where the Green's function is now non-zero for times
earlier than the time of the impulse, that is, for $t_0<t'$. The
energy gain for the reversed forced motion is therefore
\begin{equation}\label{eqn:ER}
\Upsilon(t_0,t) = \frac{m}{2\hbar \omega(t)}
\left(\omega(t)^2\xi(t_0,t)^2 + \dot{\xi}(t_0,t)^2\right).
\end{equation} For a constant frequency potential this reversed motion
results in the same energy gain as does the forward motion, and
$\Upsilon(t,t_0) -\Upsilon(t_0,t) = 0.$ However, in general when the
frequency is time-dependent, $\Upsilon(t,t_0) - \Upsilon(t_0,t) \neq
0$. The generating functions for the transitions induced in the
unforced and forced parametric oscillator which we obtain in the
next section depend precisely on the dimensionless energies
characterizing the classical energy gain of these systems.

\subsubsection{\label{sec:transition} Transition probabilities for the unforced and forced parametric oscillator}

Since we wish to solve the unforced parametric oscillator problem
first, we work in the reference frame of the minimum of the
potential used to shuttle the trapped ion. It is assumed that the
ion starts out at time $t_0$ in a pure eigenstate,
$\psi_k(s_0,t_0),$ of a harmonic oscillator of constant frequency
$\omega_0$, and that it ends up in a potential well of the same
frequency at time $t$ and position $s$ relative to the potential
minimum in some superposition of eigenstates. The connection between
the final state of the particle and its initial state can be
expressed in terms of the propagator $K(s,t|s_0,t_0), $ for the
shuttling potential. We begin by returning to the propagator for the
simple harmonic oscillator, which corresponds to a shuttling
potential moving at constant velocity and keeping a constant
frequency $\omega_0$, (see Eq.~\ref{eqn:transforsho})
\begin{eqnarray}\label{eqn:transforsho2}
K_{sho}(s,t|s_0,t_0) &=& \sqrt{\frac{\alpha^2}{2\pi i
\sin (\omega_0 \Delta t)}} \nonumber \\
&\times& \exp \left\{ \frac{i m \omega_0}{2\hbar \sin (\omega_0
\Delta t) } \left(s^2 \cos(\omega_0 \Delta t)  - 2ss_0 +
s_0^2\cos(\omega_0 \Delta t)\right)\right\},
\end{eqnarray} where $\Delta t = t - t_0.$
By comparing the functions $\sin(\omega_0\Delta t)/\omega_0$ and
$\cos(\omega_0\Delta t)$ in Eq.~\ref{eqn:transforsho2} with
solutions of the unforced parametric oscillator as given in the
limiting case of Eq.~\ref{eqn:Simplegreen} when the frequency is
constant, we can guess that the propagator $K_{sho}$ is just a
special case of the general propagator for the unforced parametric
oscillator
\begin{equation}\label{transforpo}
K_{upo}(s,t|s_0,t_0) = \sqrt{\frac{m}{2\pi i\hbar X_1(t,t_0)}} \exp
\left\{ \frac{im }{2\hbar X_1(t,t_0)} \left(\dot{X}_1(t,t_0) s^2 -
2ss_0 + X_2(t,t_0) s_0^2 \right) \right\}.
\end{equation}
Husimi (\cite{husimi}, Eq. 3.8) showed that this is indeed the case.
Substituting this propagator into
Eq.~\ref{eqn:generatingfunctionPfinal} results in a fourfold
Gaussian integral, which can be evaluated using the formula
\begin{equation}\label{eqn:multigauss}
\int_{-\infty}^\infty
\exp\left\{-\frac{\sum_{i,j}^4A_{ij}x_ix_j}{2}\right\}d^4x =
\frac{(2\pi)^{2}}{\sqrt{\det \{{\bf A}\}}},
\end{equation} where the $4\times 4$ matrix {\bf A} is symmetric
and positive-definite. The generating function for the transition
probabilities for an ion in a variable-frequency harmonic potential
is then
\begin{equation}\label{eqn:Puvpo}
P(u,v)(t,t_0) = \sqrt{2}\left\{Q(t,t_0)(1-u^2)(1-v^2) +
(1+u^2)(1+v^2) - 4uv\right\}^{-1/2},
\end{equation} where
\begin{equation}\label{eqn:Q}
Q(t,t_0) = \frac{1}{2} \left(\omega_0^2X_1^2 + \dot{X}_1^2 + X_2^2 +
\frac{1}{\omega_0^2}\dot{X}_2^2 \right).
\end{equation} The generating function in Eq.~\ref{eqn:Puvpo} is even
in the following sense:
\begin{equation}\label{eqn:symmetry}
P(-u,-v) = P(u,v).
\end{equation}
Therefore the transition probabilities are non-zero only for
beginning and ending states of the same parity, resulting in the
expected selection rule, $n -k =2 m,$ for $m$ an integer. We have
$Q\geq 1,$ where the equality holds when the frequency is constant.
In a classical parametric oscillator the quantity $Q$ represents the
proportional increase in energy due to frequency variation over an
interval of duration $T$, averaged over all possible initial
conditions having the same initial energy, $E_\omega(t_0)$, so that
(\cite{husimi}, Eq. 5.21)
\begin{equation}\label{eqn:classicalinterpofQ}
Q(t,t_0) = \frac{\langle E_\omega(T+t_0) \rangle}{E_\omega(t_0)}
\geq 1.
\end{equation}

We now seek the average final state $\langle n_k \rangle$ for an ion
in such a variable-frequency harmonic potential with negligible
inertial forcing, given that its initial state was the $k$-th
eigenstate of the initial trap. Using Eq.~\ref{eqn:avgfinalstate},
we determine that
\begin{eqnarray}\label{eqn:findEnergypo}
\sum_k u^k\sum_n n P_{nk}(t,t_0) &=& \left.\frac{\partial
P(u,v)}{\partial v}
\right|_{v=1} \nonumber \\
&=& \frac{1}{2(1-u)^2}\left((1+u)Q(t,t_0) -(1-u) \right).
\end{eqnarray} Expanding the function on the right in powers of $u$ allows
us to identify for each initial state $k$
\begin{equation}\label{eqn:nkpo}
\langle n_k \rangle = (k+1/2)Q(t,t_0) - 1/2
\end{equation}
and therefore
\begin{equation}\label{eqn:Energypo}
\langle E_\omega (T+t_0) \rangle = \left(\langle n_k
\rangle+\frac{1}{2}\right)\hbar \omega_0 = Q(t,t_0) \left(k +
\frac{1}{2}\right) \hbar \omega_0,
\end{equation} exactly corresponding to the classical result
(Eq.~\ref{eqn:classicalinterpofQ}). The distribution of the
wavefunction about the mean is found as described at the end of
Sec.~\ref{sec:generatingfunction}, and is given by
\begin{equation}\label{eqn:n2kpo}
\langle \Delta n_k^2 \rangle = 1/2(Q^2(t,t_0) - 1)(k^2+k+1).
\end{equation}
Both $\langle n_k \rangle $ and $\langle \Delta n_k^2 \rangle $ for
the parametrically-driven ion found here are functions of the
initial and final time of the shuttling through the factor
$Q(t,t_0).$ This function in turn depends on time through the
solutions $X_1$ and $X_2$. These solutions can be found analytically
for certain models of the frequency variation of the shuttling
potential (\mbox{e.g.}, Eq.~\ref{eqn:modelfreqvar}). In general,
however, they need to be evaluated numerically by integrating the
classical parametric oscillator equation over the duration of the
shuttling protocol.

The propagator for the forced parametric oscillator can be obtained
from the one for the unforced parametric oscillator by using the
fact that the propagator for each is a solution of the
Schr\"{o}dinger's equation for the corresponding shuttling
potential. Therefore, the propagator for the unforced parametric
oscillator potential (with $\ddot{x}_0=0$) is a solution of the
Schr\"{o}dinger's equation in Eq.~\ref{eqn:SchrodingerforGreen} with
$\ddot{x}_0=0$. Not only so, but it is a solution in the coordinate
system defined in Eq.~\ref{eqn:coordtrans} of the Schr\"{o}dinger's
equation found in the first line of Eq.~\ref{eqn:phisprimeandt}. The
solution to that equation, the quantum mechanical version of the
unforced parametric oscillator equation, was identified in the text
as $\chi(s',t)$. Any equation which $\chi(s',t)$ satisfies is also
satisfied by the propagator for the unforced parametric oscillator.
Therefore, the solution for the wave function of the forced
parametric oscillator obtained in terms of $\chi(s',t)$ in
Eq.~\ref{eqn:psiofsandt} can also be used to obtain the propagator
for the forced oscillator in terms of the propagator of the unforced
oscillator
\begin{equation}\label{eqn:GreenunforcedtoGreenforced}
K_{fpo}(s',t|s_0',t_0) = K_{upo}(s',t|s'_0,t_0)
e^{\frac{i}{\hbar}ms'\dot{\xi} + \frac{i}{\hbar}\int_{t_0}^t
m/2\left(\dot{\xi}^2 + \dot{x}_0^2 - \omega(t)^2\xi^2 - 2\ddot{x}_0
\xi \right) dt}.
\end{equation}
Thus, the propagator for the forced parametric oscillator is given
by
\begin{eqnarray}\label{transforforcedpo}
K_{fpo}(s',t|s'_0,t_0) &=& \sqrt{\frac{m}{2\pi i\hbar X_1}} \nonumber\\
& \times & \exp \left\{ \frac{i m }{2\hbar X_1} \left(\dot{X}_1
s'^2+ X_2 {s'_0}^2  -
2s's'_0  \right) \right\} \nonumber \\
& \times & \exp \left\{ \frac{i m}{\hbar}\left(s'\dot{\xi} + \left[
\int_{t_0}^t 1/2\left( \dot{\xi}^2 + \dot{x}_0^2 - \omega(t)^2\xi^2
- 2\ddot{x}_0 \xi \right) dt \right]\right) \right\}.
\end{eqnarray}
Using Eqs.~\ref{eqn:generatingfunctionP} and
\ref{eqn:generatingfunctionPfinal}, the generating function for the
transition probabilities of an ion in the forced parametric
oscillator potential can be obtained (\cite{husimi}, Eq. 7.13)
\begin{eqnarray}\label{eqn:Pfpouv}
P(u,v) &\equiv& \sum_{n,k=0}^\infty u^k v^n P_{nk}(t,t_0) \nonumber\\
&=& \sqrt{\frac{2}{(1-u^2)(1-v^2)Q(t,t_0)+(1+u^2)(1+v^2)-4uv}} \nonumber \\
& \times & \exp \left[
-\frac{(1-u^2)(1-v^2)\left\{\Upsilon(t_0,t)\frac{1-v}{1+v} +
\Upsilon(t,t_0)\frac{1-u}{1+u}\right\}} {\left\{
(1-u^2)(1-v^2)q(t,t_0)+(1+u^2)(1+v^2)-4uv \right\}} \right].
\end{eqnarray}
This solution was extended to the case that the initial and final
frequency of the forced oscillator is different in Perelomov
\cite{perelomov}. A simpler expression is obtained if we let $u =0,$
corresponding to the case that the forced ion was initially in the
ground state
\begin{eqnarray}\label{eqn:Pfpouzerov}
P(0,v) &\equiv& \sum_{n} v^n P_{n0}(t,t_0) \nonumber\\
&=& \sqrt{\frac{2}{Q(1-v^2)+(1+v^2)}} \exp \left[
-\frac{(1-v^2)\left\{\Upsilon(t_0,t)\frac{1-v}{1+v} +
\Upsilon(t,t_0)\right\}} {\left\{ (1-v^2)Q(t,t_0)+(1+v^2) \right\}}
\right].
\end{eqnarray}
Again, the average final motional state of the forced ion $\langle
n_0 \rangle$ can be obtained as in Eq.~\ref{eqn:avgfinalstatek0},
with the simple result,
\begin{eqnarray}\label{eqn:avgnfpo}
\langle n_0(T) \rangle  &=& \Upsilon(T,0) + \frac{1}{2} \left(Q(T,0)
- 1\right) ,
\end{eqnarray} where we have let $t_0 = 0.$
The distribution of the ion's wavefunction about its average
motional state after the total shuttling time $T$ is found as
described in Eq.~\ref{eqn:avgdistrib} and is given by
\begin{equation}\label{avgdeltanpo}
\langle \Delta n_0^2(T) \rangle = \frac{1}{2}\left(Q^2(T,0) -
1\right) + \left(2\Upsilon(T,0)Q(T,0) - \Upsilon(0,T)\right).
\end{equation}
In the case that the frequency of the potential remains constant,
these results for the final average state and state distribution
reduce to the following
\begin{eqnarray}\label{eqn:Energyfo}
\langle n_0(T) \rangle  &=& \Upsilon(T,0) , \\
\langle \Delta n_0^2(T) \rangle &=&  \Upsilon(T,0),
\end{eqnarray} since $\Upsilon(t,t_0) - \Upsilon(t_0,t) = 0$ and $Q(t,t_0) = 1$ when
$\omega = \omega_0.$

Remarkably, the impact of the frequency variation on the final
energy and dispersion of the ion is largely separable from that of
the inertial forcing due to shuttling. Therefore, we can profitably
treat the impact of each aspect of the shuttling process separately.
For the shuttling protocols outlined in
Sec.~\ref{sec:shuttlingprotocols} we can obtain closed form
expressions for $ \langle n_0(T) \rangle $ and $ \langle \Delta
n_0^2(T) \rangle$ for the shuttled ion in the cases when the
frequency is held constant or the inertial forcing is negligible. In
general, the factors $\Upsilon(T,0)$ and $Q(T,0)$ must be obtained
numerically by integrating the expressions in
Eqs.~\ref{eqn:classufho}, \ref{eqn:E} and \ref{eqn:ER} over the
entire shuttling interval. We emphasize that both of these
quantities are obtained from a classical analysis of the unforced
parametric oscillator.

\begin{table}[tb!]\label{theorytab}
\centering
\begin{tabular}{|c|c|}
\hline
Important Equations of Section 3 & Equation\\
\hline
forced oscillator, constant frequency & {}\\
\hline
propagator & 45\\
transition probability & 55\\
$\langle n_0 \rangle$ & 57\\
$\langle \Delta n_{0}^{2} \rangle$ & 59\\
\hline
unforced parametric oscillator & {}\\
\hline
propagator & 73\\
transition probability & 75\\
$Q(t,t_0)$ & 76\\
$\langle n_k \rangle$ & 80\\
$\langle \Delta n_{k}^{2} \rangle$ & 82\\
\hline
forced parametric oscillator & \\
\hline
propagator & 84\\
transition probability & 85\\
$\Upsilon(t, t_0), \Upsilon(t_0,t)$ & 68, 71\\
$\langle n_0 \rangle$ & 87\\
$\langle \Delta n_{0}^{2} \rangle$ & 90\\
\hline
\end{tabular}
\caption[]{\label{theorysummary}Table summarizing the locations of important equations in Section 3.
The propagators used to calculate the transition probabilities as well the transition
probabilities are given for forced oscillators and forced parametric oscillators as well
as unforced parametric oscillators are given. The expectation value of the oscillator motional
state as well as the variances in the motional state are given too. The subscript associated
with these quantities refer to the initial state of the ion before shuttling. In the unforced
parametric oscillator case, $Q(t,t_0)$ represents the proportional increase in energy due to frequency
variation over an interval of duration T, averaged over all possible initial conditions having
the same initial energy (see Eq. \ref{eqn:classicalinterpofQ}). In the forced parametric oscillator case,
the variable $\Upsilon(t,t_0)$ ($\Upsilon(t_0,t)$) represents the classical energy gain of the
ion due to forcing  with the sequence (reversed sequence) of the frequency variation relative to the characteristic energy of the harmonic potential. The classical motion of the ion is needed to calculate the transition
probabilities in the forced and unforced parametric oscillators, so it is important to be able to
accurately simulate the classical motion of trapped ions during shuttling. See section 2 and the
appendix for more information on simulating the classical motion of trapped ions.}
\end{table}

\subsection{\label{sec:adiabatic} Evaluation of Shuttling Protocols}

Having outlined the formalism for determining the effect of
shuttling in one dimension on the motional state of the ion, we now consider the
shuttling protocols developed in Sec.~\ref{sec:shuttlingprotocols}.
We first briefly establish several criteria for effective shuttling,
and then evaluate the relative merits of the shuttling protocols.

\subsubsection{\label{sec:adiabaticLD} Shuttling criteria}

Heating during the shuttling operation will typically occur along the
longitudinal direction (along the shuttling path). For shuttling along a
line, there is no cross talk between transverse and longitudinal heating as
the longitudinal direction corresponds to a principal axis. However,
transport of an ion through a junction will couple the two modes and our
considerations will provide an upper limit of the change of motional state
in any spatial directions after the shuttling operation.

The first and most restrictive limit on a shuttling operation is the
requirement that it produce little change in the motional state of
the ion, or
\begin{equation}\label{eqn:adiabaticlimit}
    \langle n \rangle \ll 1
\end{equation}
for an ion initially prepared in the ground state. Although this
constraint can be met by non-adiabatic processes through appropriate
phasing of the shuttling forces, it is the ultimate intention of the
adiabatic limit, and for simplicity we will call it the adiabatic
constraint.

A second and typically less restrictive limit is that the rms spread
$s_{\mbox{rms}}$ in the ion's final wavepacket remain small compared
to the relevant optical wavelengths used in the quantum information
environment. This is known as the Lamb-Dicke limit, and can be an
important criterion for the effective coupling of light fields to
the motion of trapped ions. For a coherent state or a thermal state
of harmonic motion with mean vibrational number $\langle n \rangle$,
\begin{equation}
s_{\mbox{rms}} = \sqrt{\langle \psi(s,T)\vert s^{2}\vert \psi(s,T)
\rangle}
        = \sqrt{\frac{\hbar}{2m\omega_0}( 2\langle n \rangle + 1 )} \, ,
\end{equation}
so for an ion initially in the ground state, the Lamb-Dicke
criterion can thus be written as
\begin{equation}\label{eqn:zetaEndprime}
\frac{\hbar k^{2}}{2m \omega_0} (2\langle n \rangle + 1)  \ll 1 \, ,
\end{equation}
where $k$ is the effective wave-number associated with the radiation
field in the quantum gate scheme. The Lamb-Dicke limit sets a more
meaningful limit on the required localization of the ion for many
quantum logic gate schemes \cite{molmer:1999, sackett:2000,
leibfried:2003, haljan:2005, leibfried:2005, steane:2006}.

A third and still less restrictive constraint is that the residual
motion of the ion after shuttling does not add to the diffraction
limit of the ion image.  This condition is important for schemes
that couple ion qubits through emitted photons that might be
mode-matched into an optical fiber \cite{blinov:2004,
MoehringJOSAB}.  A conservative estimate of this condition is the
usual Rayleigh criterion
\begin{equation}\label{eqn:diffraction}
s_{\mbox{rms}} \ll \frac{0.61 \lambda}{\mbox{NA}}
\end{equation}
where $\lambda$ is the radiation wavelength and $\mbox{NA} < 1$ is
the numerical aperture of the imaging objective.  This diffraction
limit condition can be written in a form similar to the Lamb-Dicke
criterion above:
\begin{equation}\label{eqn:diff2}
0.068(NA)^2\frac{\hbar k^{2}}{2m \omega_0} (2\langle n \rangle + 1)
\ll 1.
\end{equation}

The last, and typically least restrictive constraint is that the
motion of the ion remains harmonically bound in the trap. Anharmonic
wavepacket dispersion can give rise to errors in certain ultrafast
quantum gate schemes \cite{GZC}.  This condition requires that the
ion motion be localized to a region of space much smaller than the
characteristic distance from ion to trap electrode $d_{\mbox{eff}}$:

\begin{equation}\label{eqn:anh}
\frac{\hbar}{2m \omega_0 d_{\mbox{eff}}^2} \left( 2\langle n \rangle
+ 1 \right) \ll 1.
\end{equation}

Now consider those features of the shuttling process which would
make it more likely to satisfy the theoretical criteria, regardless
of the particular shuttling protocol used. From Eqs. \ref{eqn:E} and
\ref{eqn:adiabaticlimit}, the adiabatic constraint favors low mass
ions such as beryllium or calcium for a given trap frequency. The
Lamb-Dicke and diffraction criteria favor atomic ions that feature
longer-wavelength electronic transitions (Eqs.
\ref{eqn:zetaEndprime} and \ref{eqn:diff2}). In addition, for a
given atom any shuttling protocol can be made faster while making it
less likely for the ion to be placed into an excited state by
increasing the axial trap frequency, $\omega_0.$ The Lamb-Dicke and
diffraction constraints particularly benefit from such an increase.
This improvement is limited only by the risk of destabilizing the rf
transverse trap. Once an ion species and an optimal trap frequency
are chosen, however, the focus turns to the particular functional
form of the shuttling protocol and the manner with which the
frequency of the shuttling potential varies during the
implementation of the protocol.

\subsubsection{\label{sec:adiabaticatconstantomega} Shuttling in constant frequency potentials}

It is quite straightforward to imagine an experimental arrangement
in which one could perform the shuttling process so that the
frequency of the axial potential well is kept constant. One could
simply create a potential well that is quadratic over the distance
to be shuttled, and then use distant control electrodes to produce a
uniform forcing field for shuttling the ion. In contrast, it is not
possible to shuttle the ion without introducing inertial forcing on
the shuttled ion. Therefore, it is reasonable as a first
approximation to examine the effect of the shuttling process on the
final state of the ion due to the inertial forcing of the ion alone,
as given in Eq.~\ref{eqn:Energyfo}. In this case, the Green's
function for the classical forced oscillator equation is that given
in Eq.~\ref{eqn:Simplegreen}. Assuming that the ion starts out in
the ground state $k=0$, the average final state of the ion $\langle
n_0(T) \rangle$ in this idealized case becomes
\begin{eqnarray}\label{eqn:Econsom}
    \langle n_0 (T) \rangle &=& \Upsilon(T,0) \nonumber \\
    &=&\frac{m}{2\hbar \omega_0} (\omega_0^2 \xi(T,0)^2 + \dot{\xi}(T,0)^2),
\end{eqnarray}
with
\begin{eqnarray}\label{eqn:consomxixidot}
\xi(T,0) &=& -\frac{1}{\omega_0}\int_0^T \sin(\omega_0(T - t'))\ddot{x}_0(t') dt' \nonumber \\
\dot{\xi}(T,0) &=& -\int_0^T \cos(\omega_0(T - t'))
\ddot{x}_0(t')dt'.
\end{eqnarray}
In this approximation, a closed form expression for $\Upsilon(T,0)$
can be obtained for each of the shuttling protocols listed in
Eq.~\ref{eqn:timeprofiles}. For the linear and sinusoidal potential
minimum time profile, the final energy and motional state of the ion
obtained from Eq.~\ref{eqn:Econsom} are simple functions of the
distance $L$ and time $T$ of the shuttling process, as well as the
fixed frequency $\omega_0$ of the potential well. The final state
resulting from the hyperbolic tangent potential minimum time profile
can also be written down in closed form using hypergeometric
functions of the type $\left._2F_1\right.\left[a,b,c;z\right] $.
\begin{eqnarray}\label{eqn:motionsforxddot}
\langle n_0 (T)\rangle_l  &=& \frac{m L^2}{\hbar \omega_0 T^2}\left(1 - \cos(\omega_0 T)\right) ,\\ \label{eq:yep}
\langle n_0 (T)\rangle_s &=& \frac{m L^{2} \pi^{4} \omega_0
\cos^{2}\left({\frac{\omega_0 T}{2}}\right)}
{\hbar(\pi^{2}-\omega_0^{2}T^{2})^{2}} \\
\langle n_0 (T)\rangle_t &=& \left.\frac{m L^2\omega_0}{4\hbar} e^{-iT\omega_0} \right \{ 1- \coth (N) \nonumber\\
& + &  e^{iT\omega_0}\left( 1+\coth (N) -  2 \coth (N)
\left._2F_1\right.\left[1,-\frac{iT\omega_0}{4N},1-\frac{i T
\omega_0}
{4N};-e^{-2N}\right]\right)  \nonumber \\
&+& \left. 2 \coth (N) \left._2F_1\right.\left[1,-\frac{iT\omega_0}{4N},1-\frac{iT\omega_0}{4N};-e^{2N}\right] \right\} \nonumber\\
& \times & \left\{1 + \coth (N) - 2 \coth (N) \left._2F_1\right.\left[1,\frac{iT\omega_0}{4N},1+\frac{iT\omega_0}{4N};-e^{-2N}\right]\right. \nonumber \\
&\ + & \left. e^{iT\omega_0}\left(1-\coth (N) + 2 \coth (N)
\left._2F_1\right.\left[1,\frac{iT\omega_0}{4N},1+\frac{iT\omega_0}{4N};-e^{2N}\right]\right)\right\}.
\end{eqnarray}
These hypergeometric functions are defined by the integral
representation
\begin{equation}\label{eqn:hypergeom}
\left._2F_1\right.\left[a,b,c;z\right] = \frac{\Gamma(c)}
{\Gamma(b)\Gamma(c-b)} \int_0^1
\frac{t^{b-1}(1-t)^{c-b-1}}{(1-tz)^a}dt,
\end{equation} where $\Gamma(n) = \int_0^\infty t^{n-1}e^{-t} \, dt$ is
the Gamma (factorial) function for arbitrary $n$ \cite{AbramSteg}. The results for
the linear and sinusoidal shuttling profiles have also been obtained
elsewhere \cite{reichle2006, schulz:2006}. Reichle et al. \cite{reichle2006} also obtained
the corresponding result for the error function profile, while Schulz et al \cite{schulz:2006}
investigated the additional impact of anharmonicity in the potential.

These protocols will be examined in two contexts. First, it will be
assumed that these protocols are used to advance the ion in small
steps of 2.14 $\mu$m, as was done in the shuttling scheme described
later in Sec.~\ref{Linear Shuttling}. This will help illustrate some
of the basic features of the transition probabilities resulting from
each of these protocols. Secondly, the protocols will be analyzed
for a continuous shuttling operation that brings the ion from one
trapping zone to the next. Again, using the University of Michigan
trap as a template, the distance for shuttling will be taken as $L =
400 \mu m$ and the trap frequency $\omega_0/2\pi = 1.173$ MHz.
We also restrict the discussion to the case for ions starting out in
the ground state, and hence drop the subscript 0 from the average
final motional state $\langle n \rangle$ of the shuttled ions.

\begin{figure}[tb!]
\centerline{\epsfig{file=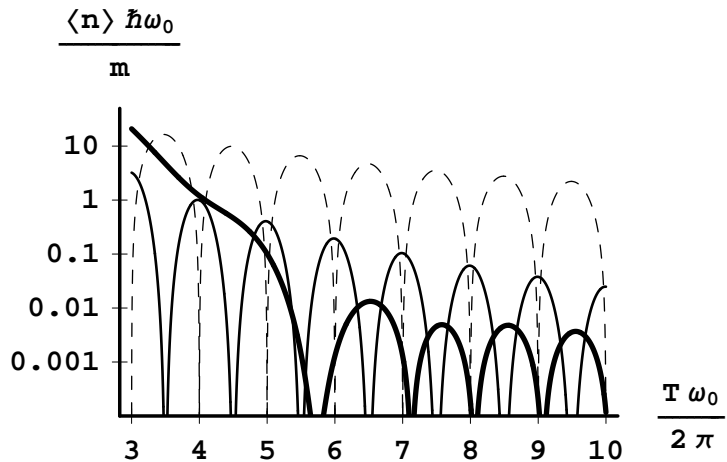, width=10cm}} 
\vspace*{13pt} \caption[]{\label{fig:linsintanhavgn} Log plot of the
average motional state $\langle n \rangle$ versus duration of the
linear (- - -), sinusoidal (\rule[0.02in]{0.25in}{0.01in}) and
hyperbolic tangent (N = 3) (\rule[0.02in]{0.25in}{0.02in}) potential
minimum time profiles. The scaled vertical axis is the energy per
ion mass given to the shuttled ion. The scaled time axis represents
the number of cycles of oscillation completed in the axial trap
during the shuttling process, where $\omega_0/2 \pi = 1.173$ MHz so
$m/\hbar \omega_0 \approx$ 1500 for $^{111}$Cd$^+$ and 120 for
$^9$Be$^+$. The distance shuttled in each case is $L=2.14$ $\mu$m ,
the standard step size in the University of Michigan shuttling
protocol described in Sec.~\ref{Linear Shuttling}. The sinusoidal
and hyperbolic tangent potential minimum time profile are seen to be
far less disturbing to the ion than the linear potential minimum
time profile when the shuttling time is greater than 3 cycles.}
\end{figure}

\begin{figure}[tb!]
\centerline{\epsfig{file=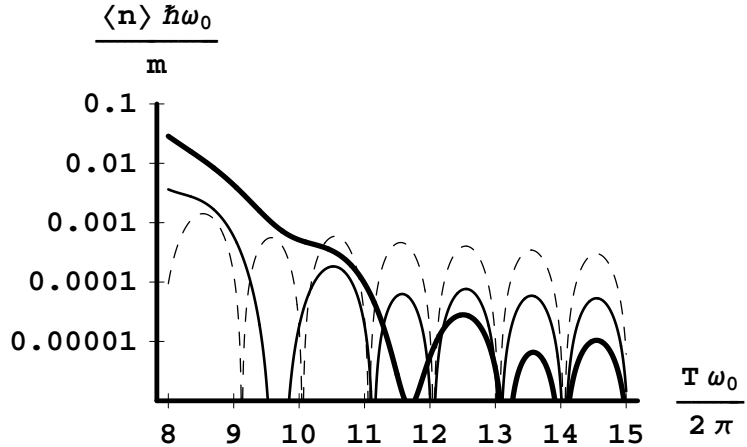, width=10cm}} 
\vspace*{13pt} \caption[]{\label{fig:tanhn303540} Log plot of the
scaled average motional state $\langle n \rangle$ versus duration of
the hyperbolic tangent potential minimum time profile for $N = 3.5$
(- - -), $N = 4.0$ (\rule[0.02in]{0.25in}{0.01in}) and $N = 4.5$
(\rule[0.02in]{0.25in}{0.02in}). The distance shuttled in each case
is for 2.14 $\mu m $. The scaled vertical axis is the energy per ion
mass given to the shuttled ion. The scaled time axis represents the
number of cycles of oscillation completed in the axial trap during
the shuttling process, where $\omega_0/2 \pi = 1.173$ MHz. Notice
that the slow time dependence comes to dominate after roughly $2N$
cycles of oscillation in the moving trap have been completed. Once
the slow time dependence becomes dominant, an increase in $N$ by one
roughly corresponds to a decrease of a factor of ten in $\langle
n\rangle$ . }
\end{figure}

When considering shuttling over a single substep of 2.14 microns,
several features of the average final motional state for all of the
proposed protocols stand out. Most noticeably, all three protocols
result in a periodically oscillating value of $\langle n \rangle$ as
a function of the duration $T$ of the shuttling operation
(Figs.~\ref{fig:linsintanhavgn} and \ref{fig:tanhn303540})
\cite{reichle2006}.

In particular, $\langle n \rangle$ becomes zero once every cycle in
the oscillation of the ion. This is the exact analogue of the phase
sensitive switching possible in a classically driven oscillator. By
timing the deceleration of the ion at the end of its motion
appropriately, it is possible to stop the ion so that it has
acquired no energy from the shuttling process. This kind of
shuttling requires the ability to switch the voltages on the
electrodes on the time scale of the secular frequency. It also
requires that the initial motional state of the ion be reasonably
well-defined. This may eventually prove to be a powerful way to
shuttle ions in a quantum information processor.

\begin{figure}[tb!]
\centerline{\epsfig{file=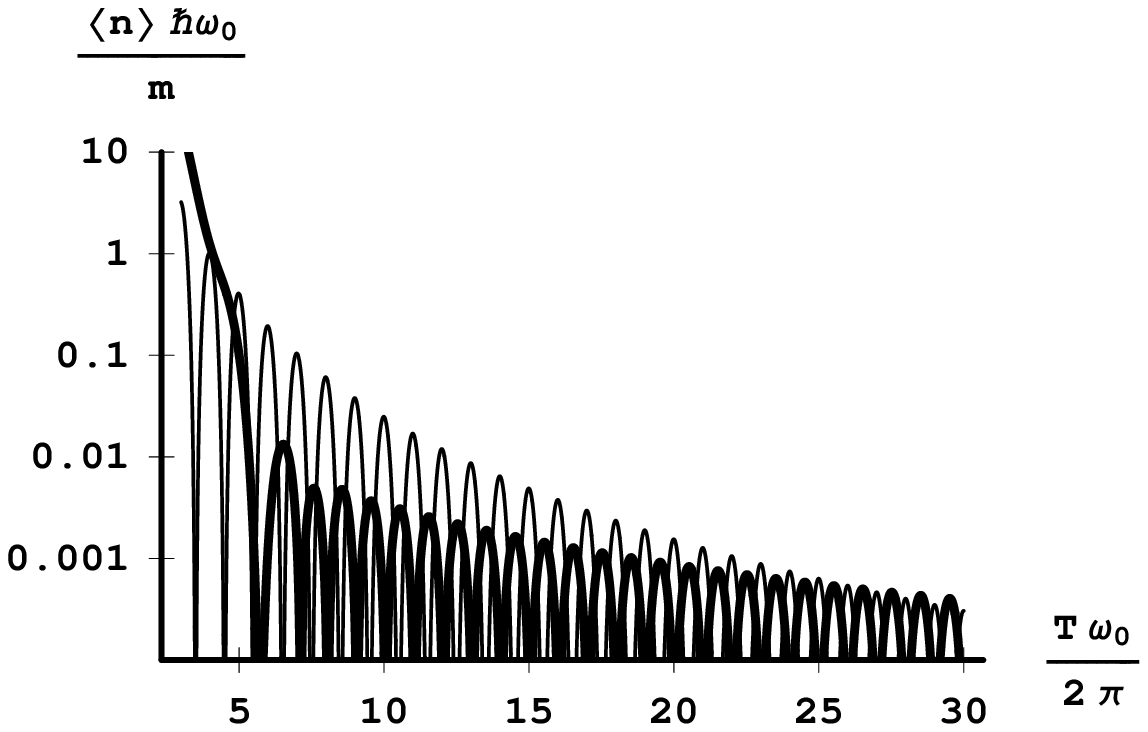, width=10cm}} 
\vspace*{13pt} \caption[]{\label{fig:sintanhn3} Log plot of the
scaled average motional state $\langle n \rangle$ versus duration of
the sinusoidal (\rule[0.02in]{0.25in}{0.01in}) and hyperbolic
tangent (N = 3) (\rule[0.02in]{0.25in}{0.025in}) potential minimum
time profiles. The distance shuttled in each case is for 2.14 $\mu m
$. The scaled vertical axis is the energy per ion mass given to the
shuttled ion. The scaled time axis represents the number of cycles
of oscillation completed in the axial trap during the shuttling
process, where $\omega_0 / 2 \pi =$ 1.173 MHz. The hyperbolic
tangent potential minimum time profile is seen to produce a far
smaller $\langle n \rangle $ for shuttling times greater than 4
cycles but less than 25 cycles. The sinusoidal potential minimum
time profile eventually exceeds any hyperbolic tangent potential
minimum time profile with fixed $N$, due to the absence of any
velocity discontinuity at the beginning and end of the protocol.}
\end{figure}

Absent the means to control the timing of shuttling protocols as
required for phase-sensitive switching, it becomes necessary to
manage the shuttling process in such a way as to minimize the value
of $\langle n (T)\rangle$. As can be seen from the expressions for
$\langle n \rangle$ in Eqs.~\ref{eqn:motionsforxddot}, the motional
state of the ion generally decreases with an increasing shuttling
time $T$ (excluding particular phasings of the shuttling time with
the trap period). The disturbance to the motional state of the ion
for the linear potential minimum time profile scales as $\langle n
\rangle \propto \frac{1}{T^2},$ while that for the sinusoidal
potential minimum time profile scales as $\langle n \rangle \propto
\frac{1}{T^4}$. Thus for the same distance shuttled, the sinusoidal
potential minimum time profile will disturb the state of the ion far
less than the linear potential minimum time profile for long
shuttling times, as seen in Fig.~\ref{fig:linsintanhavgn}. The
hyperbolic tangent potential minimum time profile has two time
scales controlling the behavior of $\langle n \rangle$
(Eq.~\ref{eqn:timeprofilesforxddot}). One time scale is the same as
for the linear potential minimum time profile, resulting from the
discontinuous jump in the speed of the potential well at the
beginning and end of the shuttling protocol. This dependence
eventually dominates the behavior of $\langle n \rangle $ over
longer shuttling times. The second time scale results in a much more
rapid drop off in the value of $\langle n \rangle$ as $T$ increases
from zero. The relative importance of these two time scales is
controlled by the parameter $N$. A larger value of $N$ results in an
$\langle n \rangle$ which for short times $T$ starts higher and
takes longer to drop off, but which drops to a lower value before
the slow time dependence takes over (Fig.~\ref{fig:tanhn303540}).
The fast time dependence of the hyperbolic tangent potential minimum
time profile makes it always possible, for a fixed shuttling
distance and shuttling time, to choose a value of $N$ that will give
a significantly smaller value of $\langle n \rangle$ than does the
sinusoidal potential minimum time profile (See
Fig.~\ref{fig:sintanhn3}). The time at which the value of $\langle n
\rangle$ resulting from the implementation of a hyperbolic tangent
potential minimum time profile with a given $N$ will reach zero is a
good indication of the time when the fast time-dependence ends, and
the slow time-dependence begins. This time is proportional to $N$
and is given roughly by $t_{\mbox{cutoff}} \approx 2 N \frac{2
\pi}{\omega_0}$.
However, the slow time dependence means that, if one shuttles for a
long enough time, the hyperbolic tangent potential minimum time
profile with a fixed value of $N$ will always result in a larger
value of $\langle n \rangle$ than the sinusoidal potential minimum
time profile.

\begin{figure}[tb!]
\centerline{\epsfig{file=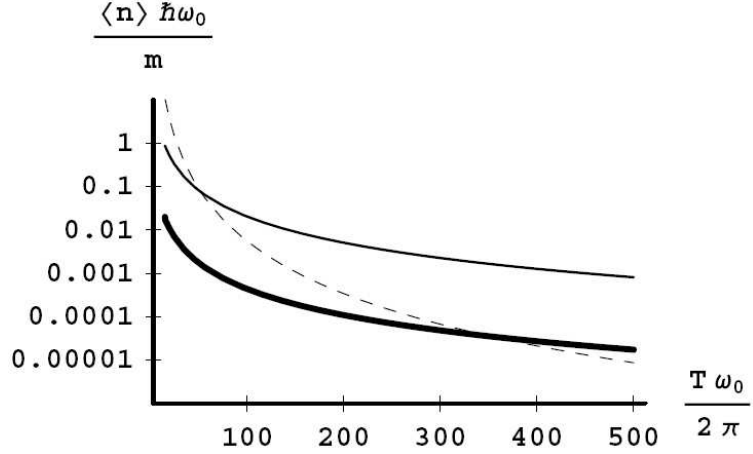, width=10cm}} 
\vspace*{13pt} \caption[]{\label{fig:sintanh3445} Plots of the
envelopes for the scaled average motional state $\langle n \rangle$
after the completion of the sinusoidal (- - -) , hyperbolic tangent
(N = 3.4) ((\rule[0.02in]{0.25in}{0.01in})) and (c) hyperbolic
tangent (N = 4.5) (\rule[0.02in]{0.25in}{0.025in}) potential minimum
time profiles. The distance shuttled in each case is for 400 $\mu$m.
The scaled time axis represents the number of cycles of oscillation
completed in the axial trap during the shuttling process, where
$\omega_0 / 2\pi = 1.173$ MHz.}
\end{figure}

These two features of the hyperbolic tangent potential minimum time
profile can also be seen by examining shuttling protocols over the
distance between two trapping zones, assumed to be $L = 400 \mu$m
(See Fig.~\ref{fig:sintanh3445}). The more rapid drop off in
$\langle n \rangle$ of the sinusoidal potential minimum time profile
with increasing shuttling time is evident when compared to the
hyperbolic tangent potential minimum time profile. However, the $N =
4.5$ protocol has a much lower value of $\langle n \rangle$ for the
shuttling times considered due to the very small discontinuity in
velocity at the beginning and end of that protocol.
Note that for the particular case of shuttling $^{111}$Cd$^+$ ions
for about 100 cycles of the oscillation (corresponding to a
shuttling time of $t = 85$ $\mu$s) the three protocols in
Fig.~\ref{fig:sintanh3445} will result in an average motional state
of less than 1, with the $N = 4.5$ hyperbolic tangent potential
minimum time profile resulting in the final state $\langle n_0
\rangle_{N = 4.5} (85$ $\mu$s$) = 0.016$. Thus, for a $^{111}$Cd$^+$
ion trapped in a potential with fixed frequency $\omega_0 = 2\pi$
(1.173 MHz), the hyperbolic tangent potential minimum time profile
with $N = 4.5$ used to shuttle the ion a distance $L = 400$ $\mu$m
over a time T $= 85$ $\mu$s is nearly adiabatic and keeps the ion in
the Lamb-Dicke limit (see section \ref{sec:adiabaticLD}), where the
extent of ion motion is much less than an optical wavelength.

It is possible to generalize the above discussion for the idealized
shuttling protocols and consider what factors determine how much
impact a given protocol has on the final motional state of the ion.
Clearly, the energy given to the ion during shuttling is
proportional to the maximum amplitude of the displacement of the ion
from $x_0$ after time $T$,
\begin{equation}\label{xiaverageddotxo}
\xi(T,0) =  -\frac{1}{\omega_0}\int_0^T \sin(\omega_0(T - t'))
\ddot{x}_0(t') dt',
\end{equation} where we have set $t_0=0$.
Recall from Eq.~\ref{eqn:generalddotx0} that the forcing term
$\ddot{x}_0(t)$ can be expressed in general as
\begin{equation}\label{eqn:generalddotx0again}
\ddot{x}_0(t) = A(t,T)\frac{L}{T}\left[\delta(t) -
\delta(t-T)\right] + B(t,T) \frac{L}{T^2}\left[H(t)-H(t-T) \right],
\end{equation}
where $A(t,T)$ characterizes the velocity discontinuity at the
beginning and end of the shuttling protocol and B(t,T) characterizes
the acceleration in the middle (see Eq. 32-34 and Eq. 35). Inserting this into the expression
for $\xi$ we have
\begin{equation}\label{eqn:generalxi}
\xi(T,0) =  -A(0,T)\frac{L}{\omega_0T} \sin (\omega_0 T) -
\frac{L}{\omega_0T^2} \int_0^T \sin(\omega_0(T - t')) B(t',T) dt',
\end{equation} where the shuttling protocol is initiated just after the
time $t_0$. By integrating the second term on the right hand side by
parts, a series expansion can be developed for $\xi$
\begin{eqnarray}\label{eqn:generalxiexpansion}
\xi(T,0) &=&  -A(0,T)\frac{L}{\omega_0T} \sin (\omega_0 T) - \frac{L}{\omega_0^2T^2} \left(B(T,T) - B(0,T)\cos(\omega_0T) \right) \nonumber \\
&-& \frac{L}{\omega_0^3T^2} \left.\frac{\partial B(t',T)}{\partial
t'}\right|_0 \sin(\omega_0 T) + \frac{L}{\omega_0^3T^2}\int_0^T
\sin(\omega_0(T - t'))
 \frac{\partial^2 B(t')}{\partial {t'}^2} dt'.
\end{eqnarray} Since each derivative of $B(t,T)$ will result in
another factor of $T$ coming into the denominator (See the second
line of Eq.~\ref{eqn:timeprofilesforxddot} for example), the
coefficients of the series are powers of the factor
$\frac{1}{\omega_0 T} $ so that we have
\begin{eqnarray}\label{eqn:generalxiexpansionTomega}
\xi(T,0) &=&  -\frac{L}{\omega_0T}A(0,T) \sin (\omega_0 T) -
\frac{L}{\omega_0^2T^2}
\left(B(T,T) - B(0,T)\cos(\omega_0T) \right) \nonumber \\
&-& \frac{L}{\omega_0^3T^3} \left(T\left.\frac{\partial
B(t',T)}{\partial t'}\right|_0 \right)
\sin(\omega_0 T) \nonumber\\
&+& \frac{L}{\omega_0^3T^3}\int_0^T \sin(\omega_0(T - t')) \left(T
\frac{\partial^2 B(t',T)}{\partial {t'}^2}\right) dt'.
\end{eqnarray}
In general, for smooth and continuous potential minimum time
profiles, the expansion can be continued by integrating by parts
repeatedly until the error term represented by the remaining
integral is arbitrarily small. As a result, regardless of the
functional form of $A$ and $B$, the series for $\xi$ can be made to
converge more rapidly and to a smaller value by shuttling for a time
$\omega_0 T \gg 1$, which is the adiabatic condition.

The leading order term in this expansion is the delta function
``kick'' associated with the starting and stopping of the shuttling
potential. It is also the term that is reduced most slowly as the
duration of the shuttling protocol is increased. It is for this
reason that the linear potential minimum time profile has the least
satisfactory behavior among the three examined here. Therefore,
shuttling protocols should be designed to start and stop as smoothly
as possible. This condition can be written down in equation form as:
\begin{equation}\label{eqn:adiabaticcondition1}
A(0,T) \ll \frac{\omega_0 T}{L }.
\end{equation} The expression $A(0,T)$ is the fraction at time $t_0=0$ of
the average speed of shuttling. The linear potential minimum time
profile has $A(0,T) = 1,$ while the hyperbolic tangent potential
minimum time profile has $A(0,T) = N \frac{\coth (N)}
{\cosh^{2}\left( N \right)}.$ This term varies from $A(0,T) =1$ for
$N = 1$ to $A(0,T) \rightarrow N e^{-2N}$ for the limit as $N$
becomes large. Hence, this factor for the hyperbolic tangent
potential minimum time profile can be made arbitrarily small by
increasing $N$. The sinusoidal potential minimum time profile has
$A(0,T)=0.$

Not only should the shuttling protocols be started and stopped as
smoothly as possible, but they should also have small accelerations
at the beginning and end of the protocol. This requirement is
expressed through a similar condition on the maximum value of the
second leading term in the expansion for $\xi$
\begin{equation}\label{eqn:adiabaticcondition2}
2 B(0,T)\ll \frac{\omega_0^2 T^2}{L},
\end{equation} where we have made the assumption that the accelerations
of the shuttling potential at the beginning and end of the protocol
will be equal in magnitude but opposite in direction
($B(0,T)=-B(T,T)$). The factor $B(0,T)$ is the fraction at time $t_0
= 0$ of the average acceleration the shuttling potential has during
the course of the shuttling protocol. The linear potential minimum
time profile of course has $B(0,T) = 0.$ The sinusoidal potential
minimum time profile gives the largest acceleration to the shuttling
potential at the beginning (and end) of the protocol, so that
$B(0,T) = \frac{\pi^2}{2}.$ The hyperbolic tangent potential minimum
time profile has $B(0,T) = \frac{4N^2} {\cosh^{2}N}.$ Therefore, its
value ranges from $B(0,T) = 4$ for $N=1$ to $B(0,T) \rightarrow
4N^2e^{-2N}$ as $N$ becomes large. Once again, by increasing the
value of $N$, one can make this term in the hyperbolic tangent
potential minimum time profile arbitrarily small. Thus, large values
of $N$ can make the hyperbolic tangent potential minimum time
profile far superior to the sinusoidal potential minimum time
profile for a given shuttling distance $L$ and shuttling time $T$,
by enforcing the condition
\begin{equation}\label{eqn:tanhbetterthansin}
\left(N + 4N^2\right) e^{-2N} < \frac{\pi^2}{2} \ll \frac{\omega_0
T}{L } + \frac{\omega_0^2 T^2}{2L}.
\end{equation}
However, it should be noted that, for a given $N$, the sinusoidal
potential minimum time profile will always eventually produce a
smaller value of $\xi$ than the corresponding hyperbolic tangent
potential minimum time profile as the shuttling time $T$ becomes
sufficiently large. This is due to the fact that the hyperbolic
tangent potential minimum time profile always has a finite value of
$A(0,T)$ and therefore, there will be a time $T$ for which
\begin{equation}\label{eqn:sinbetterthantan}
A(0,T) > \frac{\pi^2}{T\omega_0},
\end{equation} so that the leading order behavior of the hyperbolic tangent
potential minimum time profile becomes greater than the leading
order behavior of the sinusoidal potential minimum time profile.

As should be clear from this discussion, the ideal shuttling
protocol is one for which all of its derivatives at the beginning
and end of the protocol are as small as possible, in the sense
defined for $A(0,T)$ and $B(0,T)$ above. This is precisely what is
accomplished for the hyperbolic tangent potential minimum time
profile when $N$ is taken very large. A protocol defined similarly
to the hyperbolic tangent potential minimum time profile in
Eq.~\ref{eqn:timeprofiles}, but using the error function, also has
extremely small derivatives at the starting and ending points, and
as was shown by Reichle, \mbox{et al.} \cite{reichle2006} adds less
energy to a shuttled ion than a sinusoidal potential minimum time
profile. The above analysis reveals why that is the case. However,
both the hyperbolic tangent (for large $N$) and error function
potential minimum time profile require a longer shuttling time $T$
before approaching their asymptotic transient behavior. As $T
\rightarrow 0$, these functions become more and more step-like. As a
result, higher and higher order derivatives of $x_0$ are required
before the integral in Eq.~\ref{eqn:generalxiexpansion} will vanish.
Therefore, the sinusoidal potential minimum time profile is better
(gives less energy to the ion) than either the hyperbolic tangent
for fixed $N$ or error function potential minimum time profile for
very short ($\omega_0 T \sim 1$) and very long ($\omega_0 T \sim 2
\pi/A(0,T)$) shuttling times, but the sinusoidal potential minimum
time profile is worse for shuttling protocols of intermediate
duration. This may be an issue where there is some experimental
limitation on the maximum or minimum value of $N$, set for example
by the speed of the circuit governing the control electrodes.
However, it is clear that under most circumstances, shuttling
protocols like the hyperbolic tangent potential minimum time profile
can be used to shuttle ions in a given time with the least amount of
energy transferred to the ion.

\subsubsection{\label{sec:adiabaticgeneral} Shuttling in variable frequency potentials}

Finally we turn to the general case for which the shuttling
potential has a variable frequency. The average final motional state
of an ion that was in the ground state at time $t_0 =0 $ and then
shuttled for a time $T$ over a distance $L$ by a moving potential of
varying frequency is given by (Eq.~\ref{eqn:avgnfpo})
\begin{equation}
\langle n(T) \rangle = \Upsilon(T,0) + \frac{1}{2}\left(Q(T,0) -
1\right) ,
\end{equation} where $\Upsilon$ is defined as in Eq.~\ref{eqn:E} and $Q$ as in
Eq.~\ref{eqn:Q}. As in the previous section, we will begin by
working with the particular model for frequency variation introduced
in Eq.~\ref{eqn:modelfreqvar}, explore its impact on the average
final state of the shuttled ion and then see if we can make some
more general conclusions.

\begin{figure}[tb!]
\centerline{\epsfig{file=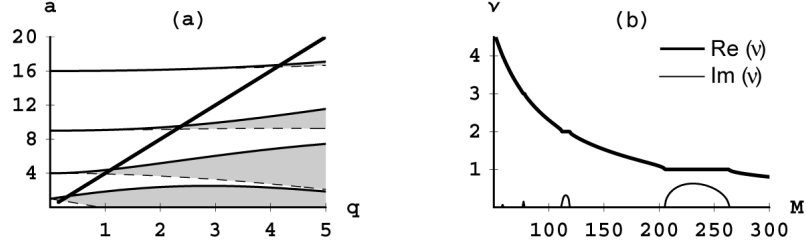, width=12cm}} 
\vspace*{13pt} \caption[]{\label{fig:stability} (a) Stability diagram
for the Mathieu's functions in a vs. q space. Unstable regions are
shaded. The dashed curves correspond to an integer value of $\nu$
for the odd Mathieu functions, $S$, and the solid curves to an
integer value of $\nu$ for the even Mathieu functions, $C$. Also
shown is the solid line plot of $a$ vs. $q$ for the frequency
variation model of Eq.~\ref{eqn:classufhomathieu} when $T = 100$ ($\approx 120\cdot 2\pi/\omega_{0}$)
$\mu$s; $g = 0.5$, $\omega_{0}/2\pi=1.173$ MHz, for $50 < M < 300.$ The largest value of $M$
corresponds to the point on the line closest to the origin. (b) Plot
of the real (heavy line) and imaginary (light line) parts of the
characteristic exponent $\nu$ as a function of $M$ for the frequency
variation model of Eq.~\ref{eqn:classufhomathieu} when $T = 100$
$\mu$s and $g = 0.5$. The imaginary values have been scaled by a
factor of 5 to make them more visible on the graph. The solutions of
Mathieu's equation are unstable when $\nu$ is complex, with integer
real part. The range of values of $M$ corresponding to unstable
solutions is larger for smaller values of the characteristic
exponent.}
\end{figure}

We consider the case for which the inertial forcing in the shuttling
potential is minimal, so that only the frequency variation has an
impact on the shuttled ion's final motional state. In this case,
$\Upsilon \approx 0,$ and to find the average final state of the ion
we need only to evaluate $Q$. Therefore, we look for the
characteristic solutions $X_1$ and $X_2$ of the unforced parametric
oscillator equation, satisfying the initial conditions
\begin{eqnarray}\label{eqn:initialconditions}
X_1(t_0) &=& 0; \hspace{0.1 in} \mbox{and} \hspace{0.1in} \dot{X}_1(t_0) = 1;\nonumber \\
X_2(t_0) &=& 1; \hspace{0.1 in} \mbox{and} \hspace{0.1in}
\dot{X}_2(t_0) = 0,
\end{eqnarray}
with the frequency dependence as given in Eq.~\ref{eqn:modelfreqvar}
\begin{equation}\label{eqn:classufhomathieu}
\ddot{X} = -\omega_0^2 (1-g\cos \left((M+1/2) \frac{2\pi
t}{T}\right))X,
\end{equation} where $g$ is the modulation depth and $M$ is related to the modulation frequency of the square of the trap frequency.
This equation can be written in the canonical form of Mathieu's
differential equation \cite{AbramSteg2},
\begin{equation}\label{eqn:mathieucanonical}
\frac{d^2X}{dz^2} + (a-2q\cos \left(2z\right))X = 0,
\end{equation}
by identifying the parameters
\begin{eqnarray}\label{eqn:mathieuparameters}
a = \left(\frac{\omega_0 T}{(M+\frac{1}{2}) \pi}\right)^2;
\hspace{.1in} q = \frac{g a}{2}; \hspace{.1in} z = \frac{\pi
t}{T}\left(M + \frac{1}{2}\right).
\end{eqnarray}
In general Eq.~\ref{eqn:mathieucanonical} has solutions which are
either even or odd in $z$, the Mathieu functions $C(a,q,z)$ and
$S(a,q,z)$, respectively. The solutions $X_1$ and $X_2$ of
Eq.~\ref{eqn:classufhomathieu} at the end of a shuttling operation
of duration $T$ satisfying the given initial conditions are then
obtained in terms of these solutions. Recalling that we chose the
time interval for our frequency model to be $[-T/2,T/2]$, we have
\begin{eqnarray}\label{eqn:X1andX2mathieu}
X_1(T/2) &=& \frac{T}{(M+1/2)\pi}\frac{C(a,q,-\tau_0)S(a,q,\tau_0) -
C(a,q,\tau_0)S(a,q,-\tau_0)}
{C'(a,q,-\tau_0)S(a,q,-\tau_0) - C(a,q,-\tau_0)S'(a,q,-\tau_0)} \\
X_2(T/2) &=& \frac{C'(a,q,-\tau_0)S(a,q,\tau_0) -
C(a,q,\tau_0)S'(a,q,-\tau_0)} {C'(a,q,-\tau_0)S(a,q,-\tau_0) -
C(a,q,-\tau_0)S'(a,q,-\tau_0)},
\end{eqnarray} where
$\tau_0 = \frac{\pi}{2}\left(M + \frac{1}{2}\right)$ and the prime
represents differentiation with respect to $z$. When $q = 0$, the
Mathieu functions reduce to $C(a,0,z) = \cos(\sqrt{a} z),$ and
$S(a,0,z) = \sin(\sqrt{a} z).$ When $q\neq0$, the behavior of the
Mathieu functions depends on the value of the characteristic
exponent, $\nu = \nu(a,q)$ \cite{AbramSteg2}. When $q$ and $a$ are
such that $\nu$ is real, the Mathieu functions are finite for all
$z$ and the solutions are stable. When $\nu$ is an integer, the
solutions are periodic functions in $z$. When $\nu$ is complex, the
functions become infinite at some value of $z$, and the solutions
are considered unstable. The stability diagram for the Mathieu
functions is shown for positive $a$ and $q$ in
Fig.~\ref{fig:stability}a. Unstable regions are shaded and bounded
by solid and dashed curves corresponding to integer values of $\nu$
for the even and odd solutions, $C$ and $S$, respectively. The
straight, heavy line in the figure displays the relationship between
$a$ and $q$ in the parametric oscillator model developed above for
$g = 0.5$.

\begin{figure}[tb!]
\centerline{\epsfig{file=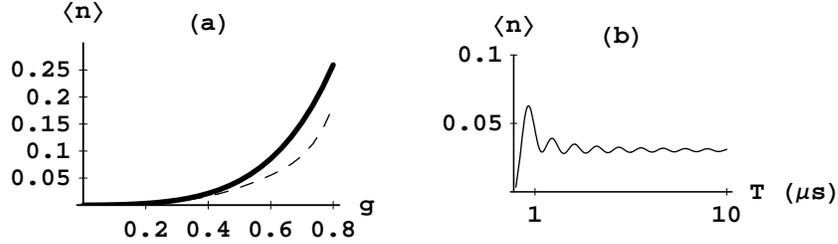, width=12cm}} 
\vspace*{13pt} \caption[]{\label{fig:freqvarM0} Figure (a) is a plot of the
numerically calculated average motional state $\langle n \rangle$ of an ion versus
the modulation depth $g$ of the trap frequency squared, with $M =
0$, for shuttling times $T = 1$ $\mu$s (---) and $T= 5$ $\mu$s (- -
-),  respectively. The distance shuttled in each case is for 2.14
$\mu m $. Figure (b) is a plot of the numerically calculated average motional state
$\langle n \rangle$ of an ion versus the shuttling time $T$ in
microseconds, with $M = 0$ and $g = 0.5$.}
\end{figure}

First, let's consider the case $M = 0,$ which is appropriate for
shuttling in a short step, so that the frequency of the shuttling
potential is decreased and then increased back to its original value
just once. Again, we'll consider the step size to be $2.14$ $\mu$m,
and the frequency of the trap at the beginning and end to be
$\omega_0/2\pi = 1.173$ MHz. Ignoring the impact of any inertial
forcing, the contribution to the average final state of the shuttled
ion from the frequency variation of the potential will therefore be
a function of the shuttling time and the modulation depth $g$. For
shuttling times much less than one trap oscillation period, the
impact of frequency variation of this type on the motional ion state
is negligible, as expected (Fig.~\ref{fig:freqvarM0}). As the
shuttling time increases above the oscillation period of the trap,
the average final state of the ion reaches a maximum and then
rapidly decreases to an asymptotic value for all values of the
modulation depth, $g$. For shuttling times greater than roughly 5
trap oscillation periods, $\langle n \rangle$ no longer depends on
the shuttling time and becomes a simple function of $g$, as seen in
Fig.~\ref{fig:freqvarM0}. In general, the impact of this type of
perturbation on the ion's final state is minimal for any reasonable
frequency squared modulation depth, $g$ and shuttling time, $T$.

The second case to be considered is for longer shuttling times and
distances, with a series of rises and falls in the shuttling
potential frequency. This model might represent the frequency
variation a shuttled ion would experience as the ion passes through
areas in the trap array where the trap depth is successively weaker
then stronger due to periodic arrangement of the electrodes in the
trap array. Even in the case where the shuttling potential is kept constant along the ion's trajectory, frequency variations in the transverse trap potential may be occurring periodically, feeding energy into the transverse modes of the ion's motion through a subharmonic resonance. It might also represent the fluctuations in the axial
trap frequency resulting from fluctuating voltages on the control
electrodes. We now consider a fixed shuttling time of 100 $\mu s$
over a distance of 400 $\mu$m. For these parameters, it was shown in
Sec.~\ref{sec:adiabaticatconstantomega} that a hyperbolic tangent
shuttling potential minimum time profile with $N=4.5$ produces
minimal inertial forcing on a Cd$^+$ ion, and so we can focus on the
impact of the frequency variation on the ion's final motional state.
As discussed above, the solutions for the parametric oscillator are
stable or unstable, depending on the value of the characteristic
exponent, $\nu$. Now, as can be seen from
Eq.~\ref{eqn:mathieuparameters}, for a particular amplitude of
frequency variation given by the parameter $g$, the relationship
between $a$ and $q$ for the shuttled ion will be fixed along a line
in $a-q$ space given by $a = 2 q/g$ (Eq.~
\ref{eqn:mathieuparameters}). This curve is shown for the specific
value $g = 0.5$ in Fig.~\ref{fig:stability}a by a heavy line.
Smaller values of $M$ are to the right and up, and larger values of
$M$ are to the left and down, corresponding to values of $a$ and $q$
located closer to the origin. If the amplitude $g$ of the variation
were to increase from this value, then the slope of the line in
Fig.~\ref{fig:stability} would decrease. It is then clear from the
stability diagram in Fig.~\ref{fig:stability}a that this line would
pass through larger and larger sections of the unstable regions.
Therefore as $g$ increases, the likelihood of an unstable solution
will also increase, as expected.

From Eq.~\ref{eqn:mathieuparameters} we see that both dimensionless
parameters $a$ and $q$ are proportional to the factor
\begin{equation}\label{eqn:tfreqvar}
T_{fv} = \left(\frac{\omega_0 T}{(M+\frac{1}{2}) \pi}\right) .
\end{equation}
Since $\omega_0 T/2\pi$ is the duration of the shuttling protocol
measured in periods of the ion's secular motion, $T_{fv}$ is the
ratio of the effective shuttling time to the number of cycles of
frequency variation of the shuttling potential. Therefore, we call
$T_{fv}$ the period of frequency modulation. If the beginning and
ending frequencies of the shuttling potential are equal, $M$ is an
integer. Therefore, the values of $a$ and $q$ for this model of
parametric frequency modulation are not continuous functions of
$T_{fv},$ but are discrete points along the line in
Fig.~\ref{fig:stability} for a given value of $g$. Those points are
relatively farther apart for small values of $M$, and closer for
larger values of $M$, when $a$ and $q$ are both small. The relative
spacing between points on the line is a function of the total
shuttling time. For longer shuttling times, the points all along the
line will be closer, and for shorter times, the points will be
placed farther apart.

This has a significant impact on whether a given frequency variation
will result in a catastrophic shuttling protocol. For example, the
line drawn in Fig.\ref{fig:stability}a corresponds to the values of
$a$ and $q$ for a frequency variation of 50\% over a time of $100$
$\mu$s $\approx 120(2\pi/\omega_0).$ The line passes through all
four regions of instability drawn on the diagram. These regions
correspond to integer values of the real part of $\nu$ from 1 to 4
as one moves away from the origin. The corresponding values of
$T_{fv}$ are roughly 1, 2, 3 and 4. The range in $M$ for which
the shuttled ion's motion is catastrophically unstable grows
markedly for higher values of $M$, corresponding to smaller values
of $T_{fv}.$ This is partially accounted for by the fact that the
extent of the regions of instability through which the line in
Fig.~\ref{fig:stability}a passes does shrink as one moves up and to
the right along the line (corresponding to smaller values of $M$).
This effect is enhanced by the fact that the spacing between
successive values along the line is increasing in that direction as
well. As a result, the range of values of $M$ for which the
parametric driving is unstable grows significantly for increasing
$M$. This is illustrated in Fig.~\ref{fig:stability}b, where the
imaginary part of $\nu$ is scaled by a factor of 5 to make evident
which values of $M$ will result in unstable parametric oscillations.
As seen in the figure, only one or two values of $M$ at 57 and 78
correspond to unstable solutions with the real part of $\nu = 4$ or
$3$, while 60 values of $M$ from 205 to 265 are unstable when the
real part of $\nu = 1.$ Since the regions of instability for a given
value of $g$ are determined by the value of $T_{fv},$ a shuttling
protocol that takes 10 times as long to complete would require 10
times the number of frequency variations to observe the same
unstable behavior. But it would also expand the range of values of
$M$ for which the solution is unstable by a factor 10.

We find that longer shuttling times can result in unstable behavior
at larger values of $T_{fv}$, corresponding to slower rates of
frequency variation, while none was observed for the shorter
duration shuttling protocols. This points to a new potential source
of ion heating and possible trap loss during shuttling: in the rest
frame of the ion electric field inhomogeneities along the ion
trajectory will appear as electric field noise \cite{turchette:2000,
deslauriers:2006}.

\begin{figure}[tb!]
\centerline{\epsfig{file=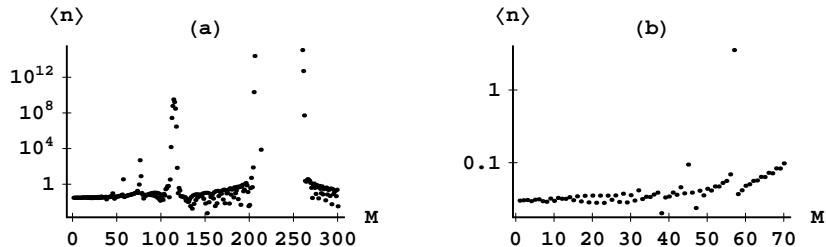, width=12cm}} 
\vspace*{13pt} \caption[]{\label{fig:freqvarMvar} (a) Plot of the calculated
average motional state $\langle n \rangle$ of an ion versus $M$
(given by the modulation frequency of the trap). The shuttling time
is 100 $\mu s$ and the modulation depth is $g = 0.5$. For  $M > 80
$, resonances can be observed arising from unstable regions in the
Mathieu a-q plane. (b) Plot $\langle n \rangle$  versus $M$, with $T
= 100 \mu s$ and $g = 0.5.$ For $M < 70,$ corresponding to a period
of frequency variation $T_{fv} > 3.3,$ the impact on the final
motional state of the ion is minimal.}
\end{figure}

The impact of these instabilities on the average final motional
state of a shuttled ion can be seen in Fig.~\ref{fig:freqvarMvar}a.
They are indeed catastrophic. On the other hand, if $M$ is kept to
values much smaller than the number of ion oscillations during the
shuttling protocol, so that $T_{fv} \gg 3$, the solutions $X_1$ and
$X_2$ generally remain stable, and the
impact on $\langle n \rangle$ is quite small, even for the large
modulations ($g=0.5$) in the shuttling potential frequency assumed for this
calculation, as seen in Fig.~\ref{fig:freqvarMvar}b. Although resonances are predicted for all integer values of $T_{fv}$ , they become increasingly narrow and less significant at larger values. This is particularly true given the fact that the modulation of the frequency will not likely be exactly sinusoidal, and the resonances correspondingly dampened. Therefore, as a rule of thumb, we can say from the above analysis that a shuttling
protocol that has fractional variations in the trap frequency
squared of less than 20\% ($g < 0.2$) and a period of frequency
variation much greater than one,($T_{fv} \gg 3$)  typically will not
perturb the motional state of the shuttled ion.


\section{\label{sec:trap design}Shuttling and trap design}

In this section, we discuss the problem of shuttling atomic ions
through junctions and in multiple dimensions. Shuttling in two
dimensions is useful for the realization of simple quantum
algorithms, and may also be indispensible for the implementation of
quantum error correction \cite{shorsteane}.  By shuttling many
atomic ion qubits simultaneously, error correction can be perfomed
in a massively parallel fashion, thereby lowering the fault-tolerant
error thresholds \cite{Steane2003}. The ability to shuttle
effectively is intertwined with the design of trap architectures.
Linear shuttling has been previously implemented in a series of
experiments employing extended linear rf traps \cite{barrett:2004,
shuttle error correction, chiaverini:2005, rowe, T paper,
stick:2006}. Multidimensional shuttling was accomplished in a
T-junction, where the rf nodal trapping pathways of three linear
traps are joined at the junction.  In this trap, ions were shuttled
between traps through the junction, and the positions of two ions
were swapped by executing a ``three point turn" through the junction
\cite{T paper}.

\begin{figure}[tb!]
\centerline{\epsfig{file=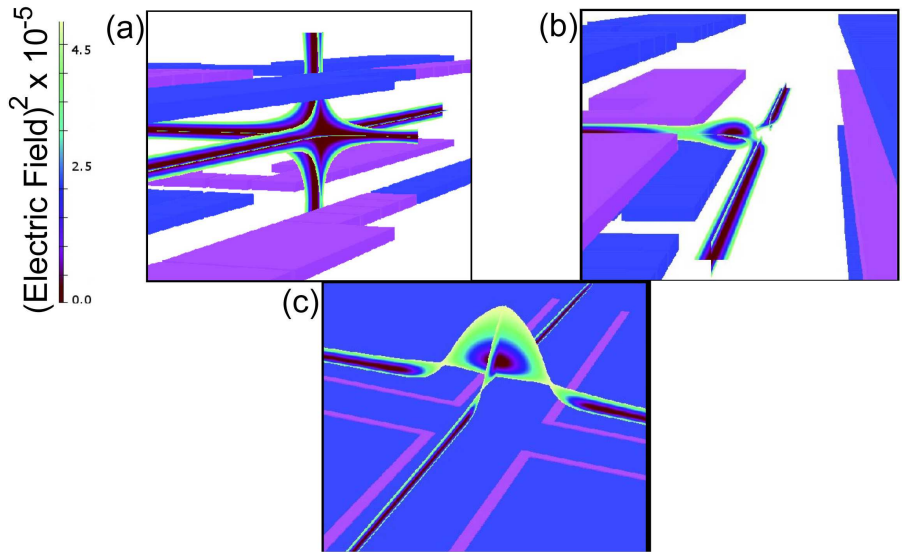, width=12 cm}} 
\vspace*{13pt} \caption[]{\label{fig:threeinrow2}Plots of the rf
pseudopotential for an asymmetric, cross-junction trap and two
symmetric T-junction traps having two and three electrode layers,
respectively. The electrodes are separated by 80 $\mu$m in the
asymmetric trap, and 200 $\mu$m in the symmetric traps. The rf
electrodes are shown in light pink, while the grounded layers are
blue. In all three simulations, the rf voltage was set to 1.0 V. The
contour plots are of the Electric Field amplitude squared, going
from dark red to light yellow as the values range from 0 to $4.90
\times 10^5 \left(V/m\right)^2.$ Notice the rf barriers along the
ion pathways for the asymmetric and three layer traps, while the two
layer trap has an rf hole at its center. The rf hump in the
pseudopotential for the asymmetric trap is also clearly seen.}
\end{figure}

Near a junction, the trapping potential is no longer strictly a
linear trap. The ions may encounter large rf fields in this region.
This gives rise to a repulsive ponderomotive axial force pushing the
ions away from the junction. The additional static forces necessary
to confine the ion in the presence of theses new axial rf forces can
weaken or destabilize the trap, allowing the ions to escape.
Therefore, care must be taken when designing the trap junctions so
that sufficient confinement is maintained while ions are transferred
from one pathway to another. In order to make our discussion of the
nontrivial features of the rf ponderomotive potential near a
junction more precise, we define three terms:
\begin{itemize}
    \item An rf \textit{hole} describes the occurrence of an unstable (non-trapping)
region of the rf ponderomotive potential near a trap junction,
typically out of the plane of the ion pathway (Fig.
\ref{fig:threeinrow2}a).
    \item An rf \textit{barrier} is a region of non-zero ponderomotive potential along
the ion pathway as it approaches a junction (Fig.
\ref{fig:threeinrow2}a,c).
  \item An rf \textit{hump} is a displacement of the minimum of the
ponderomotive potential in a direction perpendicular to two or more
merging ion pathways (Fig. \ref{fig:threeinrow2}c)\footnote{Note
that in Ref. \cite{T paper}, the term ``rf hump" was used in the
context of three-layer traps to describe and energetic hump or rf
barrier along the axial minimum of the ponderomotive potential. The
definitions above are provided to make a clear distinction among the
features of the rf ponderomotive potential in the different types of
traps being discussed in the ion-trapping community}\,\,.
\end{itemize}

These three features of the ponderomotive potential are illustrated
in Fig. \ref{fig:threeinrow2}. The rf barrier, which occurs in all
multidimensional junction traps explored to date, represents a
region near the junction in which the rf ponderomotive potential
impedes ion shuttling. An rf hole can occur in some trap geometries,
and must be either avoided either by design, or the ions must be
steered around the hole.  An rf hump is typically present near
junctions in asymmetric traps, described below.

\subsection{Trap geometries}

Trap designs that address the need for two-dimensional control
of ions presently fall into two broad categories, symmetric and
asymmetric traps, which are characterized by the degree of symmetry
of the trap electrodes in the third dimension perpendicular to the
plane in which the ion trajectory is confined. Both designs have advantages and disadvantages
of their own related to issues of heating, and of electrical and
optical access. The next sections compare these two geometries with
special attention to multidimensional shuttling.

Symmetric ion traps feature rf electrodes that are located
symmetrically around the ion pathway. Symmetric traps may be more
difficult to fabricate because they typically require multiple
layers, but they also feature high trap depths and efficient pushing
forces on ions. Symmetric ion traps include two-layer traps, which
have the rf electrodes arranged in two planes above and below the
plane of the ion pathway \cite{turchette:2000, rowe, martin paper},
and three-layer traps, where the rf electrode is confined to a
single layer in the plane of the ion pathway and two outer layers
held at rf ground consisting of segmented electrodes that carry
static voltages \cite{Louis2004, T paper}.

Two-layer symmetric traps have drawbacks.  Compensation of
uncontrolled external static fields in three dimensions (potentially
giving rise to large amounts of micromotion \cite{Pseudopotential})
requires extra electrodes or the application of static voltages to
the rf electrodes. More importantly, the junction region in
two-layer traps is complicated by the fact that the rf electrodes
are not in the same plane as the rf nodal pathways (See Fig.
\ref{fig:threeinrow2}). This lack of lateral symmetry usually
produces an rf hole at the junction center, and the ions must then
be carefully steered around the hole and through sizable rf barriers
for successful transit. A variation of the two-layer trap with small
bridges extending diagonally across the junction has also been
proposed to close that hole, at the cost of introducing rf barriers
along the ion pathways \cite{2 layer}.

Three-layer traps \cite{Louis2004, T paper} can have both vertical
symmetry (perpendicular to the multiple channels forming a junction)
and transverse symmetry, and allow for complete three-dimensional
compensation of background static fields exclusively through the
control electrodes. Shuttling through a junction is also simpler in
three-layer traps, as the added symmetry avoids rf holes at the
junction.

Asymmetric traps, also referred to as single-layer traps or surface
traps, have all their electrodes located in one or more planes below
the ion pathway \cite{slusher:2005,britton:2006, seidelin:2006}.
This offers clear advantages in the context of large-scale
fabrication, as the electrical lead-ins can be fed from the
underside of the trap electrode surface, at the expense of more
restricted optical access.

It is possible to greatly reduce the effect of the
rf barrier in both the symmetric and asymmetric trap junctions
by tapering the (surface) electrodes in the approach to the
junction \cite{slusherTAPER}. This serves to increase the size of
the region along the ion pathway in the linear portion of the trap
having a non-zero axial electric field component and effectively
spreads out the rf barrier that the ion must pass through.  While
the total work necessary to transport the ion through the junction
would not change, the force required to advance the ion at any point
along the pathway can be significantly reduced. For the asymmetric
trap this also reduces the characteristic trap size near the junction
and pushes the ponderomotive minimum closer to the trap surface,
minimizing the rf hump. However, the rf barrier also effectively
reduces the depth of the trap over the range of its spatial extent,
which could be an issue for shallow traps.

\subsection{Shuttling through junctions}

Shuttling through junctions, where three or more linear trap axes
join, is greatly complicated by the presence of rf barriers leading
into the junction. These barriers can result in added kinetic energy
to the ions that is dificult to control, and in some cases loss of
the ion from the trap altogether. The increasing complexity of the
trap design and the greater demands on precise knowledge of the
trap fields means that accurate numerical simulation of the fields
as outlined earlier in this paper becomes essential.

Although rf barriers are present in all trap junctions considered
here, we will examine their influence and characterize their
features in a simple, three-layer T-trap. The spatial extent of
these rf barriers and the potential gradient along each side of the
barrier determine how the ion can be reliably carried through the
junction. For the ion to make it through the barrier in a controlled
fashion, the potential energy gradient of the guiding control fields
should be at least as great (and opposite in sign) as the gradient
in the rf pseudopotential barrier. Otherwise, the ions must be given
enough kinetic energy to make it through. If the gradient is
canceled only on one side of the barrier, the ions will still be
accelerated by the gradient in the rf ponderomotive potential as the
ion enters the junction, again giving the ions unwanted kinetic
energy. Thus it is most desirable to reverse the sign of the
potential energy gradient on both sides of the barrier, so that the
ion remains in a smooth axial trap throughout its motion. This
suggests a strong relationship between the spatial extent of the rf
axial barrier, and the maximum size of the control electrodes used
to shuttle the ions. Tapered electrodes at the junction serve to relax this maximum size restriction. The strength of the rf barrier is also related
to the trap depth perpendicular to the plane of the ions' motion at
the junction, and if the quasistatic control forces are too high,
then they will destablize the trap out of the plane of the junction.
In sum, the four most important features of the rf barriers are
their spatial extent, gradient, strength and the trap depth out of
the plane at the barrier location.

As dimensional analysis suggests and numerical simulations show, all
of these features are controlled by the characteristic distances of
the trap architecture near the junction. These include the channel
width of the trap, $a$, defined by the perpendicular distance
between the rf electrodes and the rf nodal pathway, the aspect ratio
of the trap, $\alpha$, defined by the ratio of the channel width,
$a$, with the vertical separation, $d$, between the rf and control
electrodes, and finally, the ratio, $\delta$, which is defined by
the distance between the electrodes, $h$, to the thickness of the
electrodes, $w$. This last ratio is significant only in the case of
very large aspect ratios. The salient features of the rf barrier for
most traps are set primarily by the channel width $a$. The rf
barrier results from unbalanced fields produced by the electrodes
across the junction, as well as from the corners of the nearest
electrodes. Therefore, as a practical matter, it is best to design
the trap so that the control electrodes are segmented into pieces
having a width less than or equal to $a$. This guarantees that
quasistatic control field gradients can be generated to overcome the
rf barrier gradient. This simple rule of thumb was shown in
numerical simulations to be sufficient to reliably shuttle the ions
around the corner of both two and three layer traps. If the control
electrode segments are much larger than $a$ (such as in the
T-junction array of Ref. \cite{T paper}), the shuttling procedure
involves control electrodes that are far removed from the ion's
position, and very high voltages are required to produce static
potential gradients that cancel those of the rf barrier. Such a
protocol may result in the ion acquiring a significant amount of
kinetic energy that will need to be mitigated via laser cooling,
sympathetic cooling, or phase sensitive switching of the trapping
potentials.

\begin{figure}[tb!]
\centerline{\epsfig{file=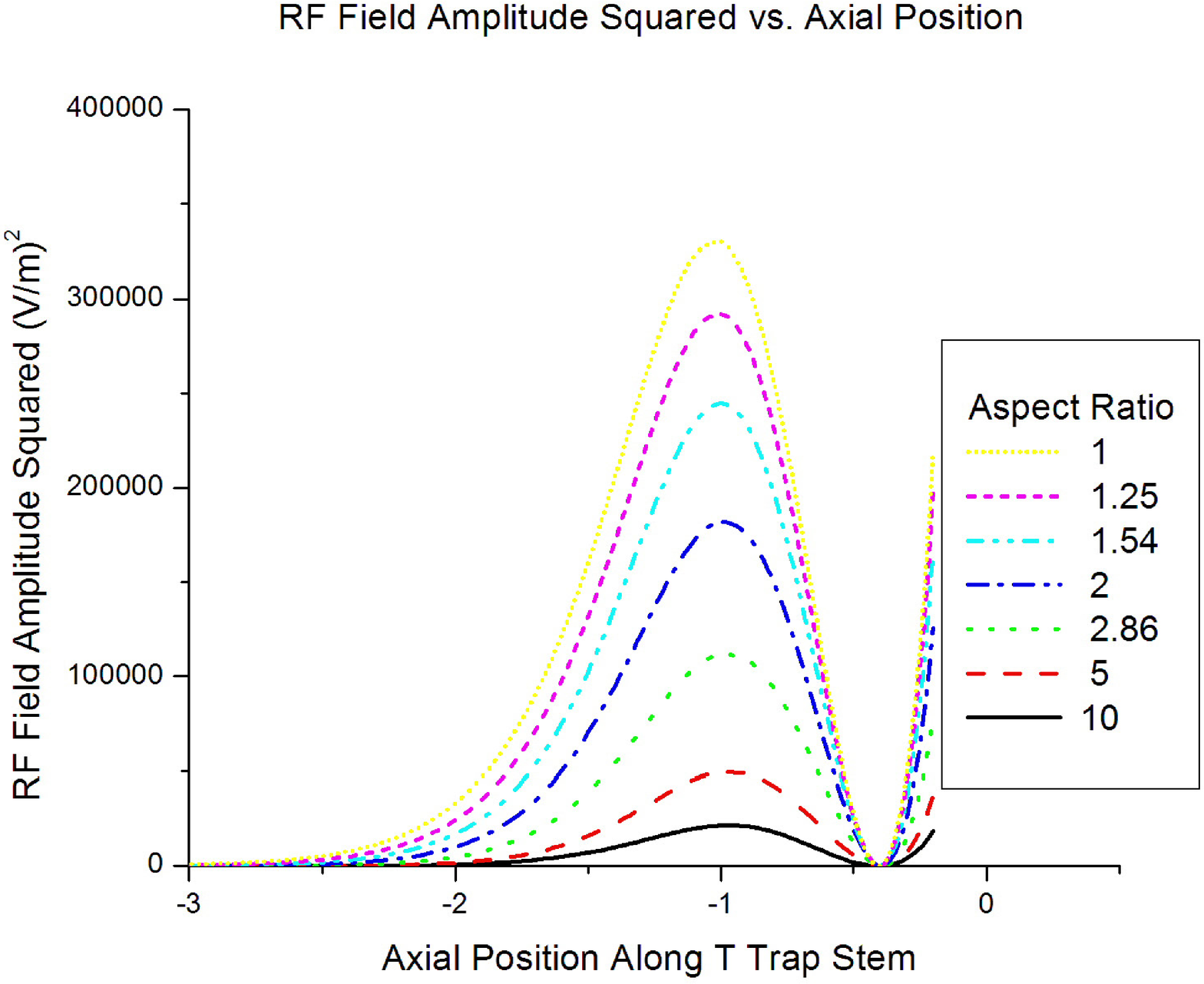, width=8.2 cm}} 
\vspace*{13pt} \caption[]{\label{fig:rf_barrier_strength_and_aspect}A
plot of the rf electric field amplitude squared vs. axial position
along the axis of the stem of a three layer T trap with one volt
applied to the rf electrodes. The axial position is scaled with the
channel width, $a = 100 \mu m$. The origin is at the center of the T-junction. The electrode thickness was modeled
to be $d= 20 \mu m$, and aspect ratios, defined as the ratio of the distance
from rf electrode tip to rf electrode tip to the distance between the rf electrodes
and the control electrodes (see Fig. \ref{fig:rf_depthstrength_vs_aspect}), $\alpha$, were chosen
ranging from 1 to 10.}
\end{figure}

\begin{figure}[tb!]
\centerline{\epsfig{file=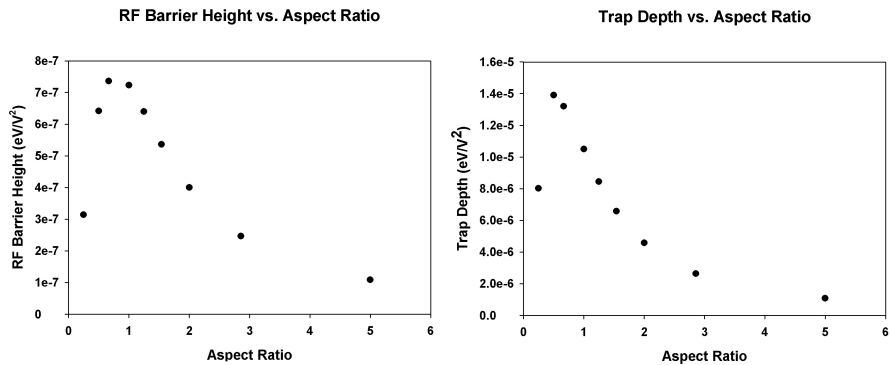, width=12cm}} 
\vspace*{13pt} \caption[]{\label{fig:barrier_and_height}Plot of the
rf barrier strength and trap depth at the barrier maximum versus the
aspect ratio in a three layer symmetric T-trap array with 1 volt
applied rf. The aspect ratio is defined to be the ratio of the rf
layer channel width to the tip to tip separation of the rf and
control electrode layers. It is possible to decrease the rf barrier
height at the expense of trap depth by increasing the trap aspect
ratio.}
\end{figure}

\begin{figure}[tb!]
\centerline{\epsfig{file=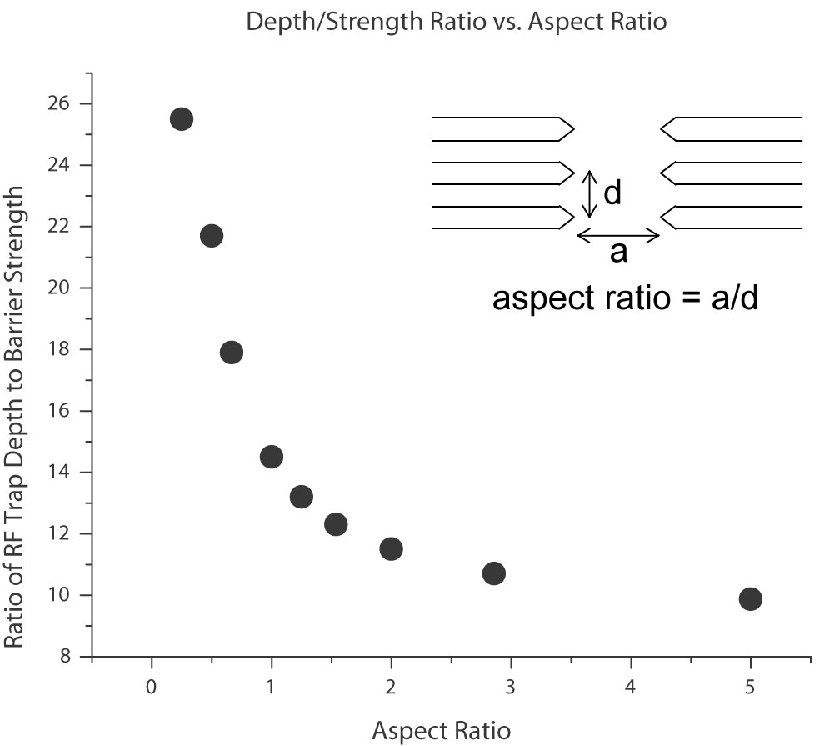, width=8.2 cm}} 
\vspace*{13pt} \caption[]{\label{fig:rf_depthstrength_vs_aspect} A
plot of the ratio of rf trap depth and barrier strength vs. aspect
ratio in a three layer T trap. The T has channel width $a = 100 \mu
m$ and electrode thickness $d= 20 \mu m$. }
\end{figure}

The rf barrier strength for a three layer trap of width $a= 100$
$\mu$m and electrode thickness $w = 20$ $\mu$m is shown for various
trap aspect ratios $\alpha$ in Fig.
\ref{fig:rf_barrier_strength_and_aspect}.
Note that the spatial extent of the barrier is largely unaffected by
aspect ratio. However, the barrier strength decreases dramatically
as the aspect ratio increases, as the figure illustrates. This can
be understood by considering the effect of the control electrodes as
they come closer to the rf electrodes, drawing the electric field
lines from rf to control electrodes and reducing the electric field
amplitude uniformly out in the channel. Aspect ratio also has a
direct impact on trap depth (Fig. \ref{fig:barrier_and_height} (b)),
so that trap depth increases with decreasing aspect ratio, up to the
limit of an aspect ratio of $\approx$ 1/2. Below that limit, the geometry
approaches the situation of an rf layer with ground at infinity, and
the depth therefore begins to drop again to its asymptotic value. By
taking the ratio of rf trap depth to barrier strength, we obtain
Fig. \ref{fig:rf_depthstrength_vs_aspect}. This figure shows that
the trap depth remains at least ten times as great as the barrier
strength, even for large aspect ratios, while it increases to twenty
five for an aspect ratio of 0.25. By decreasing the trap aspect
ratio in three layer junctions, the trap depth at the junction can
be made much larger than the rf barrier strength, providing a large
margin of safety for reliably shuttling through the junction.  More
importantly, by increasing the aspect ratio of a three-layer
junction, the rf barrier can be made arbitrarily small while
retaining a strong trapping potential inside the junction region.

\section{\label{sec:T example}Practical implementation of shuttling operations}
\subsection{\label{sec:shuttling}Introduction}

In order to realize most quantum computing architectures with
trapped ions, it may be necessary to shuttle ions between memory
(storage) zones and interaction (entanglement) zones. To bring any
two ions together in an entangling zone, it is necessary to be able
to sort a linear chain of ions into any desired order. This requires
the successful implementation of four key protocols: separating two
ions that are located in the same trap, linearly shuttling two ions
that are in the same trap, recombining two ions together into one
trap, and shuttling ions around corners through a junction on an
individual basis \cite{barrett:2004, shuttle error correction,
chiaverini:2005, rowe, T paper}. The combination of all four
elementary protocols allows for arbitrary control of trapped ions in
two dimensions.

\begin{figure}[tb!]
\centerline{\epsfig{file=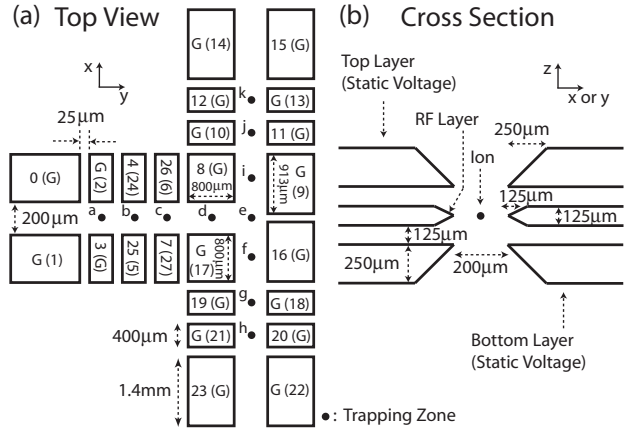, width=8.2cm}} 
\vspace*{13pt} \caption[]{\label{ion trap schematic}Top view and
cross section of the Michigan T-junction ion trap array showing all
11 trapping zones a-k \cite{T paper}. The control electrodes are
numbered with bottom-layer control electrodes in parentheses. G
indicates a grounded control electrode.}
\end{figure}

The simulation of ion trap potentials via the method of basis
functions has thus far been completely general. The power of this
method and its utility for simulating potentials will be shown by
analyzing the shuttling protocols used in experiments at the
University of Michigan \cite{T paper} where ions are shuttled
linearly, around a corner, and are swapped in an 11-zone, three
layer T-junction ion trap array (see Fig. \ref{ion trap schematic}).
Discussing the practical implementation of shuttling operations in
this particular geometry will serve as an instructive example and
will provide recipes to develop shuttling protocols for arbitrary
geometries. As discussed in Sec. \ref{sec:trap design}, there are
ways to design optimal ion trap geometries. However, fabrication
constraints (such as the longitudinal extension of the corner
electrode inside a junction) may result in non-optimal trap
geometries. The strategies that will be introduced in order to
overcome such constraints for the T-junction array discussed here
are therefore of general interest for designing shuttling protocols
in any two-dimensional ion trap.

\begin{figure} [tb!]
\centerline{\epsfig{file=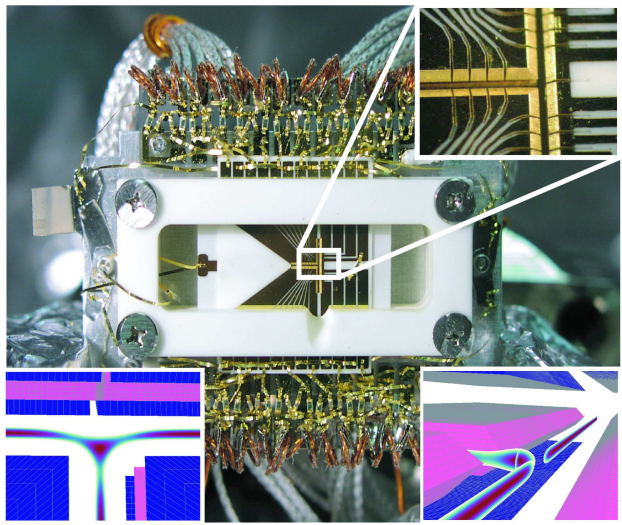, width=8.2cm}} 
\vspace*{13pt} \caption[]{\label{fullmodel}T-junction ion trap with
associated RC filters, wirebonds, and wires. The T-junction ion trap
is an eleven zone ion trap with 48 DC electrodes of which 28 are
connected to variable external voltages. The fabrication of the ion
trap array required over 500 wire bonds. The inset in the upper
right corner shows a magnified view of the junction region. The
segmented linear DC control electrodes have an axial extent of
400$\mu$m while the corner electrodes have an axial extent of
$\sim$800$\mu$m. The graph in the lower left is a plot of the rf
pseudopotential near the T-junction and shows the RF barriers that
impede entry to the junction region from all three directions. The
picture in the lower right is a perspective view of the potential
looking down the channel in the top of the T-junction.}
\end{figure}

The Michigan T-junction ion trap array \cite{T paper} has 49
electrodes and a sufficient number of trapping zones to swap the
positions of two ions. The central layer contains a T-shaped
channel; the electrode is formed by depositing gold around the
channel with an electron beam evaporator on an alumina substrate.
Gold-coating of the 24 control electrodes on each of the two outer
layers is accomplished with dry-film photolithography and
wet-chemical etching. Here, electrodes and tracks are formed by
depositing 0.015 $\mu$m of titanium followed by 0.4 $\mu$m of gold.
Two thin alumina spacer plates are inserted between each outer layer
and the central rf layer substrate. All three substrates are held
together via rectangular alumina mount bars. Chip capacitors and
resistors are ribbon-bonded onto a gold coated quartz plate that is
mounted adjacent to the alumina substrates (top and bottom of figure
\ref{fullmodel}). To isolate the control electrodes from external
noise and from induced rf from the nearby rf electrode, each of the
28 non-grounded control electrodes is immediately shunted to ground
via a 1 nF capacitor and then connected in series to a 1 k$\Omega$
resistor leading to the vacuum feedthrough.

The trap array has a plane of symmetry along the $xy$ plane through
the rf layer, and a second in the $yz$ plane which divides the trap
along the stem of the T, with the origin at the center of the
junction of the T. The trap array was modeled in Vector Fields'
Opera by centering it in a bounding box extending 20,000 $\mu$m in
all three directions. For simulations done on an Intel 4, 2.8 GHz
processor with 1.0 GB of RAM running Windows XP, the number of nodes
in the problem ranged from 1 to 2 million. When calculations for the
rf pseudopotential were carried out, the node spacing was made
tightest in the junction region of the T, where it reached a minimum
of 10 by 10 by 1.5 $\mu$m. When calculations for the control
electrode basis functions were performed, the node spacing was kept
at 10 by 10 by 3.0 $\mu$m along the electrodes where the potential
changed most rapidly. The electrodes themselves were excluded from
the simulation volume.

The rf pseudopotential used to trap ions in the T depends on the
square of the electric field amplitude produced when the rf layer is
at the maximum voltage and the two outer layers of electrodes are
held at rf ground. To estimate the error in the calculated electric
field, a comparison was made between the simulation used to
determine the rf pseudopotential, and an identical model with the
mesh density doubled throughout. (Because of memory limitations,
this higher mesh density model took one month to run on a  3.2 GHz
dual-processor PC workstation with 1.0 GB of RAM, running Windows
XP.) The voltage used on the rf layer in both cases was 1.0 V. The
field was evaluated for both models along a grid of points along the
entire length of the top channel of the T, and the fractional
difference in the field was evaluated at each point. The average
fractional difference was equivalent to a 0.15\% error. This average
was sharply skewed upward because of the presence of the nodal lines
along the center of channel due to field cancelation. Along that
line, fractional errors of 60\% were reached. The difference in
actual values of the field along the nodal line, however,
corresponded to less than 1 V/m. This is insignificant when compared
to the fields of thousands of V/m just microns off axis. The
electrostatic potential calculations done with the same mesh spacing
had significantly smaller errors than even these. Once these fields
have been determined numerically, it becomes possible to construct
the basis functions necessary to calculate and analyze the required
shuttling protocols.

\subsection{\label{Linear Shuttling}Linear Shuttling}
The most elementary shuttling protocol is moving an ion along a
linear path from one position to another.  In a segmented linear rf
Paul trap, this procedure implies shuttling the ion along the rf
node between trapping zones.  Several notable experiments have
already utilized linear shuttling protocols, and have demonstrated
essentially unit probability of success \cite{barrett:2004, rowe}.
Here we describe the design and the implementation of a shuttling
protocol for linear ion transport that holds the secular frequency
constant. This constraint allows us to use the simplified analysis
of section 3.3.1.

The simulations of trap voltages via the basis functions described
in Sec. \ref{CaluculatingIonDynamics} allow us to determine both the
trap frequencies and the position of the trap. The design of the
shuttling protocol begins by determining the start and end locations
of the trap; for instance, zones c (start) and b (end) in Fig.
\ref{ion trap schematic}. The total shuttling distance may then be
broken up into any number of steps, where the finite time needed to
change the voltages on the electrodes establishes an upper bound for
the total steps required. The frequency of the potential should be
kept constant throughout this process. In order to determine the
voltage changes required for each step in the experiment, two
actions are undertaken in the numerical simulations. The first is to
iteratively adjust/produce an asymmetry between the electrodes
serving as the endcaps of the current trapping zone, thereby moving
the minimum of the potential in the axial direction by the
designated amount (to within some arbitrary choice of error).  The
change in voltage will usually alter the axial trap frequency as
well. The second action seeks to compensate for this by multiplying
all the voltages by some scale factor, iteratively adjusted until
the previous secular frequency is once again obtained. While
multiplication by this scale factor appears to have negligible
effect on the position of the potential minimum, any shift could
always be accounted for by iteratively repeating these two previous
actions. This method is very robust in that each step is some small
shuttling distance, which may be used to produce any time dependent
behavior (\mbox{e.g.} linear, sinusoidal, or transcendental
potential minimum time profile) for the potential minimum, $x_0(t)$,
in the total shuttling protocol as described in Sec.
\ref{sec:shuttlingprotocols}. The final product is an array of
voltages in which the axial trap frequency is kept nearly
constant\footnote{The amount of fluctuation in the secular frequency
can be minimized, as additional computation time allows iteration on
a finer scale.}\,\,.

\begin{figure}[tb!]
\centerline{\epsfig{file=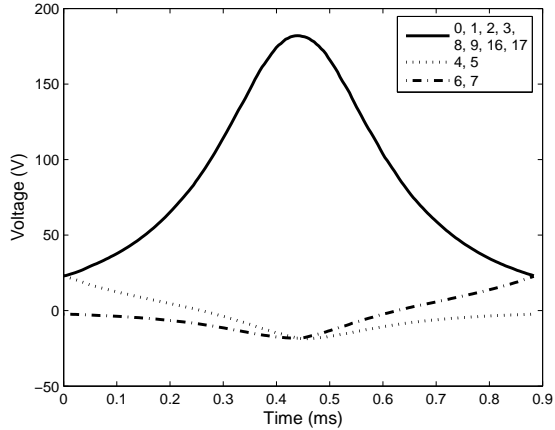, width=8.2cm}} 
\caption[]{\label{fig:linearshuttlefig}This shows the voltages on the
various electrodes of the trap while the ion is linearly shuttled
from zone c to b.}
\end{figure}

Using the scheme described above, we generated a shuttling protocol
consisting of 200 discrete time steps that shuttled the ion from
zone c to b, a distance of $425$ $\mu$m in $1$ ms. The ion was moved
a constant distance at each step, $2.14$ $\mu$m, resulting in a
linear potential minimum time profile. The axial trap frequency was
kept at 1.173 MHz with a tolerance of 0.5\%. The voltage sequence is
shown in Fig.~\ref{fig:linearshuttlefig}. The experiment was carried
out such that a single $\mbox{Cd}^{+}$ ion was laser-cooled,
shuttled, detected, and then laser-cooled once again. The success
rate was observed to be $>99.999$\% (100000 attempts). Hence, we
have demonstrated a linear shuttling protocol that has virtually
unit probability of success.

In order to show that the shuttling protocol is adiabatic, it is
necessary to cool the ion to the ground state and measure its
temperature before and after the shuttling operation. A somewhat
weaker test would be to shuttle the ion back and forth many times
between two trapping zones without the application of any laser
cooling. If the boil-out time (the time it takes for the ion to be
heated out of the trap) remains unchanged with and without the
shuttling operation, weak evidence for adiabaticity is obtained.
This test would constitute a necessary but insufficient condition
for the shuttling process to be adiabatic.


\subsection{\label{corner shuttling}Corner Shuttling}


\subsubsection{\label{corner shuttle development}General considerations}
There are many challenges associated with shuttling an ion around a
corner. First of all, a suitable control electrode geometry should
be chosen to give maximum control over the ion in the junction
region. Specifically, the control electrode widths should be no
larger than the channel width of the ion trap array. It is still
possible to shuttle ions around a corner using electrode sizes that
are larger than the channel width \cite{T paper}, but it becomes
difficult to control the motion of an ion in the junction region.
Smaller control electrodes also make it easier to overwrite rf
barriers near junction regions without destabilizing the trap in the
transverse direction, however practical fabrication and operation
constraints will typically place certain limitations on the
electrode geometry. In addition for the case of asymmetric (planar
ion traps) one also needs to account for the presence of rf humps
(see Sec. \ref{sec:trap design}).

Secondly, large electrodes near a junction region can make it
difficult to maintain constant trap frequencies when shuttling an
ion into a junction region without destabilizing the trap in the
transverse direction. Maintaining constant trap frequencies may be
an important factor in minimizing acquired kinetic energy when
shuttling ions around a corner as is discussed in Section
\ref{sec:adiabatic}. This presents a unique challenge when shuttling
an ion around a corner because the weak axis of the trap rotates.
For example, when shuttling from zone d to zone i (see Fig. \ref{ion
trap schematic}), the weak axis is originally in the y-direction
when an ion is in zone d. After an ion is shuttled around the corner
to zone i, the weak axis is in the x-direction. Smaller control
electrodes should help maintain a constant trap frequency along the
weak axis.

A third challenge to corner shuttling which may arise is the
accurate simulation of an ion's motion through an ion trap array.
The advantages of the Bulirsch-Stoer method are discussed in Section
\ref{sec:overall comparison} and the Appendix. The accurate
simulation of an ion's classical motion plays a key role in
calculating  the average motional state of the ion during shuttling.
A limitation to the accuracy of numerical solutions of an ion's
trajectory can arise in the form of a local maximum in the plane of
the trap array when trying to overwrite the rf barrier. For example,
in the T-trap, there is a small local maximum in the pseudopotential
at $t=25$ $\mu$s in the xy plane during corner shuttling near the
center of the junction near zone e (see Fig. \ref{ion trap
schematic}) that slopes down toward zone i and zone f. A local
maximum in the pseudopotential means that the calculated motion is
sensitive to perturbations in the initial conditions or phase of
oscillation of an ion and can cause errors in an ion's calculated
trajectory in the junction region. The issue of having a local
pseudopotential maximum is related to control electrode size;
smaller control electrodes should allow a control electrode
potential to overwrite any small local potential maximum.

Another issue important to the accuracy of the numerical simulation
is whether to use the pseudopotential approximation when simulating
the rf trapping field. As was discussed earlier, it is possible to
simulate the ion's motion either with the time averaged
pseudopotential (Eq. \ref{Pseudo-approx}) or the actual rf potential
(Eq. \ref{Exact Potential}). In any of the junction geometries
considered in this paper, it is impossible to maintain an ion near
an rf node throughout a corner shuttling procedure, making the
impact of micromotion on the ion trajectory potentially significant.
In addition, if the shuttled ion does pick up a significant amount
of kinetic energy, it will likely explore regions in the final
trapping configuration where micromotion is potentially large.  In
our case study of the T-junction array, the ion typically picks up
~0.5 eV of energy due to micromotion alone.  This is comparable to
the final kinetic energy of the large scale secular motion of the
ion. Therefore, despite the computational speed-up (in our case, a
factor of 2) by using the pseudo-potential, in order to fully
understand the ion's motion through a junction, one should simulate
the ion's motion using the sinusoidally varying rf potential from
Eq. \ref{Exact Potential} if there is any uncertainty whether the
process is fully adiabatic.

In order to provide a detailed overview on how such a corner
shuttling operation is achieved, we discuss the particular example
of shuttling an ion around the corner inside a T-junction ion trap
array \cite{T paper}. This particular array is limited by the fact
that the control electrodes at the corners of the junction are too
large (4 times larger than the channel width) to achieve adiabatic
shuttling throughout the junction. In contrast, the ions in this
trap are pushed through the rf barriers by applying very high
control electrode voltages on those electrodes (labeled 6, 7, 26,
27, 10, 11, 18 and 19 in Fig. \ref{ion trap schematic}) that are
nearest neighbors to the corner electrodes of the T. These large,
but distant, voltages produce a sufficiently positive gradient in
the potential at the position of the rf barriers so that the ions
can be pushed into the junction region and then directed through one
of the rf barriers in the top of the T to be trapped in one of the
zones labeled f through k. In doing so, an ion gains $\approx$1 eV
of kinetic energy. The large extent of the corner electrodes and the
finite trap depth in the junction of the T means that performing
detailed numerical simulations was crucial for determining those
shuttling protocols which would successfully move ions through the
junction of the trap array. Constraints in future trap designs is
likely to remain significant as the demands on the performance of
these traps increase as well. An ion trap array that contains many
qubits may require many thousands of electrodes to carry out any
needed shuttling operations. It is inevitable that fabrication
constraints may result in non-optimal geometries. Therefore, the
numerical analysis of the T-junction ion trap array \cite{T paper}
and the comparison of the simulated ion dynamics with experiment
during the operation of this trap provide a good case study when
considering the process for scaling up current trap technology to
meet the demands of quantum information processors and other ion
trap applications.

\subsubsection{\label{forward}Stem-to-top corner shuttling in a T-junction array}

\begin{figure}[tb!]
\centerline{\epsfig{file=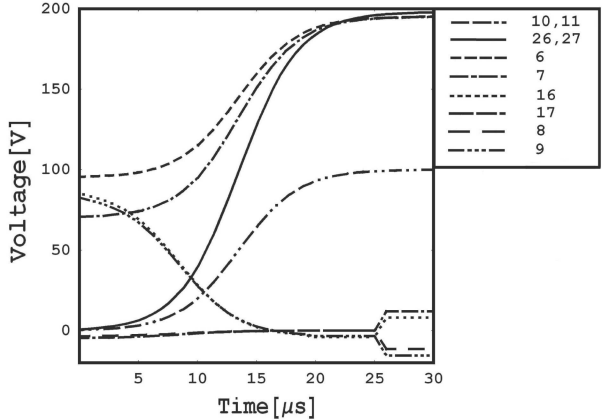, width=8.2cm}} 
\caption[]{\label{voltage sequence of voltage profile}The control
electrode voltage profile for the corner shuttling protocol. The
figure shows the voltages applied to the relevant electrodes to
execute the corner shuttling protocol.}
\end{figure}

Here we describe the process of stem-to-top corner shuttling using
the specific example of a T-junction ion trap array \cite{T paper}.
An ion is initially trapped in zone d (see Fig. \ref{ion trap
schematic}) with trap frequencies ($\omega_{x}/2\pi$,
$\omega_{y}/2\pi$, $\omega_{z}/2\pi$) = (5.0 MHz, 0.7 MHz, 4.9 MHz).
Figure \ref{voltage sequence of voltage profile} shows the voltage
time profile used to shuttle an ion around the corner.
The voltages of control electrodes 6, 7, 26, 27 are raised to
$\sim$200 V in $\sim$ 20 $\mu$s using hyperbolic tangent voltage vs.
time profiles to push the ion toward the junction region. Note that
this voltage time profile does not imply that the potential minimum of the
trap follows a hyperbolic tangent in time (see Eq. \ref{eqn:movingmin}). It
is possible to specify a particular potential minimum versus time
profile \cite{hansel:2001}. Simultaneously, the voltages of control electrodes 8 and 17 are
raised from approximately -4 V to 0 V while control electrodes 9 and
16 are lowered from $\sim$80 V to -3 V. The net effect of these
voltage changes over the first $\sim$ 20 $\mu$s is to push the ion
through the rf potential energy barrier.

\begin{figure}[tb!]
\centerline{\epsfig{file=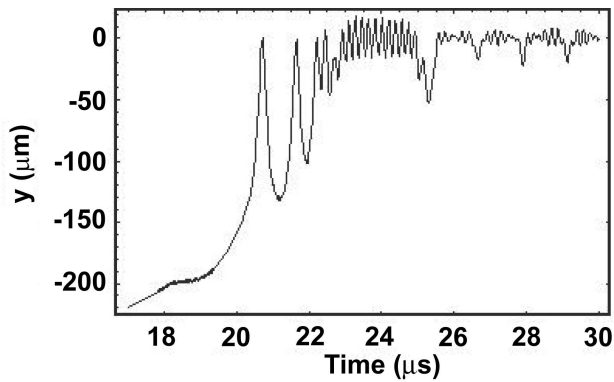, width=8.2cm}} 
\caption[]{\label{ycomponent of forward shuttle} Simulation
(including micromotion) of the ion's position along the y-axis (axis
of the stem of the T) in and around the junction region during
corner turning. Note the accentuated micro-motion from $t=18$ $\mu$s
to $t=20$ $\mu$s as the ion traverses the rf barrier.}
\end{figure}

\begin{figure}[tb!]
\centerline{\epsfig{file=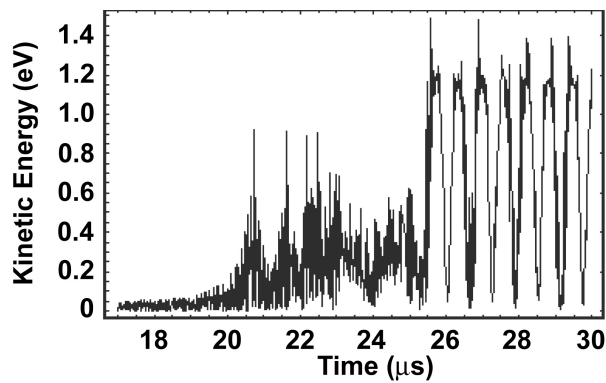, width=8.2cm}} 
\caption[]{\label{KE of forward shuttle}Simulation of the ion kinetic
energy in and around the junction region during the corner turning
protocol. Note the rise in kinetic energy after the ion crosses the
rf barrier at $18$ $\mu$s and the abrupt rise in energy at the $25$
$\mu$s when the ``catch'' step is implemented.}
\end{figure}

\begin{figure}[tb!]
\centerline{\epsfig{file=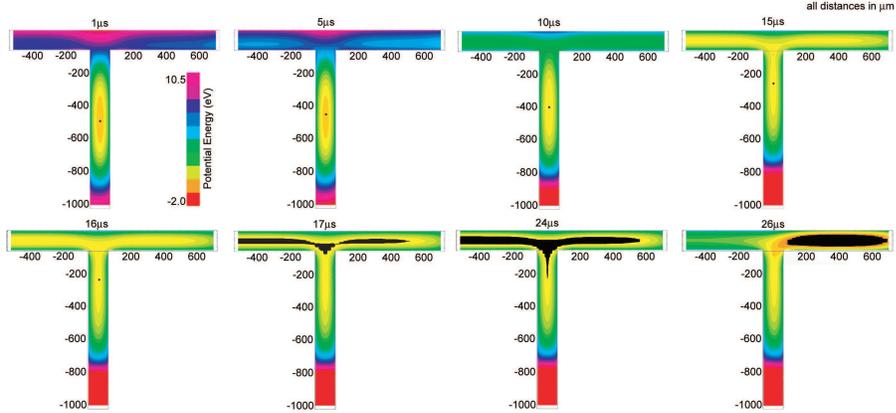, width=12cm}} 
\vspace*{13pt} \caption[]{\label{pottable}Potential energy map (in
the pseudopotential approximation) as an ion is shuttled around a
corner of the T-junction. The region energetically accessible to the
ion is indicated in black. The kinetic energy of the ion is
initially small, so the region of the T that the ion can
energetically traverse is small. As the ion goes through the
 rf barrier, it
acquires $\sim$1 eV of energy, so the region that the ion can
energetically traverse is much larger.}
\end{figure}

After going through the rf barrier, the ion is {\em accelerated} by
the potential gradient on the backside of the rf barrier from
$y=-220$ $\mu$m to $y=0$ $\mu$m in $\sim2$ $\mu$s, acquiring
$\sim$0.5 eV of kinetic energy in the process (see Fig
\ref{ycomponent of forward shuttle} - \ref{pottable}). The ion
continues in the y-direction past zone e toward the rf layer in the
top of the T trap. The ion is pushed back by the rf potential toward
zone e. Thus, the first step in the corner-shuttling protocol is to
successfully move an ion into the junction region and then keep it
trapped there until the next stage of the protocol is implemented.
This is made difficult by the relatively weak trap in the junction
region, and the ion's relatively large kinetic energy acquired
during its traversal of the rf barrier into the junction. Note that
in the shuttling protocol as recorded in Fig.~\ref{voltage sequence
of voltage profile}, the voltages on the control electrodes 8, 17, 9
and 16 all converge nearly to ground while the ion was in the
junction region.  High voltages on these electrodes can easily make
zone e anti-trapping. There was a delicate tradeoff when designing
the corner shuttling protocol for this trap. A sufficiently high
potential gradient needed to be applied to push the ion through the
rf barrier. At the same time, the control electrode voltages near
the junction region needed to be kept near ground to prevent the
junction region from becoming anti-trapping. The key to successfully
shuttling an ion around the corner through the junction region
proved to be finding the right balance between these two
requirements.

When the ion is in the junction region, the four corner electrodes
are used to guide the ion towards zone i. When the ion has reached
the desired final trapping position at zone i, the control electrode
voltages change to their final values in $\simeq 1$ $\mu$s. This
rapid change in voltage is timed so that the ion crosses the final
trap minimum as the voltages ramp up. Changing the voltages in this
way helps to minimize any further contribution to the ion's kinetic
energy. Specifically, the voltages of control electrodes 16 and 17
are ramped to approximately $+10$ V while the electrodes 8 and 9 are
lowered to approximately -10 V in 1 $\mu$s, ``catching" the ion in
zone i. Simulations incorporating micromotion show that the amount
of kinetic energy added in this step is still considerable: 0.7 eV
(see fig \ref{KE of forward shuttle}). The final trap position in
zone i is defined by these voltages, with the resulting secular
frequencies ($\omega_{x}$, $\omega_{y}$, $\omega_{z}$) = (0.5 MHz,
5.5 MHz, 4.3 MHz).

When the ion enters zone e, there is a local potential maximum in
zone e. This local maximum places a limitation on the accuracy of
numerical solutions because small perturbations in the initial
position of the ion can cause the ion to go in radically different
directions in the junction region. The optimization of the ``catch"
step described in the preceding paragraph is to ensure that when an
ion does go past the potential maximum in the correct direction, the
ion will be caught with little added kinetic energy. However, if the
ion is guided by the local maximum toward the opposite direction,
then the ion will gain considerably more kinetic energy as the
voltages suddenly ramp up to form the final trap. In addition, there
is some evidence that this is an effect that can be seen
experimentally. Occasionally the ion was observed to crystallize in
zone i one full second after being shuttled from zone d, a crystallization time several orders of
magnitude longer than the crystallization time after linear shuttling. This could
be due to the fact that the ion occasionally makes an excursion
toward zone f which means that when the voltages suddenly change,
the ion gains a lot more kinetic energy and takes longer to cool.

The corner shuttling protocol was used to shuttle the ion from zone
d to zone i in the T-trap in 26 $\mu$s with a success rate of
greater than $99\%$ (881 out of 882 attempts) \cite{T paper}. The
speed of the shuttling protocol was limited by the RC filters and
the speed of the analog output cards that supply voltage to high
speed op amps that are connected to the trap electrodes.

Despite the high success rate of the shuttling protocol from zone d
to zone i, we were unsuccessful in shuttling the ion from zone d to
zone f using a voltage protocol mirrored at the y-z plane through
the center of the stem of the T. This discrepancy may be attributed
to static bias fields or a misalignment observed in the three
electrode layers due to the manual assembly process. The use of
semi-conductor etching techniques to build multi-layer ion trap
arrays is one way to avoid such misalignments and asymmetries in the
trap \cite{stick:2006, brownnutt:2006}. Instead, a composite shuttling protocol was
used.  The protocol begins with shuttling the ion from zone d to i
as described above, the ion is laser cooled and then shuttled from
zone i to f in a linear fashion. The shuttling from i to f is not
optimized to maintain constant secular frequency, however, the
success rate was 100\% (50 attempts).

\subsubsection{Top-to-stem corner shuttling in a T-junction array}
The shuttling sequence from trapping zone i to trapping zone d is
similar to the shuttling protocol from zones d to i in that the ion
must be pushed through the rf barrier without causing the trap to
become destabilized in the transverse direction while the ion is in
the junction region. However, there is an important difference
between the two protocols in the T-junction trap array.  In general,
it is not possible to simply reflect the voltage profile that
shuttled an ion from stem-to-top (zone d to i) in time in order to
shuttle from top-to-stem (zone i to d). While Newton's equations of
motion are time-symmetric, the initial conditions of an ion being
shuttled from zone i (for example after laser cooling) to zone d
are not the same as the final conditions of the ion after it has
acquired $\sim$1 eV of kinetic energy after being shuttled from zone
d to zone i. Therefore, to use the time reflected stem-to-top
control voltage profile in order to shuttle from top-to-stem, one
would need to prepare the ion in zone i with initial conditions
equal to the final conditions after the shuttling operation from d
to i. The complexity of the geometry will determine how well the two
sets of conditions will need to be matched. As this may be difficult
to achieve one may have to design a new voltage control sequence.
Per definition an adiabatic shuttling protocol can always be time
reflected for a successful reverse shuttling operation.

\begin{figure}[tb!]
\centerline{\epsfig{file=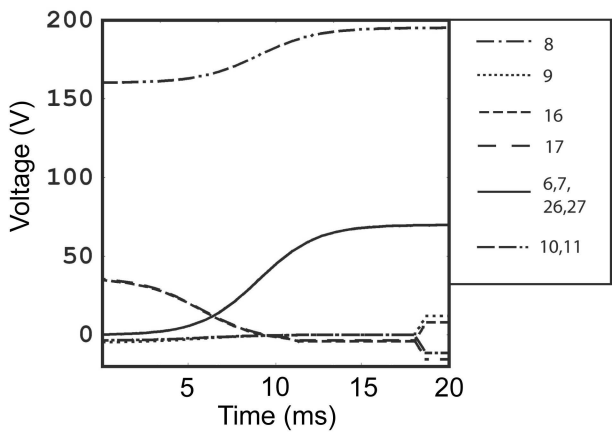, width=8.2cm}} 
\vspace*{13pt} \caption[]{\label{ReverseVoltageFile}Control electrode
time profile for corner shuttling from the top of the T to the stem
of the T. A plot of the voltage vs. time for the electrodes used to
shuttle the ion from the top of the T to the stem of the T. }
\end{figure}

Even though simply reversing the stem-to-top protocol was
unsuccessful, it served as the basis for developing a working
routine.  In order to realize the top-to-stem protocol, the roles of
the following electrodes are swapped:
\begin{eqnarray}\label{swapelectrodes}
9 &\leftrightarrow& 17 \\10 &\leftrightarrow& 6 \nonumber
\\11&\leftrightarrow&  7\nonumber
\end{eqnarray}
while electrodes 8 and 16 keep the same time dependent voltage
profile. The successful shuttling protocol from zone i to zone d
\cite{T paper} starts with an ion in zone i with trap frequencies
($\omega_{x}$, $\omega_{y}$, $\omega_{z}$) = (0.7 MHz, 4.5 MHz, 4.2
MHz). The voltages on control electrodes 10 and 11 are changed to
200 V via a hyperbolic tangent function while simultaneously raising
the voltages on control electrodes 8 and 9 from $-4$ V to ground and
lowering the voltages on electrodes 16 and 17 from $35$ V to $-4$ V.
This provides enough potential gradient to push the ion through the
rf barrier into the junction region (zone e). Again, the ion gains
$\sim$1 eV of kinetic energy after going through the rf barrier, and
once the ion is in the junction region, it makes large oscillations
in the relatively flat trap of zone e. Then the voltages on the
control electrodes are suddenly ramped, as can be seen in Fig.
\ref{ReverseVoltageFile}, to form a trap in zone d with
($\omega_{x}$, $\omega_{y}$, $\omega_{z}$) = (5.0 MHz, 0.6 MHz, 3.6
MHz). The timing of this step is done in an attempt to have the
voltage ramp occur when the ion is near the potential minimum to
minimize the amount of  kinetic energy gained.

The control voltage scheme in Fig.~\ref{ReverseVoltageFile}
successfully shuttled an ion from zone i back to zone d in the
T-junction ion trap with a success rate of 98\% \cite{T paper} (118
attempts) with a total shuttling time of $30$ ms.  The reliability
of the shuttling protocol drops off dramatically if we increase the
shuttling speed. We failed to observe any instance of successful
shuttling if the shuttle time was less than 20 ms; this is 3 orders
of magnitude slower than the shuttling scheme between zone d to zone
i. If misalignments in the trap electrodes and stray electric fields
are incorporated into simulations, one should be able to optimize
the voltage control scheme so that an ion may be shuttled
successfully from zone i to zone d just as fast as an ion may be
shuttled from zone d to zone i.

\subsubsection{\label{stability plot}Characterization and optimization of corner shuttling protocols}
Once a shuttling protocol is created that successfully transports a
single ion around a corner inside an array, a number of steps can be
carried out in order to optimize the original shuttling protocol
depicted in Fig. \ref{voltage sequence of voltage profile}.
Indicators for the quality of the protocol consist of the success
rate of the protocol and estimations of the kinetic energy the ion
acquires during the shuttling process based on numerical analysis.
The corner shuttling protocols being discussed here can be refined
using numerical simulations in which the relevant voltages on the
corner electrodes are systematically perturbed. In order to
illustrate this process we give a detailed discussion of such
refinements that were carried out for the forward shuttling protocol
used in the T-junction array \cite{T paper}.

\begin{figure}[tb!]
\centerline{\epsfig{file=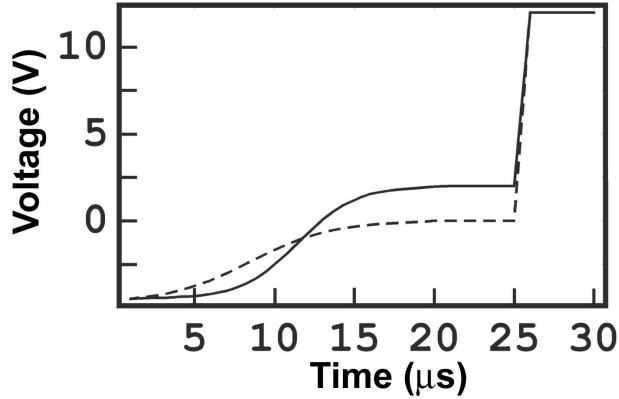, width=8.2cm}} 
\caption[]{\label{originalandperturb} Example of the perturbation of
the corner shuttling protocols. A time profile of the voltage on
electrode 17 from the original forward shuttling protocol (dashed
line) and the perturbation $V_{17,+2}$(t) (solid line) to that
profile.}
\end{figure}

The process of shuttling ions around the corners of the T-trap
depended most sensitively on the voltages applied to the four
junction control electrodes (electrodes 8, 9, 16, and 17) of the
T-trap.  The figure of merit to optimize in a corner shuttling
protocol is the acquired kinetic energy during shuttling. In this
case, the protocol was perturbed by changing the final electrode
voltages of the corner electrodes before the sudden ramp in the
voltages that defines the final trap in zone i as seen in Fig.
\ref{originalandperturb}. We define V$_{i,j}(t)$ to be the time
dependent voltage of the i-th electrode, where j is the target
voltage for the electrode just before the sudden ramp at t= 25
$\mu$s (see Fig. \ref{originalandperturb}). The simulations were
examined to determine which set of voltages produced a successful
shuttling operation, and gave the smallest gain in kinetic energy to
the ion.

\begin{figure}[tb!]
\centerline{\epsfig{file=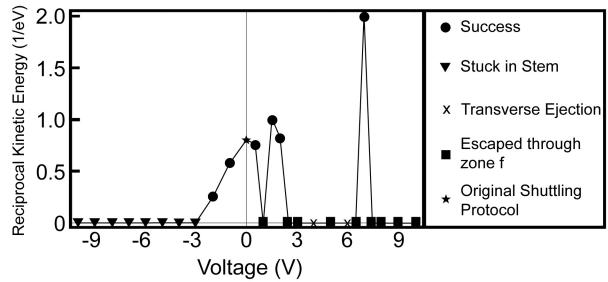, width=8.2cm}} 
\caption[]{\label{DC8stabplot}Stability plot for electrode 8. The
plot shows the reciprocal kinetic energy gained by the ion as a
function of the target voltage (as defined in Fig.
\ref{originalandperturb}) on electrode 8. The vertical line
highlights the original protocol shown in Fig. \ref{voltage sequence
of voltage profile}. All failed shuttling operations are arbitrarily
plotted with 0 reciprocal kinetic energy.}
\end{figure}

\begin{figure}[tb!]
\centerline{\epsfig{file=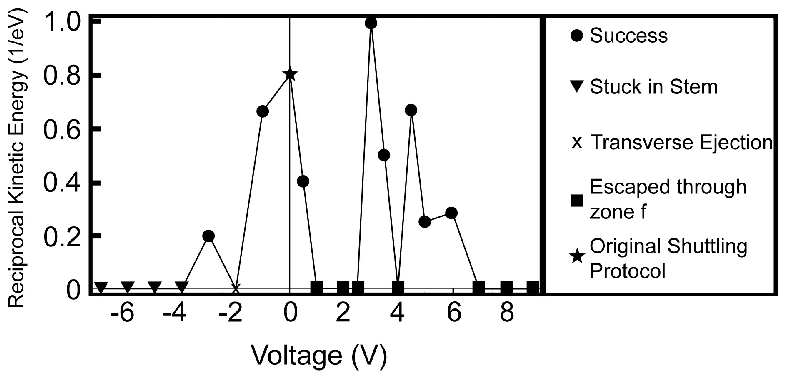, width=8.2cm}} 
\vspace*{13pt} \caption[]{\label{DC17stabplot}Stability plot for
electrode 17. The plot shows the reciprocal kinetic energy gained by
the ion as a function of the target voltage (as defined in Fig.
\ref{originalandperturb}) on electrode 17. The vertical line
highlights the original protocol shown in Fig. \ref{voltage sequence
of voltage profile}. All failed shuttling operations are arbitrarily
plotted with 0 reciprocal kinetic energy.}
\end{figure}

\begin{figure}[tb!]
\centerline{\epsfig{file=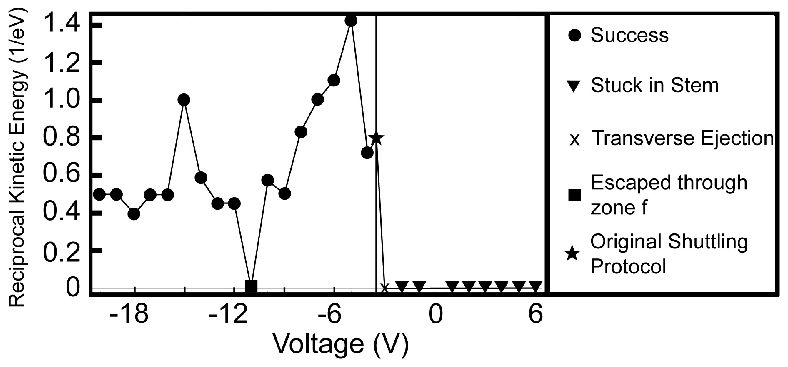, width=8.2cm}} 
\vspace*{13pt} \caption[]{\label{DC9stabplot}Stability plot for
electrode 9. The plot shows the reciprocal kinetic energy gained by
the ion as a function of the target voltage (as defined in Fig.
\ref{originalandperturb}) on electrode 9. The vertical line
highlights the original protocol shown in Fig. \ref{voltage sequence
of voltage profile}. All failed shuttling operations are arbitrarily
plotted with 0 reciprocal kinetic energy.}
\end{figure}

\begin{figure}[tb!]
\centerline{\epsfig{file=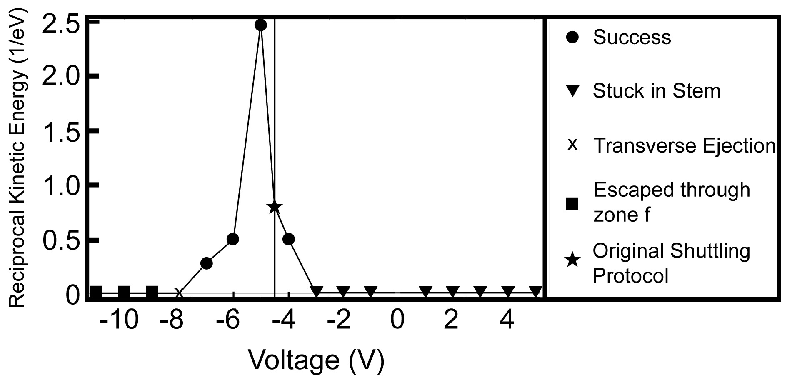, width=8.2cm}} 
\vspace*{13pt} \caption[]{\label{DC16stabplot}Stability plot for
electrode 16. The plot shows the reciprocal kinetic energy gained by
the ion as a function of the target voltage (as defined in Fig.
\ref{originalandperturb}) on electrode 16. The vertical line
highlights the original protocol shown in Fig. \ref{voltage sequence
of voltage profile}. All failed shuttling operations are arbitrarily
plotted with 0 reciprocal kinetic energy.}
\end{figure}

It is also important to consider the different ways a shuttling
protocol might fail. There are three basic ways in which shuttling
from zone d to zone i can fail. The ion might not make it through
the rf barrier and becomes stuck in the original trapping zone.
Alternatively, the ion could be ejected from the junction region
because the trap becomes too weak in the z-direction. Finally, the
ion could go too far toward zone f so that when the final voltage
ramp occurs, the ion is ejected in the negative x-direction.
Therefore, we examined the effect of the various perturbations on
the control electrode voltages by simulating the shuttling process
and recording the success or failure of the operation. The success
was further characterized by the gain in kinetic energy, while the
failure was characterized by the mode of failure observed. The
results of this analysis are shown in the various stability plots in
Figs.~\ref{DC8stabplot} - \ref{DC16stabplot}. Each plot show the
reciprocal kinetic energy gained by the ion as a function of the
target voltage on the corner electrodes. Unsuccessful operations are
arbitrarily plotted with a reciprocal kinetic energy of 0 eV$^{-1}$
with the type of failure marked.

The corner shuttling stability plots show that there is a cutoff
voltage lower than which the ion does not move out of the stem of
the T-junction, but remains stuck near zone d. By not raising the
voltages on control electrodes 8 and 17 high enough (Fig.
\ref{DC8stabplot}, \ref{DC17stabplot}), the control electrodes
cannot provide enough of a potential gradient to push the ion
through the rf barrier. Similarly, by raising the voltages on
control electrodes 9 and 16 too high (Fig. \ref{DC16stabplot},
\ref{DC9stabplot}), the shuttling sequences also fail.  Perturbing
the voltage on control electrode 9 by a large amount compared to the
other three corner electrodes will still successfully shuttle an ion
around the corner from zone d to zone i. By lowering the target
voltage on electrode 9 before the sudden voltage ramp, the control
electrode strongly attracts the ion toward zone i. However, by
making the voltage too negative, the ion is significantly pulled
away from the rf minimum making micromotion severe. Secondly, the
larger negative voltage results in a steeper potential gradient
leading from the junction region into zone i which results in the
ion acquiring more kinetic energy by roughly a factor of 2, as can
be seen in Fig.~\ref{DC9stabplot}. On the other hand, raising the
voltage on control electrode 16 to high positive values does not
yield successful shuttling (Fig. \ref{DC16stabplot}) as the high
potential prevents the ion from overcoming the rf barrier.

In Fig. \ref{DC9stabplot} and Fig. \ref{DC16stabplot}, there is a
well defined range of perturbations of electrodes 9 and 16 that
successfully shuttle an ion from zone d to zone i. On the other
hand, the perturbations of the voltages on control electrodes 8 and
17 do not exhibit a clear pattern. Notably, the failure of $V_{8,1}$
and the success of $V_{8,7}$ in Fig. \ref{DC8stabplot} and the
failures of $V_{17,1}$ , $V_{17,2}$, $V_{17,2.5}$ in Fig.
\ref{DC17stabplot} indicate that the perturbation of the voltages on
electrodes 9 and 16 (Fig. \ref{DC9stabplot}, \ref{DC16stabplot})
seems to have less of an impact on successfully shuttling an ion
from zone d to zone i than perturbing the electrode voltages of
electrodes 8 and 17 (Fig. \ref{DC8stabplot}, \ref{DC17stabplot}).
The reason for this is that once the ion has crossed the rf barrier,
the confinement in the x-direction abruptly decreases. Since the ion
is closer to control electrodes 8 and 17 at this point in its
motion, perturbations of the voltages on these electrodes may cause
the ion to go toward zone f.

The working range of voltage perturbations of control electrode 8
(Fig. \ref{DC8stabplot}) is smaller than that of control electrode
17 (Fig. \ref{DC17stabplot}). This is to be expected as the target
voltage of electrode 8 has two competing requirements. The electrode
voltage needs to be raised high in order to provide the potential
gradient to overcome the rf barrier, yet setting the voltage too
high will repel the ion away from the zone i (the ion's destination)
to zone f. Hence in designing corner shuttling sequences, special
attention should be paid to the voltage on the electrode(s) on the
inside of the turn.

In general, one way to optimize ion shuttling in a trap array with
arbitrary geometry junctions is first to pick how the potential
minimum of a trap should be moved in time according to the
considerations in section \ref{sec:theory}. Next, a numerical method
for calculating the classical motion of a trapped ion should be
selected according to section \ref{sec:overall comparison}. We found
that the Bulirsch-Stoer method was the most effective for our
geometry. The acquired kinetic energy is then plotted as a function
of the voltage perturbation. Finally, the voltage sequence is
optimized by choosing the protocol that minimizes the acquired
kinetic energy. These plots also indicate how sensitive an ion's
acquired kinetic energy is to small voltage deviations from the
ideal control sequence. Control sequences that are not overly
sensitive to voltage perturbations may have a better chance of
success (an ion is shuttled between two trapping zones) in the
presence of any stray, background electric fields. We found that
perturbing the control electrode voltages used to shuttle an ion
from zone d to zone i by $\sim$1 volt does not appreciably affect
the success rate nor does it affect the acquired kinetic energy by
more than 30\% in the specific example of a T-junction ion trap
array.

We have shown that the use of the calculated basis functions in
conjunction with numerical ODE solvers are invaluable tools for both
designing and refining shuttling protocols in two-dimensional ion
trap arrays. They help identify reasons for success and failure of
shuttling protocols, and identify key points of control for
shuttling operations. However, they are not yet precise enough to
make exact predictions about ion behavior, or guarantee that a
voltage sequence will or will not successfully shuttle an ion
through the trap. This gap between experiment and simulation can be
closed as the technology for building traps becomes more refined and
trap geometries are more reliably modeled.


\subsubsection{\label{section2}Adiabatic Corner Shuttling}

As discussed in previous sections, shuttling around the corner
inside the Michigan T-junction array \cite{T paper} is not
adiabatic. More specifically, the ion gains a significant amount of
kinetic energy during corner shuttling so it does not stay in the
same motional state and most likely will not stay in the Lamb-Dicke
regime. The main hurdle to shuttle an ion adiabatically through a
junction region is the existence of rf barriers near the junction
region.

It is always possible to overwrite any ponderomotive barrier with a
static potential so that the barrier is not present in the resulting
effective potential. This can be done by using large enough control
voltages, however, the confinement of the ion in the direction
perpendicular to the plane of the junction may be compromised for
such control voltages. Geometric static potential efficiency factors
that are unique for a particular electrode geometry will determine
whether confinement in this direction can be maintained for suitable
control voltages. If the width of the junction electrode is of
similar size or smaller than the tip-to-tip electrode separation
(200 $\mu$m for the T-junction trap, see Fig. \ref{ion trap
schematic}), we expect to obtain suitable static potential
efficiency factors for arbitrary symmetric three-layer geometries. A
more general discussion can be found in Sec. \ref{sec:trap design}.

\subsection{Ion Separation and Recombination}
The final components necessary to achieve arbitrary two-dimensional
control of the trapped ion system is the ability to separate ions
initially confined in the same trap to two different zones, and to
combine ions initially in two different zones to one trapping area.
Shuttling protocols that enable the separation and recombination of
ions in a linear trap array have been experimentally demonstrated
with near unit efficiency \cite{barrett:2004, rowe}. In addition,
theoretical design considerations have been investigated
\cite{home:2006}. Therefore, in this section we briefly review some
of the essential aspects of these protocols, present the results of
our experiments and numerical simulations, and comment on possible
methods for improvement.

Following the discussion by Home, et al. \cite{home:2006}, the
process of separating two ions may rely on the creation of an
octupole or ``double-well'' potential.  Consider two ions in a
potential of the form
\begin{equation}
\label{eq:separationpotential}
    V = 2 e \alpha x^2 + 2 e \beta x^4 + \frac{e^2}{4 \pi \epsilon_0 (2 x)}
\end{equation}
where the distance between the two ions is $(2x)$.  The first two
terms on the right-hand side are contributions of the trapping
potential $(V_T)$, while the third term is the result of the Coulomb
repulsion of the ions.  Initially, the ions are confined in a
single, nearly harmonic trap, corresponding to having $\alpha >>
\beta > 0$.  Here the contribution of the quartic term is negligible
for the relevant range of $x$, and the distance between the two ions
is given by the extrema points of Eq.~\ref{eq:separationpotential},
found to be
\begin{equation}
    2 x \approx 2^{1/3} \left( \frac{e}{32 \pi \alpha \epsilon_0} \right)^{1/3}
\end{equation}
and the trap frequency of the center of mass mode is
\begin{equation}
    \omega = \sqrt{\frac{1}{m} \frac{\partial^2 V_T}{\partial x^2}} \approx \sqrt{\frac{2 e \alpha}{m}}
\end{equation}
The formation of the octupole, which separates the two ions into
disparate regions, requires $\beta > 0 > \alpha$.  Thus, at some
point during this process, $\alpha = 0$, and it is here that the
trap frequency of the center of mass mode is minimized, as the only
contributing portion is the quartic term.  Solving for the extrema
$x_s$ of Eq.~\ref{eq:separationpotential} when $\alpha = 0$ yields
\begin{equation}
    2 x_s = \left( \frac{e}{2 \beta \pi \epsilon_0} \right)^{1/5}
\end{equation}
as the distance between the two ions, and
\begin{equation}
    \omega_s = \sqrt{\frac{1}{m} \left. \frac{\partial^2 V_T}{\partial x^2} \right|_{x_s}} =
    \sqrt{\frac{3e}{m}} \left( \frac{e}{2 \pi \epsilon_0} \right)^{1/5} \beta^{3/10}
\end{equation}
as the corresponding trap frequency of the center of mass mode
\cite{home:2006}.

\begin{figure}
\centering
\includegraphics[width=0.80\textwidth]{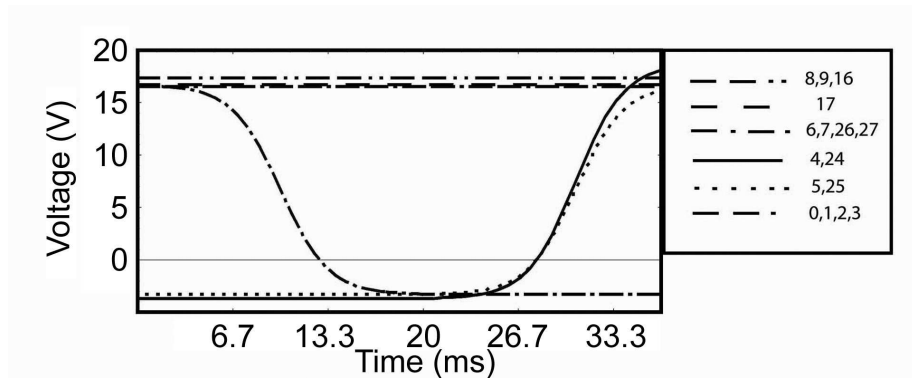}
\caption[]{Control Electrode Voltage Profile to Separate Ions:  Time
dependant voltages applied on control electrodes in order to
separate two ions initially held in a single harmonic trap at zone
$b$ into two separate harmonic traps, one at $a$ and the other at
$c$.  Asymmetries in the voltage profile about $x = 0$ are to
compensate for misalignments in the electrode layers or stray
fields.  A time reversal of the above voltage profile allows us to
combine two ions, one in trapping zone $a$ and the other at $c$,
into a single harmonic trap at zone $b$.}
\label{fig:separationvoltageprofile4}
\end{figure}

The separation protocol implemented in the Michigan T-junction ion
trap is illustrated in Fig.~\ref{fig:separationvoltageprofile4}
 \cite{T paper}. The recombination protocol was a simple reversal of this
 voltage profile. Two ions were initially confined to a common trap at zone b. The
trap is then weakened to extend over zones a, b, and c. Finally, the
double-well (octupole) potential and separation of the ions is
achieved by increasing the voltage on the electrodes that define the
center of zone b (electrodes 4, 5, 24, and 25). Numerical
simulations of this procedure indicate that the smallest trap
frequency of the center of mass mode obtained is $\sim 27$
\mbox{kHz}.  At this point, the distance between the two ions is
found to be $\sim 64$ $\mu$m.  Experimentally, the success rate of
this separation protocol was $\sim 58$ \% (64 attempts) for a total
shuttling time of 10 ms.

Undoubtedly, the axial extent of the electrodes ($400$ $\mu$m)
hindered the ability to efficiently separate ions given the 200
$\mu$m channel width. In general, the electrode widths should be of
order the channel width. Others have reported near unit efficiency
in ion separation in traps where the minimum axial extent of the
electrodes was $400$ $\mu$m  with a channel width of 400 $\mu$m
\cite{rowe}. The advent of micro-fabricated traps may allow for even
finer control of the trapped ions \cite{britton:2006, stick:2006}.
However, as indicated above, separation procedures appear to demand
that the center of mass trap frequency be reduced for the creation
of the octupole potential. Since motional heating has been shown to
be inversely proportional to the trap frequency
\cite{deslauriers:2006, turchette:2000} separation protocols may
cause some additional heating as well.


\subsection{\label{composite protocols}Composite Protocols For Arbitrary Two Dimensional Control
of Trapped Ions}
\begin{figure}[tb!]
\centerline{\epsfig{file=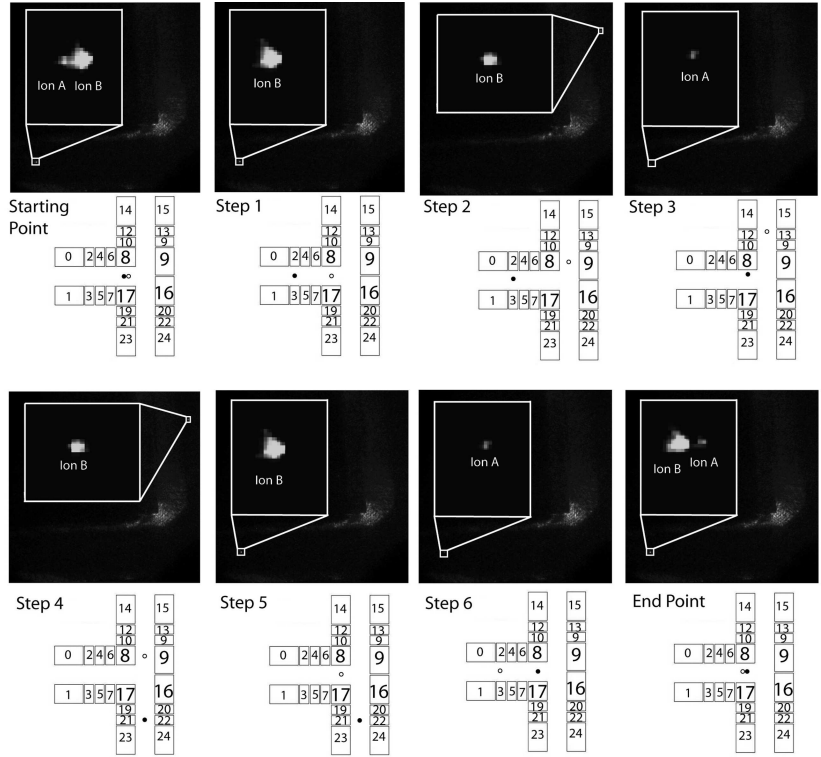, width=12cm}} 
\vspace*{15pt} \caption[]{\label{fig:swapping}Images of two ions
during the swapping protocol with schematics that indicate the ions'
positions in the trap. Two ions are initially trapped in zone d.
They scatter a different number of photons because they are
different isotopes and thus have two slightly different S$_{1/2}$ to
P$_{3/2}$  transition frequencies due to the isotope shift. The ions
make a three point turn by sending one ion around the corner one way
and the other ion around the corner in the other direction after
being separated inside the stem of the T. The two ions are brought
back in the opposite order, effectively swapping the positions of
the ions. This protocol makes use of linear shuttling, corner
shuttling, separation, and recombination.}
\end{figure}
Linear shuttling, corner shuttling, separation, and recombination
protocols may be combined to swap the positions of two ions that are
initially trapped in the same trapping zone. The ability to perform
all of these shuttling protocols in an ion trap array allows any two
arbitrary ions to be brought together because a string of ions can
be arbitrarily sorted. An application of this protocol would be the
entanglement of two arbitrary ions inside a large ion string via
two-ion quantum gates.

The step-wise process for the experimental implementation of a
swapping protocol in the T-junction ion trap array is depicted in
Fig. \ref{fig:swapping} \cite{T paper}. Two ions are initially
trapped in zone d. In order to distinguish the two ions, different
isotopes are used. The two different isotopes have different
S$_{1/2}$ to P$_{3/2}$ resonance frequencies, so the ions scatter a
different number of photons when a detection laser is incident. The
difference in photon scattering can be seen in the first panel of
Fig. \ref{fig:swapping}. The ions are shuttled to zone b where they
are separated as described in the previous section. One ion (ion A)
is shuttled to zone a while the other (ion B) is shuttled back to
zone d where it is laser cooled. Ion B is then shuttled around the
corner of the T-junction from zone d to zone i after which it is
linearly shuttled to zone k (step 3 of Fig. \ref{fig:swapping})
while ion A is shuttled from zone a to zone d where it is laser
cooled. Ion A is then shuttled to zone i (where the ion is again
laser cooled) after which it is shuttled through the junction to
zone f. This three point turn is required since shuttling ions from
zone d directly to zone f has not
 yet been accomplished (see Sec.~\ref{forward}). Ion A in zone
f is then linearly shuttled to zone h. Ion B is shuttled back from
zone k to zone i and then to zone d (step 5) where it can be laser
cooled. Finally, ion A is shuttled to zone a. Ion A is then shuttled
back from zone h to zone i where it can be laser cooled, and then
ion A is shuttled to zone d. The two ions (in zones a and d) are
then recombined in zone b and shuttled together to zone d where they
are laser cooled and imaged. The net effect is that the ions have
swapped places by executing a three point turn in the T-junction ion
trap array.

This process is carried out in successive 10 ms steps and has an
overall success rate of only 24\% (51 attempts), but this low rate
is mainly due to the the 58\% success rate of the initial separation
attempts and the final recombination attempts. Excluding the
separation step at the beginning of the swapping protocol, the
success rate of the remaining steps is 82\% (34 attempts). Other
than a failed separation, the main cause of a failed swapping
protocol is that ions can swap places during the recombination step.
Note that this protocol is carried out in successive 10 ms steps
instead of as a continuous process. The reason that the swapping
procedure was tested in a stepwise fashion was to ensure the success
of each individual step of the protocol. The quoted 82\% success
rate may also be strongly dependent on laser cooling the ions
whenever they are in zones d or i. Laser cooling serves to dissipate
any additional energy and/or spread in the energy given to the ions
during the shuttling procedures. Future work could characterize the
success rate these protocols without implementing laser cooling
during the swapping process.


\section{Ion Transport in Three Dimensions}
\begin{figure}[tb!]
\centering
\centerline{\epsfig{file=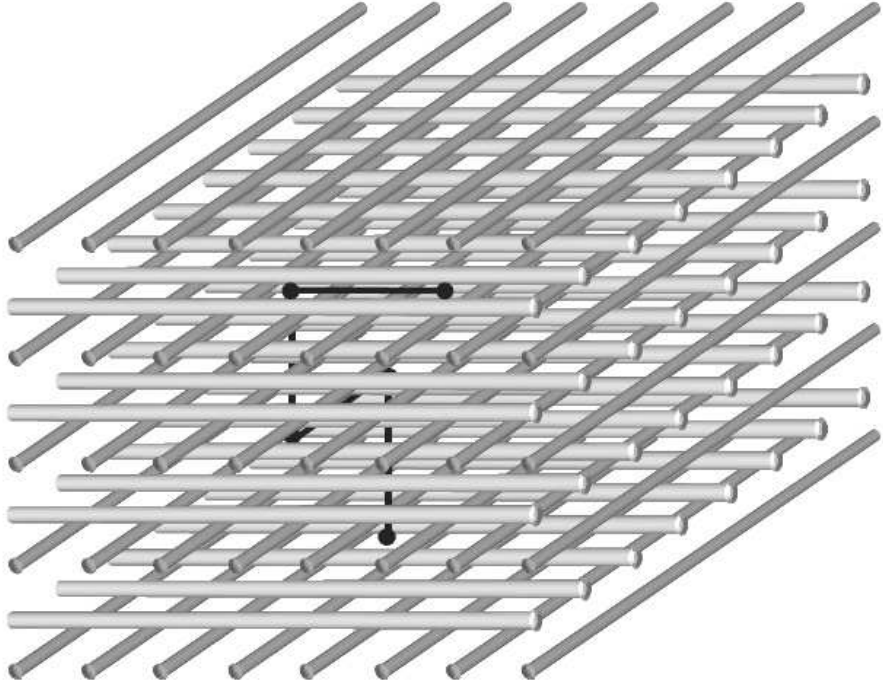, width=12cm}}
\caption[]{\label{3DTrap}Schematic of a 3-dimensional ion trap array.  Alternating layers of sets of
parallel linear conductors run perpendicular to each other.  One set of linear electrode
layers (say, the lighter shaded electrodes running left-to-right) are meant to carry rf
 potentials, while the other (darker) electrode layers are meant to carry static potentials.
Ions can be transported between any lattice point in between the grid of electrodes, in all
three dimensions.  Such a trap has been demonstrated with macroscopic charged dust particles,
with 3-dimensional shuttling through a single layer and between layers.}
\end{figure}
In principle, trapped atomic ions can be transported in three dimensions, given a suitable
trap geometry.  The main advantage here is that space is used more efficiently, and for
very large numbers of ions, using the third dimension would relax many constraints on
the physical size of the collection of trapped ions. However, there are several practical
issues concerning 3-dimensional trap channels.  First, it may be difficult to gain optical
access to submerged layers of trapping zones.  Second, a 3-dimensional junction appears to
have considerably more complexity than the 2-dimensional junctions considered earlier.
Finally, the fabrication and electrical hookups in such a design may be particular challenging,
especially if electrodes are completely submerged within the trap array.  In this section,
we nevertheless highlight a simple geometry amenable to 3-dimensional shuttling for completeness.

One example of a 3-dimensional array of ion trap zones is shown in
Fig \ref{3DTrap}. Each layer is composed of a set of parallel
line-conductors as electrodes, with alternating layers running in
perpendicular directions.  For the simplest case of three layers,
the middle layer all carry equal rf potentials, and the outer layers
carry static potentials.  For more layers, this pattern is simply
repeated.  In this geometry, ions would be confined in the spaces
between rf rails, as indicated in the figure.  In addition to linear
shuttling in a single layer, the ions can be transported through the
gaps in between the static electrodes, and shuttled in 3-dimensions.
Three-dimensional shuttling of macroscopic charged dust particles in
such a trap was demonstrated at the University of Michigan
\cite{Liz}.


\section{\label{sec:conclusions}Conclusion}

We have discussed a wide range of issues that arise when considering
the transport of atomic ions in linear and multi-dimensional ion
trap arrays. We consider methods for calculating the dynamics of
single atomic ions in the complex effective potentials arising from
the shuttling process, and justify the use of classical trajectory
treatments. We introduce the method of basis functions to
efficiently account for the contribution from each of the trap
electrodes to the electrostatic and ponderomotive potentials
experienced by an ion being shuttled in such an array. We then show
how these derived potentials can be used to obtain numerical
simulations of the classical motion of a shuttled ion, using
numerical methods such as the Bulirsch-Stoer method. Next, we
provide a general theoretical framework for the shuttling process.
This includes analytical expressions for the energy gain of the
shuttled ions, accounting for both the inertial forcing of the ion
due to changes in acceleration of the shuttling potential, and
parametric forcing resulting from changes in the frequency of the
shuttling potential. In order to minimize the added kinetic energy
imparted to the shuttled ion, these considerations lead us to
shuttling protocols that keep the shuttling potential at a fixed
frequency and transport the ion following smooth trajectories such
as a hyperbolic tangent path in time.

We pay particular attention to the implementation of junctions in
both symmetric and asymmetric electrode geometries. When shuttling
ions in two or more dimensions, an important issue we identified is
the transport of ions through junction regions. The existence of rf
holes, humps and energy barriers results in the need for
sophisticated shuttling protocols. In order to reliably guide the
ion through junctions, we find that the control electrodes near to
the junction should be smaller than the characteristic channel
width. This makes it possible to override the rf potential energy
barriers in the junction region, allowing for the use of
near-constant frequency shuttling potentials throughout the
shuttling protocol.

We use experimental results on the Michigan T-junction trap \cite{T
paper} array and theoretical analysis of the process by which ions
were shuttled through this T-junction in order to illustrate general
principles in designing advanced shuttling protocols. All
complicated shuttling processes are typically created from
elementary sequences such as linear shuttling, corner turning,
separation and recombination. We discuss the implementation of all
of these processes in the particular case of this three-layer
T-junction ion trap array and describe the implementation of a
specific composite protocol, the three-point turn, along with a
theoretical analysis.

The experiments in the T-junction trap array resulted in large
amounts of kinetic energy following shuttling.  Future work should
allow shuttling in the strict adiabatic limit, where the quantum
state of motion does not change appreciably after shuttling. This
would involve either a more appropriate design of the ion trap
electrodes to better control the ion through the junction, or the
use of sympathetic cooling or phase sensitive switching of the
trapping voltages.  At the same time, we note that adiabaticity (or
even confinement to the Lamb-Dicke limit) is not strictly necessary
in certain quantum logic gate methods.

Shuttling ions in large scale arrays may form an important backbone
for the implementation of an large-scale ion trap quantum processor.
Progress in this field is lively, with several groups now working
with various ion trap array geometries.  In the future, the
manipulation of hundreds or thousands of ions in large ion trap
arrays will pose many more challenges such as appropriate device
engineering, control electronics, motional control issues and ways
to provide adiabatic but fast shuttling sequences. Mastering these
challenges will require significant time and resources, but it is
encouraging that none of the challenges seem to be of a fundamental
nature.


\section{Acknowledgments} We acknowledge useful discussions
with M. Acton, J. Burress, M. Madsen, E.J. Otto, R. Slusher, D.
Stick, J. Sterk, and D. Wineland.  This work is supported by the
National Security Agency and the Disruptive Technology Office under
Army Research Office contract W911NF-04-1-0234, the National Science
Foundation Information Technology Research (ITR) and Physics at the
Information Frontier (PIF) Programs, and the UK Engineering and
Physical Sciences Research Council (EP/E011136/1).


\appendix
\noindent The purpose of this appendix is to illustrate the workings
of numerical solver methods that may be useful in numerically
solving Newton's equations of motion in multizone ion trap arrays.
Of particular importance is the Bulirsch-Stoer method which was used
to develop the appropriate control electrode voltage profiles to
guide ions around the corner of a T-junction ion trap \cite{T
paper}.

\section{\label{ERK}Explicit Runge-Kutta Methods}
In general, numerical differential equation solvers try to solve
first-order ordinary differential equations of the type
\begin{equation} \label{ODEformapp}
\frac{d x}{d t} = f(x,t),
\end{equation}
There are many different classes of ODE solvers with their own
merits. As a point of comparison, we introduce the familiar class of
Explicit Runge-Kutta (ERK) methods. ERK methods are useful when only
a low accuracy solution is required.  In addition, if the potential
gradient is rough or only roughly known so that the right hand side
of Eq. \ref{ODEformapp} is not smooth, ERK Methods are usually more
efficient \cite{numRecERK}. In general, a Runge-Kutta method is of
the form
\begin{equation}\label{ERK form}
x_{n+1}=x_{n}+h\Lambda(f,h,x_{n})
\end{equation}
where $\Lambda$ is a function of $x_{n}$, the step size $h$ and the
right hand side of equation \ref{ODEformapp}.

The simplest version of an ERK method is Euler's method. Euler's
method is given by
\begin{equation}\label{Euler2}
x_{n+1}=x_{n}+hf(t_{n},x_{n})
\end{equation}
The interpretation of this method is simple; assuming the step size
$h$ is small enough, the average rate of change of x over the time
interval $(t_{n},t_{n+1}=t_{n}+h)$ is nearly equal to the rate of
change at the start of the interval $\frac{dx(t_{n})}{dt}$. To
characterize the local error of the Euler method, first consider the
Taylor Expansion of $x$ about $t=t_{n}$ where $\alpha_{i}$ are the
Taylor coefficients.
\begin{equation}\label{Taylorx}
x(t_{n+1}=t_{n}+h)=x(t_{n})+\alpha_{1}h+\alpha_{2}h^{2}+...
\end{equation}
If $x_n$ is the exact solution to the differential equation then the
numerical estimation $x_{n+1}$, of $x(t_{n}+h)$ is given by
\begin{equation}\label{error}
x_{n+1}=x_{n}+h\frac{dx}{dt} =x(t_{n})+\alpha_{1}h
\end{equation}
The error that we get from a single step is
\begin{equation}\label{error2}
\epsilon=x(t_{n}+h)-x_{n+1}=\alpha_{2}h^{2}+\alpha_{3}h^{3}+...
\end{equation}
This is the local error as defined in section \ref{sec:overall
comparison}. Euler's method is accurate to first order because the
lowest order contribution to the error is proportional to $h^2$.

A possible refinement to Euler's method is to use the value of the
derivative at the middle of the time interval at
$t=\frac{t_{n}+t_{n+1}}{2}$.  However, to calculate $f$ at
$t=\frac{t_{n}+t_{n+1}}{2}$, it is necessary to calculate
$x(\frac{t_{n}+t_{n+1}}{2})$, hence we use Equation \ref{Euler2} to
take a trial step to estimate $x(t=\frac{t_{n}+t_{n+1}}{2})$.
\begin{equation}\label{MM1}
x(\frac{t_{n}+t_{n+1}}{2})=x_{n}+\frac{h}{2}f(t_{n},x_{n})=z_{1}
\end{equation}
We then use this intermediate point to calculate the rate of change
and thus estimate the next node.
\begin{equation}\label{MM2}
x_{n+1}=x_{n}+hf(t_{n}+h/2,z_{1})
\end{equation}
Note that $z_{1}$ is the result of an intermediate calculation that
can be discarded once $x_{n+1}$ is obtained. By using the Taylor
expansion of $x$, the leading term of the local error is
proportional to $h^{3}$ \cite{numRecERK}.  Therefore, this method is
accurate to second order and is called the Second Order Runge-Kutta
Method (RK2) or the Midpoint Method.

A general prescription to derive higher order Runge-Kutta methods is
found in the literature \cite{Leader}. One commonly used method is
the Fourth Order Runge-Kutta Method (RK4). This method incorporates
four intermediate steps in the process of obtaining the value of
$x_{i+1}$:
\begin{eqnarray}\label{scalarERK}
x_{i+1}&=&x_{n}+\frac{h}{3}(\frac{g_{1}}{2}+g_{2}+g_{3}+\frac{g_{4}}{2}) \\
g_{1}&=&f(t_{n},x_{n})\\
g_{2}&=&f(t_{n}+\frac{h}{2},x_{n}+\frac{g_{1}}{2}) \\
g_{3}&=&f(t_{n}+\frac{h}{2},x_{n}+\frac{g_{2}}{2}) \\
\label{ERK4}g_{4}&=&f(t_{n}+h,x_{n}+g_{3}) 
\end{eqnarray}
This method is 4th order accurate because the leading term in the
error is proportional to $h^{5}$ \cite{Leader}. In each step, RK4
does four evaluations of the function $f$ and is accurate to 4th
order.  By comparison, the Euler method does 1 evaluation of $f$ per
step and is accurate to 1st order, and RK2 does 2 evaluations of $f$
and is accurate to 2nd order.    ERK methods with order greater than
4 are therefore considered computationally inefficient because the
number of evaluations of $f(x,t)$ is greater than the leading term
in the error. RK4 has a good tradeoff in terms of accuracy and the
computational efficiency since it is the highest order ERK method in
which the accuracy order is the same as the number of evaluations of
$f$ per step \cite{Leader}, and therefore has become the standard
method for the numerical evaluation of ODEs.

Newton's equations of motion for trapped ions are a set of second
order ODEs, so in order for us to use the ERK method (or other ODE
numerical solvers), the equations need to be reformulated into a
system of 2k first order ODEs where k is the number of ions in the
system.
\begin{eqnarray}\label{def y}
\frac{d\mathbf{x_{j}}}{dt}&=&\mathbf{v_{j}}\\
\label{def
y2}\frac{d\mathbf{v_{j}}}{dt}&=&\mathbf{a}(\mathbf{x_{1}},...,\mathbf{x_{k}},t)
\end{eqnarray}
To simplify the notation, we define the following 3k-dimensional
vectors where $\mathbf{X}$ denotes the positions of the k ions,
$\mathbf{V}$ denotes the velocities of the k ions, and $\mathbf{A}$
denotes the accelerations of the k ions.
\begin{equation}\label{defXVandA}
\mathbf{X}(t)=\left(\begin{array}{c} \mathbf{x}_{1}(t)
\\\vdots
\\\mathbf{x}_{k}(t)
\end{array}\right); \hspace{.1in}
\mathbf{V}(t)=\left(\begin{array}{c} \mathbf{v}_{1}(t)
\\\vdots
\\\mathbf{v}_{k}(t)
\end{array}\right); \hspace{.1in}
\mathbf{A}(\mathbf{x},t)= \left(\begin{array}{c}
\mathbf{a}_{1}(\mathbf{x},t)
\\\vdots
\\\mathbf{a}_{k}(\mathbf{x},t)
\end{array}\right).
\end{equation}
The equations to calculate the next node are similar to Eq.
\ref{scalarERK} with the exception that the scalar quantities are
replaced by vector quantities.
\begin{equation}\label{RK4}
\left(\begin{array}{c} \mathbf{V_{n+1}}
\\\mathbf{X_{n+1}}
\end{array}\right)\\
=\left(\begin{array}{c} \mathbf{V_{n}}
\\\mathbf{X_{n}}
\end{array}\right)\\
+\frac{h}{3} \left[ \frac{1}{2} \left(\begin{array}{c}
\mathbf{g_{1}}
\\\mathbf{f_{1}}
\end{array}\right)\\+
\left(\begin{array}{c} \mathbf{g_{2}}
\\\mathbf{f_{2}}
\end{array}\right)\\+\left(\begin{array}{c}
\mathbf{g_{3}}
\\\mathbf{f_{3}}
\end{array}\right)\\+ \frac{1}{2}\left(\begin{array}{c}
\mathbf{g_{4}}
\\\mathbf{f_{4}}
\end{array}\right) \right],
\end{equation}
\[ \mbox{where} \hspace{.1in}
\left(\begin{array}{c} \mathbf{g_{1}}
\\\mathbf{f_{1}}
\end{array}\right)\\ =\left(\begin{array}{c}
\mathbf{A}(\mathbf{X_{n}},t_{n})
\\\mathbf{V_{n}}
\end{array}\right),
\]
\[
\left(\begin{array}{c} \mathbf{g_{2}}
\\\mathbf{f_{2}}
\end{array}\right)\\ =\left(\begin{array}{c}
\mathbf{A}(\mathbf{X_{n}}+\frac{1}{2}h\mathbf{f_{1}},t_{n}+\frac{h}{2})
\\\mathbf{V_{n}}+\frac{h}{2}\mathbf{g_{1}}
\end{array}\right) ,
\]
\[
\left(\begin{array}{c} \mathbf{g_{3}}
\\\mathbf{f_{3}}
\end{array}\right)\\ =\left(\begin{array}{c}
\mathbf{A}(\mathbf{X_{n}}+\frac{1}{2}h\mathbf{f_{2}},t_{n}+\frac{h}{2})
\\\mathbf{V_{n}}+\frac{h}{2}\mathbf{g_{2}}
\end{array}\right) ,
\]
\[
\left(\begin{array}{c} \mathbf{g_{4}}
\\\mathbf{f_{4}}
\end{array}\right)\\ =\left(\begin{array}{c}
\mathbf{A}(\mathbf{X_{n}}+h\mathbf{f_{3}},t_{n}+h)
\\\mathbf{V_{n}}+h\mathbf{g_{3}}
\end{array}\right)  .
\]
An adaptive step size algorithm is almost always used in conjunction
with the Runge-Kutta Method \cite{Leader}. Therefore, it is
necessary to estimate the local error in making each time-step. The
error of the nodal value $\mathbf{V_{n+1}}$, $\mathbf{X_{n+1}}$ is
estimated after each step $h$ via the method of RK pairs
\cite{Leader} which determines the difference between 4th and 5th
order RK method results for $\mathbf{V_{n}}$ and $\mathbf{X_{n}}$.
If the difference is within a pre-defined error goal, then the
solver moves on to calculate the next node. Otherwise, the step size
is reduced and the algorithm repeated until the error goal is met.
In addition, more complicated versions of the Explicit Runge-Kutta
solver also adaptively alter the order of the Runge-Kutta solver
method. For example, one step may use the RK pair RK4 and RK5, while
the next step may use RK2 and RK3.

\section{\label{BSM}Bulirsch-Stoer Method} The Bulirsch-Stoer
method is an ODE solver that yields high accuracy solutions
efficiently \cite{Bulirsch-Stoer}.  However, if a low accuracy
solution is desired or if the simulated forces on the particle are
rough or discontinuous, this method is not as effective as the ERK
\cite{numRecERK}. The Bulirsch-Stoer method proved to be the most
efficient and accurate ODE solver method for our purposes (see
Section \ref{sec:overall comparison}) and therefore, this method was
our workhorse solver.

The Bulirsch-Stoer method seeks to obtain an accurate approximation
of the nodal point $x_{i+1}$ from $x_{i}$ by first evaluating the
derivatives of the solution at $n$ points evenly spaced throughout
the time-step of size $H$, using the so-called Modified Midpoint
method. The Bulirsch-Stoer method differs from the ERK class of
methods in that this process is repeated for two or more different
values of $n$, and therefore for several different sub-step sizes,
$h = H/n$. As a result, several estimations $\chi_{i+1}(h=H/n)$ are
made for the nodal point $(x_{i+1},t_{i} + H)$, each one
characterized by the sub-step size, $H/n$. It is obvious that the
smaller the sub-step size, the more accurate the approximation for
the nodal point will be, with a limiting and presumably exact value,
$\chi_{i+1}(0),$ obtained when the number of sub-steps goes to
infinity. This limiting value would be prohibitively time-consuming
and expensive to reach directly. However, by treating the various
$\chi_{i+1}(h)$ as points on a polynomial function plotted versus
$h^2=H^2/n^2$, it is possible to make a very accurate estimation of
the value $\chi_{i+1}(0)$ by determining the y-intercept of this
polynomial. This is done through the use of Richardson Extrapolation
\cite{Bulirsch-Stoer}. The Modified Midpoint method is well-suited
for this approach because its local error is expressible as a power
series in $h^2$ rather than $h$, meaning that $\chi_{i+1}(h)$ can be
thought of as a function of $h^2$. Therefore, a reduction in
sub-step size of 1/2 will result in obtaining a point in the
polynomial four times closer to the y-intercept, making the estimate
of $\chi_{i+1}(0)$ obtained by the Richardson Extrapolation far more
accurate.  Not only so, but the fact that $\chi_{i+1}(h)$ is an even
function of the sub-step size $h$ means that this method is
inherently time-reversible. This guarantees that the solutions
obtained in this way satisfy an important constraint on solutions of
Newton's equations.

The Modified Midpoint method generates the initial crude estimates
of the nodal point $\chi_{i+1}(h)$ that will be used for the
Richardson extrapolation in order to calculate $x(t_i +H)$. The
Modified Midpoint method advances across the time interval
$(t_{i},t_{i+1})$ in a series of $n$ uniform sub-steps of size
$h=H/n$.  The equations for this method are
\begin{eqnarray}\label{ScalarMidpoint}
z_{0}&=&x_{i} \nonumber \\
z_{1}&=&z_{0}+hf(t_{i},z_{0}) \nonumber \\
z_{m+1}&=&z_{m-1}+2hf(t_{i}+m h,z_{m}) \nonumber \\
\label{ScalarMMM14}\chi_{i+1}(h)&=&\frac{1}{2}[z_{n}+z_{n-1}+hf(t_{i}+H,z_{n})]
\end{eqnarray}
Every sub-step of the Modified Midpoint method calculates a value
$z_{m}$.  These values are intermediate calculations and will be
discarded after finding the value of $x_{i+1}$. The Modified
Midpoint method outlined above is similar to the Midpoint method
given in Eq. \ref{MM2}, because in order to advance from the point
$(t_{i}+mh,z_{m})$ to the point $(t_{i}+(m+2)h, z_{m+2})$, we use
information from the derivative at the middle of the time interval,
\mbox{i.e.} $f(t_{i}+(m+1)h,z_{m+1})$, to calculate the next point.
However, the Modified Midpoint method is faster than the Midpoint
method.  If we had used the Midpoint method as our base ODE solver
with the same sub-step size $h$, we would need to do $\frac{2
H}{h}=2n$ evaluations of $f$.  In contrast, the Modified Midpoint
method only does $\frac{H}{h}+1=n+1$ such evaluations. As
evaluations of $f$ are computationally expensive, the Modified
Midpoint method is more efficient. The Modified Midpoint method has
an additional advantage as the error contains only even powers of
the step size $h$ \cite{gragg:1965}, i.e.
\begin{equation}\label{evenError}
\chi_{i+1}(h)-x(t_{i}+H)=\sum_{j=1}^{\infty}c_{j}\left(h
\right)^{2j},
\end{equation}
where the term $\sum_{j=1}^{\infty} c_{j}\left(\frac{H}{n}
\right)^{2j}$ is the error, and $x(t_{i}+H)$ is the exact solution.

A significant improvement in accuracy can be obtained by employing
the Modified Midpoint method several times over the Bulirsch-Stoer
time step, $H$, each time with a different sub-step size, and then
using each of those values to estimate the result were it to be run
with an infinitesimally small sub-step size, $h \rightarrow 0$. For
example, suppose that we have run the method $j$ times and as such
have obtained $j$ points as follows
\begin{equation}\label{npoints}
\{(h_{1}^{2},\chi(h_{1})),(h_{2}^{2},\chi(h_{2})),...,(h_{j}^{2},\chi(h_{j}))\}
\end{equation}
There is a unique polynomial of order $j-1$ that passes through each
of the $j$ points.  For example, with two points, there is a unique
linear polynomial, for three points, there is a unique quadratic,
etc. This polynomial is given by Lagrange's Formula
\begin{eqnarray}\label{Lagrange Formula}
P(h^{2})&=&\frac{(h^{2}-h_{2}^{2})(h^{2}-h_{3}^{2})...(h^{2}-h_{j}^{2})}{(h_{1}^{2}-h_{2}^{2})(h_{1}^{2}-h_{3}^{2})...(h_{1}^{2}-h_{j}^{2})}\chi(h_{1})+
\frac{(h^{2}-h_{1}^{2})(h^{2}-h_{3}^{2})...(h^{2}-h_{j}^{2})}{(h_{2}^{2}-h_{1}^{2})(h_{2}^{2}-h_{3}^{2})...(h_{2}^{2}-h_{j}^{2})}\chi(h_{2})+...\nonumber
\\&+&\frac{(h^{2}-h_{1}^{2})(h^{2}-h_{2}^{2})...(h^{2}-h_{j-1}^{2})}{(h_{j}^{2}-h_{1}^{2})(h_{j}^{2}-h_{3}^{2})...(h_{j}^{2}-h_{j-1}^{2})}\chi(h_{j})
\end{eqnarray}
In order to complete the Richardson extrapolation, we need only read
off the value $P(0)$ from this formula to obtain the next node. We
note that equation \ref{Lagrange Formula} is a well-defined
polynomial in $h^{2}$ due to equation \ref{evenError}. The advantage
in doing so is that it gives more weight to the points with smaller
$h$ in determining the value of $P(0)$ and greatly improves the
accuracy of that calculation.

However, Eq. \ref{Lagrange Formula} has two disadvantages. First, it
provides information about the entire polynomial whereas we are only
interested in the value of the polynomial at $h^{2}=0$.  Secondly,
Eq. \ref{Lagrange Formula} provides no error estimate
\cite{Neville}. Therefore, Neville's Algorithm \cite{Neville} is
used instead of Lagrange's formula in the Bulirsch-Stoer algorithm
to extrapolate the value of $\chi_{i+1}$ at $h^2 = 0$. Given our set
of $j$ points, Neville's Algorithm finds the value of $P(0)$
directly from the points obtained through the use of the Modified
Midpoint Method, making Neville's algorithm faster. In addition,
Neville's algorithm also provides a simple estimate of the  error in
the resulting value.

To illustrate Neville's algorithm, we will fit the following three
points to a second order polynomial and find the value of that
polynomial at $h^{2}=0$. The three points are {$(0.02250, 1.34838),
(5.625\times 10^{-3}, 1.34948), (2.500 \times 10^{-3}, 1.34969)$}.
We begin by defining $P_{(1)}$ to be the value at $h^{2}=0$ of the
unique zeroth-order polynomial (horizontal line) passing through the
first point $P_{(1)}=1.34838$.  $P_{(2)}$ and $P_{(3)}$ are defined
similarly, so that $P_{(2)}=1.34948$ and $P_{(3)}=1.34969$. Next we
define $P_{(1)(2)}$ to be the value at $h^2=0$ of the unique
first-order polynomial that passes through the first two points.
This value can be obtained from $P_{(1)}$ and $P_{(2)}$ using the
following iterative equation:
\begin{equation}\label{Neville Alg}
P_{(i)(i+1)...(i+m)}=\frac{-h_{i+m}^{2}P_{i(i+1)...(i+m-1)}+h_{i}^{2}P_{(i+1)(i+2)...(i+m)}}{h_{i}^{2}-h_{i+m}^{2}}
.\end{equation} From Eq. \ref{Neville Alg} we can also derive the
value of $P_{(2)(3)}$, the value at $h^2=0$ of the first order
polynomial that passes through the second and third points. Our
final value, $P_{(1)(2)(3)}$, is the value at $h^{2}=0$ of the
unique second-order polynomial that passes through all three points.
Again, $P_{(1)(2)(3)}$ can be derived from $P_{(1)(2)}$ and
$P_{(2)(3)}$ via equation \ref{Neville Alg}. An error estimate comes
from the difference between the highest order estimate of $P(0)$ and
the next highest order estimate of $P(0)$. For example, the error
estimate of $P_{(1)(2)(3)}$ is given by
\begin{equation}\label{error estimation}
\epsilon = \frac{max\{\vert P_{(1)(2)(3)}-P_{(1)(2)}\vert ,\vert
P_{(1)(2)(3)}-P_{(2)(3)}\vert \}}{\vert P_{(1)(2)(3)}\vert } =6.4
\times 10^{-6}.
\end{equation} Another advantage to Neville's algorithm is that if we need
to fit a new point into our polynomial, we need not recalculate all
the values of $P$ from scratch.  For example, if we add a fourth
point to the above three points, we would only need to calculate
$P_{(4)}$, $P_{(3)(4)}$, $P_{(2)(3)(4)}$ and $P_{(1)(2)(3)(4)}$.
This improves the overall efficiency of the Bulirsch-Stoer Method.

\begin{figure}[tb!]
\centerline{\epsfig{file=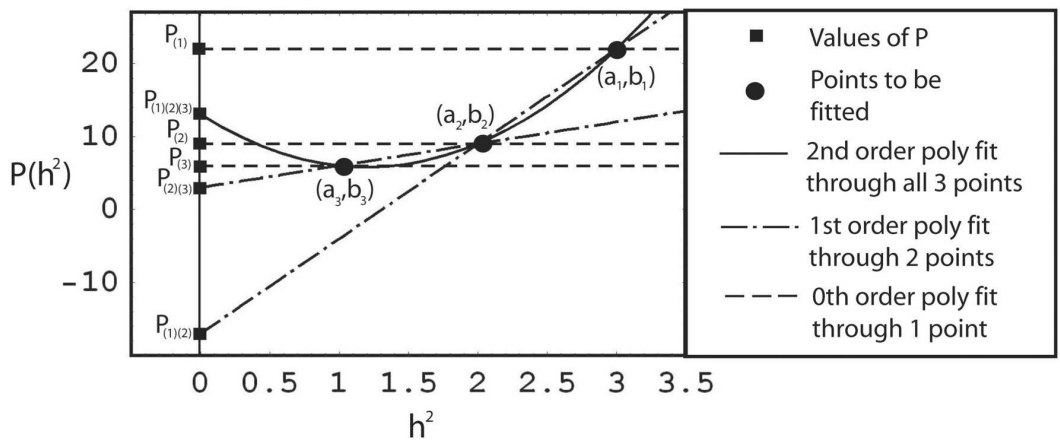, width=14cm}} 
\vspace*{15pt} \caption[]{\label{NevilleExample}We demonstrate
Neville's algorithm for three random points
$\{(a_{1},b_{1}),(a_{2},b_{2}),(a_{3},b_{3})\}$.  The y intercept of
the six polynomials give us the various $P$'s.  Finally, Neville's
algorithm does not actually calculate the interpolating polynomials
but only the y intercepts of the polynomials.  The interpolating
polynomials are added for illustrative purposes.}
\end{figure}

We are now ready to employ the Bulirsch-Stoer method to solve
ordinary differential equations. We begin by finding rough estimates
of $x(t_{i+1})$ with the Modified Midpoint Method using $n=2$
sub-steps and then $n=4$ sub-steps (Eq. \ref{ScalarMMM14}). Next, we
use Neville's algorithm to calculate $\chi_{i+1}(0),$ and estimate
the error (Eqs. \ref{Neville Alg} and \ref{error estimation}). If we
are within our error goals, we record the result and go on to
calculate the next node. Otherwise, we obtain a third point,
$(H^2/6^2,\chi_{i+1}(H/6)$ using the Modified Midpoint method and
repeat Neville's algorithm using a second-order polynomial and check
to see whether the error goal has been met. The sequence of numbers
of sub-steps is $n=2,4,6,8,10,12,14,16,\ldots$. If the solution does
not meet the error goals beyond a certain value of $n$, this would
indicate that there is some unusual behavior within the time
interval of the Bulirsch-Stoer step. Therefore, this sequence is
usually terminated at the 8th iteration which corresponds to $n=16$
sub-steps. At that point, $H$ is reduced (usually halved) and  the
above procedure is repeated. A more detailed discussion of the
adaptive step size algorithm is provided in the literature
\cite{numRecERK}.

We now illustrate the Bulirsch-Stoer method by numerically solving a
simple differential equation $\dot{x}(t)=x(t)$ with $x(0)=1$ so that
$f(t,x)=x$. The exact solution to this differential equation is
$x(t)=e^{t}$. For our example, we will take one large Bulirsch-Stoer
step from $t=0$ to $t=0.3$, so that $H=0.3$.  The result of this
calculation will thus be a numerical estimate of the exact solution
$x(t=0.3)$. We note that since the first node is given by initial
conditions, $(t_{0}=0,x_{0}=1)$, we are trying to calculate the
value $x_{1}\approx x(t=0.3)$.

Following the three-step algorithm that was outlined above, we first
use the Modified Midpoint method to find $\chi_{1}(H/2) = 1.34838$
and $\chi_{1}(H/4)=1.34948$, both evaluated to six significant
figures. Note that the values used in our example of Neville's
algorithm are taken directly from this sample problem. Applying
Richardson Extrapolation to these points, we obtain our first
solution $x_{1}=P_{(1)(2)}=1.34985$ to six significant figures. We
know that $e^{0.3}=1.34986 $, to six significant figures, so the
fractional difference of $\chi_{1}(H/2)$ from the exact solution is
$1.1 \times 10^{-3}$, while the fractional difference of
$\chi_{1}(H/4)$ from the exact solution is $2.8\times 10^{-4}$.
However, by applying Richardson Extrapolation to these two points to
estimate $\chi_{1}(0)$, an extrapolated value is obtained which has
a fractional difference from the exact result of $6.4\times10^{-6}$.
Thus, the extrapolation yields a result that is forty times more
accurate than our raw estimates from the Modified Midpoint Method.
Using the error estimation scheme given in Eq. \ref{error
estimation}, we estimate the fractional error to be
$1.1\times10^{-3}$. If this error is not acceptable, we then proceed
to find the value of $\chi(H/6)=1.34969$ (six significant figures).
The value $\chi_{1}(H/6)$ has a fractional difference of
$1.2\times10^{-4}$ from the exact answer. But when we apply
Richardson Extrapolation to all three points, we obtain a value for
$\chi_1 (0)$ of $1.34986$, (six significant figures). This has a
fractional difference from the exact value of $1.5\times10^{-8}$.
This result is roughly 10,000 times more accurate than our best
$\chi_{1}$ value obtained directly from the Modified Midpoint
Method, $\chi_{1}(H/6)$. The fractional local error estimate is
given again by Eq. \ref{error estimation} and is $6.4\times
10^{-6}$. We note that in order to reach an accuracy of one part in
$10^{8}$, the Bulirsch-Stoer algorithm only performs 12 evaluations
of $f$. In order to reach the same accuracy using the Euler Method,
we would need $3\times 10^6$ evaluations of $f$!
\begin{table}[tb!]
\centering
\begin{tabular}{|c|c|c|}
\hline
Substep size & Fract. diff.        & Fract. diff after \\
{}           & from Mod. Mid. Meth.  & Richardson Extrapolation \\
\hline
$h=H/2$&$1.1\times10^{-3}$&NA \\
\hline
$h=H/4$&$2.8\times10^{-4}$&$6.4\times10^{-6}$\\
\hline
$h=H/6$&$1.2\times10^{-4}$&$1.5\times10^{-8}$ \\
\hline
\end{tabular}
\caption[]{The first column refers to the fractional difference
obtained in the numerical estimation of $x(t=0.3)$ and the exact
solution.  The second column shows the fractional difference after
Richardson Extrapolation is applied.  Notice that Richardson
Extrapolation provides a huge increase in accuracy}
\end{table}

To generalize the Bulirsch-Stoer method to solve Newton's Equation
for multiple ions (Eq. \ref{Equation to be solved}), we replace
scalar quantities with vector quantities as was done for the ERK method.
Given a node $(t_{i},\mathbf{V_{i}},\mathbf{X_{i}})$, we would like
to calculate the next node
$(t_{i+1}=t_{i}+H,\mathbf{V_{i+1}},\mathbf{X_{i+1}})$ with
$\mathbf{V_{i+1}}$ approximating the exact velocities
$\mathbf{V}(t_{i}+H)$ and $\mathbf{X_{i+1}}$ approximating the exact
ion positions $\mathbf{X}(t_{i}+H)$.
\begin{equation}\label{MMM1}
\left(\begin{array}{c} \mathbf{\upsilon_{0}}
\\\mathbf{\Xi_{0}}
\end{array}\right)\\=\left(\begin{array}{c} \mathbf{V}(t_{i})
\\\mathbf{X}(t_{i})
\end{array}\right)\\
\end{equation}
\begin{equation}\label{MMM2}
\left(\begin{array}{c} \mathbf{\upsilon_{1}}
\\\mathbf{\Xi_{1}}
\end{array}\right)\\
=\left(\begin{array}{c} \mathbf{\upsilon_{0}}
\\\mathbf{\Xi_{0}}
\end{array}\right)\\+h\left(\begin{array}{c}
\mathbf{A}(\mathbf{\Xi_{0}},t_{i})
\\\mathbf{\upsilon_{0}}
\end{array}\right)\\
\end{equation}
\begin{equation}\label{MMM3}
\left(\begin{array}{c} \mathbf{\upsilon_{m+1}}
\\\mathbf{\Xi_{m+1}}
\end{array}\right)\\=\left(\begin{array}{c} \mathbf{\upsilon_{m-1}}
\\\mathbf{\Xi_{m-1}}
\end{array}\right)\\
+2h\left(\begin{array}{c} \mathbf{A}(\mathbf{\Xi_{m}},t_{i}+mh)
\\\mathbf{\upsilon_{m}}
\end{array}\right)
\end{equation}
\begin{equation}\label{MMM4}
\\\left(\begin{array}{c}
\mathbf{\tilde{X}_{i+1}}(H^{2}/n^{2})
\\\mathbf{\tilde{V}_{i+1}}(H^{2}/n^{2})
\end{array}\right)=\frac{1}{2}\left[\left(\begin{array}{c} \mathbf{\upsilon_{n}}
\\\mathbf{\Xi_{n}}
\end{array}\right)\\+\left(\begin{array}{c} \mathbf{\upsilon_{n-1}}
\\\mathbf{\Xi_{n-1}}
\end{array}\right)\\
+h\left(\begin{array}{c} \mathbf{A}(\mathbf{\Xi_{n}},t_{i}+H)
\\\mathbf{\upsilon_{n}}
\end{array}\right)\right ]
\end{equation}
$\mathbf{A}$ refers to the acceleration of the $k$ ions, and
$\mathbf{\Xi_{i}}$, $\mathbf{\upsilon_{i}}$ are intermediate vectors
that the modified midpoint method calculate at each sub-step.  Once
the calculation of each $\mathbf{X_{i+1}}$ and $\mathbf{V_{i+1}}$ is
complete, the values $\mathbf{\Xi_{i}}$, $\mathbf{\upsilon_{i}}$
will be discarded.

$\mathbf{\tilde{X}_{i+1}}(H^{2}/n^{2})$ is the desired approximation
to the exact solution $\mathbf{X}(t=t_{i}+H)$ and
$\mathbf{\tilde{V}_{i+1}}(H^{2}/n^{2})$ is the desired approximation
to the exact solution $\mathbf{V}(t=t_{i}+H)$.
$\mathbf{\tilde{X}_{i+1}}$ and $\mathbf{\tilde{V}_{i+1}}$ are
functions of $h^{2}$ where $h=H/n$ is the Modified Midpoint Method
step size. Polynomial interpolation is carried out using Neville's
Algorithm (Eq. \ref{Neville Alg}). However, each $P(h^2)$ is now a
$6k$-dimensional vector representing intermediate estimates of the
position and velocity vectors. Finally, we need to generalize the
error estimation scheme in order to implement our adaptive step size
algorithm.
\begin{eqnarray}
\mathbf{\epsilon}=\frac{max\{\vert\mathbf{P}_{(i)(i+1)\ldots
(i+m)}-\mathbf{P}_{(i)(i+1)(i+m-1)}\vert   , \vert
\mathbf{P}_{(i)(i+1)\ldots (i+m)} - \mathbf{P}_{(i+1)(i+2)\ldots
(i+m)}\vert \}    } {\vert \mathbf{P}_{(i)(i+1)\ldots (i+m)}\vert}
\end{eqnarray}


\section{\label{BDF}Stiff Systems and Backward Difference Formulas}
The differential equation for a system of ions moving in an array of
ion traps may be stiff as was discussed in Section
\ref{CaluculatingIonDynamics}. One way that stiffness can occur is
if there are two very different time scales that govern the
evolution of the solution to a differential equation. An example of
a system of ODEs that is stiff is given by Press \cite{numRecERK}:
\begin{eqnarray}\label{Stiff System}
u^{\prime}(t)&=&998u+1998v \nonumber \\
v^{\prime}(t)&=&-999u-1999v
\end{eqnarray}
The exact solutions to this system of ODEs with initial conditions
$u(0)=1$ and $v(0)=0$  are
\begin{eqnarray}
u(t)&=&2e^{-t}-e^{-1000t} \nonumber \\
v(t)&=&-e^{-t}+e^{-1000t}
\end{eqnarray}
The two terms in the solutions for $u(t)$ and $v(t)$ have vastly
different timescales. Just after $t=0$, the $e^{-1000t}$ term
dominates the evolution of the system, but near $t=1$, this term
becomes negligible.

In order to illustrate the difficulties an explicit ODE solver has
when handling a stiff system, we use the Euler Method, Eq.
\ref{Euler2} to solve Eq. \ref{Stiff System}. In matrix form, we may
write this as
\begin{equation}\label{Euler}
\left(\begin{array}{c} u_{n+1}
\\v_{n+1}
\end{array}\right)\\
=\left(\begin{array}{c} u_{n}
\\v_{n}
\end{array}\right)\\
- h \left(\begin{array}{cc} -998&-1998
\\999&1999
\end{array}\right)\\
\left(\begin{array}{c} u_{n}
\\v_{n}
\end{array}\right)\\
\equiv(\mathbf{1}-h\mathbf{C})\left(\begin{array}{c} u_{n}
\\v_{n}
\end{array}\right)\\
\end{equation}
This simplifies to
\begin{equation}\label{Eulersimp}
\left(\begin{array}{c} u_{n}
\\v_{n}
\end{array}\right)\\=(\mathbf{1}-h\mathbf{C})^{n}\left(\begin{array}{c} u_{0}
\\v_{0}
\end{array}\right)\\
\end{equation}
Since the solutions for $u(t)$ and $v(t)$ both approach zero as a
steady state solution, the numerical solution should also approach
zero as a steady state solution.  If the numerical solution exhibits
any behavior that is qualitatively different, the numerical ODE
solver is clearly be an unstable method.  In order for the numerical
solutions of Eq. \ref{Stiff System} to approach 0,
$(\mathbf{1}-h\mathbf{C})^n$ must converge to the zero matrix as $n
\rightarrow \infty$ or equivalently \cite{numRecERK}
\begin{equation}\label{h upperbound}
h<\frac{2}{|\lambda_{max}|}=\frac{2}{1000}
\end{equation}
where $\vert \lambda_{max} \vert$ is the largest absolute value of
the eigenvalues of the matrix $\mathbf{C}$.  The expression in Eq.
\ref{h upperbound} provides a strict upper-bound for the step size
to be used when obtaining a numerical solution to the problem that
is due entirely to stability concerns and not those of local error
tolerances. For example, at $t=10$, we set our local error goal (the
relative numerical error from one numerical step to the next, see
equation \ref{error goal}) to be $\epsilon=10^{-6}$.  To estimate
the largest step size that we could take, we calculate a numerical
estimate for $u$ and $v$ at $t=10+h$.  This is derived from taking a
single Euler step of size $h$ from the exact solution of Eq.
\ref{Stiff System} at $t=10$, i.e.
\begin{equation}\label{EulerLoCalErrorStuff}
\left(\begin{array}{c} u(10+h)
\\v(10+h)
\end{array}\right)\\
\approx\left(\begin{array}{c} u(10)
\\v(10)
\end{array}\right)\\
- h \left(\begin{array}{cc} -998&-1998
\\999&1999
\end{array}\right)\\
\left(\begin{array}{c} u(10)
\\v(10)
\end{array}\right)\\
\end{equation}
The LHS of Equation \ref{EulerLoCalErrorStuff} is the exact solution
whilst the RHS is a numerical estimate of the exact solution at
t=10+h. Therefore the difference between the LHS and RHS of Equation
\ref{EulerLoCalErrorStuff} is the local error. Using equation
\ref{EulerLoCalErrorStuff}, we find that the largest step size
allowable is $h=0.1$. However, equation \ref{h upperbound} implies
that we \emph{cannot} take such a large step size as the qualitative
behavior of the numerical solution (in our example, the long term
behavior) will vastly differ from the exact solution and hence the
Euler method will be unstable. Therefore, stiffness results in a
loss of computational efficiency.

One way to to address this problem is to use implicit numerical
methods\footnote{An implicit numerical method is one where the node
to be calculated does not depend explicitly on previously determined
quantities,  i.e. we \emph{do not} use a function $\Lambda$ such
that
$x_{i+1}=\Lambda(f;x_{i},x_{i-1},...,x_{0},t_{i},t_{i-1},...t_{0})$}\,\,
\cite{Shampine Gear SIAM review}. Although implicit numerical
methods are generally more stable \cite{numRecERK}, it is more
difficult to solve an implicit equation and this increases
computational cost. For example, the Implicit Euler Method in one
dimension is defined as follows
\begin{equation}\label{Imp Euler 1d}
x_{n+1}=x_{n}+hf(t_{n+1},x_{n+1})
\end{equation}

To illustrate how an implicit method is more robust than a typical
explicit method, let us use the Implicit Euler Formula, \ref{Imp
Euler 1d}, to solve the example given in \ref{Stiff System}
\begin{equation}\label{Imp Euler}
\left(\begin{array}{c} u_{n+1}
\\v_{n+1}
\end{array}\right)\\
=\left(\begin{array}{c} u_{n}
\\v_{n}
\end{array}\right)\\
- h \left(\begin{array}{cc} -998&-1998
\\999&1999
\end{array}\right)\\
\left(\begin{array}{c} u_{n+1}
\\v_{n+1}
\end{array}\right)\\
\end{equation}
which simplifies to
\begin{equation}\label{ImpEulersimp}
\left(\begin{array}{c} u_{n}
\\v_{n}
\end{array}\right)\\=(\mathbf{1}+h\mathbf{C})^{-n}\left(\begin{array}{c} u_{0}
\\v_{0}
\end{array}\right)\\
\end{equation}
The eigenvalues of the matrix $(\mathbf{1}+h\mathbf{C})$ are
$1/(1+h\lambda)$ which is always less than unity regardless of $h$,
thus the matrix $(\mathbf{1}+h\mathbf{C})^{-n}$ will converge to the
zero matrix as $n \rightarrow \infty$ regardless of $h$.  Therefore,
the Implicit Euler Method is more robust in this example. It also
turns out that implicit methods give better stability for general
ODEs \cite{numRecERK}. However, the price to be paid for such stable
behavior is that at every step, one needs to solve an implicit
equation.

The Implicit Euler Method is the simplest member of the class of ODE
solvers known as Backward Difference Formulae. The essential idea of
the Backward Difference Formulae is to use polynomial extrapolation
on previously calculated nodes to estimate the next node.  For
example, the second order Backward Difference Formula as used to
solve equation \ref{Equation to be solved} is
\begin{equation}\label{BDF2}
\left(\begin{array}{c} \mathbf{V_{n+1}}
\\\mathbf{X_{n+1}}
\end{array}\right)\\
=\frac{4}{3}\left(\begin{array}{c} \mathbf{V_{n}}
\\\mathbf{X_{n}}
\end{array}\right)\\
- \frac{1}{3}\left(\begin{array}{c} \mathbf{V_{n-1}}
\\\mathbf{X_{n-1}}
\end{array}\right)\\
+\frac{2}{3}  \left(\begin{array}{c}
\mathbf{A}(\mathbf{X_{n+1}},t_{n}+h)
\\\mathbf{V}_{n+1}
\end{array}\right)\\
\end{equation}
Because it is implicit, each Backward Difference Formula step is
usually more computationally expensive than a step in an explicit
ODE solver like the Modified Midpoint Method.  Nevertheless, when
solving  a stiff system it is usually better to use an implicit
method rather than an explicit solver as the implicit method
requires far fewer steps. The reduction in the number of steps has a
stronger impact than the increased computational expense of an
implicit method for each step.  Therefore, when stiffness in an ODE
is detected or suspected, the designer should switch to the Backward
Difference Formulae. For example, Mathematica's ``NDSolve" uses the
Adams Predictor Corrector Method by default and switches to a
Backward Difference formula with an adaptive step-size and an
adaptive order when stiffness is detected.
\end{document}